\documentclass[prd,aps,tightline,twocolumn,
nofootinbib,superscriptaddress]{revtex4}
\usepackage{amsmath}
\usepackage{booktabs}
\usepackage{amssymb}
\usepackage{latexsym}
\usepackage{wrapfig}

\usepackage[pdftex]{graphicx,color}
\usepackage{epstopdf}
\usepackage[bookmarks=true,bookmarksnumbered=true,bookmarkstype=toc]{hyperref}

\usepackage{fancybox}
\usepackage{color}

\newcommand{\lsim}
{\raise0.3ex\hbox{$\;<$\kern-0.75em\raise-1.1ex\hbox{$\sim\;$}}}
\newcommand{\gsim}
{\raise0.3ex\hbox{$\;>$\kern-0.75em\raise-1.1ex\hbox{$\sim\;$}}}

\begin{document}

\begin{flushleft}
STUPP-13-215, KEK-TH-1667 \hspace{50mm} 
\today
\end{flushleft}

\title{ 
A first evidence of the CMSSM is appearing soon
}

\author{Yasufumi Konishi}
\affiliation{Department of Physics, Saitama University, 
     Shimo-okubo, Sakura-ku, Saitama, 338-8570, Japan}

\author{Shingo Ohta}
\affiliation{Department of Physics, Saitama University, 
     Shimo-okubo, Sakura-ku, Saitama, 338-8570, Japan}
     
\author{Joe Sato}
\affiliation{Department of Physics, Saitama University, 
     Shimo-okubo, Sakura-ku, Saitama, 338-8570, Japan}
     
\author{Takashi Shimomura}
\affiliation{Department of Physics, Niigata University, Niigata, 950-2181, Japan} 
\affiliation{Max-Planck-Institut fur Kernphysik, Saupfercheckweg 1, D-69117 Heidelberg, Germany}

\author{Kenichi Sugai}
\affiliation{Department of Physics, Saitama University, 
     Shimo-okubo, Sakura-ku, Saitama, 338-8570, Japan}    

\author{Masato Yamanaka}
\affiliation{Theory Center, Institute of Particle and Nuclear Studies,
KEK (High Energy Accelerator Research Organization),
1-1 Oho, Tsukuba 305-0801, Japan}
\affiliation{Department of Physics, Nagoya University, Nagoya 464-8602, Japan}

\vskip 0.15in

\begin{abstract}
We explore the coannihilation region of the constrained minimal 
supersymmetric standard model (CMSSM) being consistent with 
current experimental/observational results. 
The requirements from the experimental/observational results are 
the 125GeV Higgs mass and the relic abundances of both the dark 
matter and light elements, especially the lithium-7. 
We put these requirements on the caluculated values, and thus 
we obtain allowed region.  
Then we give predictions to the mass spectra of the SUSY particles, 
the anomalous magnetic moment of muon, branching fractions of 
the $B$-meson rare decays, the direct detection of the neutralino 
dark matter, and the number of SUSY particles produced in 14TeV 
run at the LHC experiment.  
Comparing these predictions with current bounds, we show the 
feasibility of the test for this scenario in near future experiment. 
\end{abstract}

\maketitle

\section{Introduction}  

The challenges of the LHC are to discover of the Higgs boson and to search for 
new physics beyond the standard model (SM). The discovery of the Higgs boson 
was reported by the ATLAS~\cite{2012gk} and the CMS collaborations~\cite{2012gu}. 
Meanwhile no signals of new physics have been observed so far from the LHC. 
However the nature indicates the existence of physics beyond the SM that accounts for the 
shortcomings in the SM, e.g., no candidate of dark matter, the origin of neutrino 
mass, the baryon asymmetry of the universe, and so on.

It is a challenge to confirm the new physics from only experimental/observational 
data. Theoretical studies in advance are essential to identify the signatures of new 
physics. We now have measurement data of the Higgs boson, the relic 
abundance of the dark matter, and so on. What we should do, therefore, is to precisely 
extract probable parameter space, and predict the signatures in each scenario of new 
physics by using the measurement data.

One of the leading candidates of the new physics is supersymmetric (SUSY) extension. 
A scenario in the extension is the constrained minimal supersymmetric SM (CMSSM) 
which is simple but phenomenologically successful framework~\cite{King:1995vk, 
Buchmueller:2011ab, Kadastik:2011aa}.  
In the CMSSM, all of the observables are described by only five parameters, and these 
parameters are tightly connected with the property of the Higgs boson. The 
reported mass of the Higgs boson, $m_h \simeq 125$GeV, suggests heavy SUSY 
particles~\cite{Okada:1990vk, Ellis:1990nz, Haber:1990aw, Casas:1994us}. 
This suggestion is consistent with a null signal of exotics at the LHC. 
The heavy SUSY particles imply the heavy dark matter in this framework, because 
the lightest SUSY particle (LSP) works as the dark matter.

Cosmological and astrophysical measurements confirmed the existence of dark matter, 
and numerical simulations suggest that weakly interacting massive particles (WIMPs) are 
the most feasible candidate for the dark matter~\cite{Hinshaw:2012fq, Clowe:2006eq, 
Navarro:1995iw, Springel:2008cc, BoylanKolchin:2009nc, Guo:2010ap}. 
In the CMSSM with R-parity conservation, the bino-like neutralino 
is the LSP and consequently is a WIMP dark matter candidate. 
The measured abundance of the neutralino dark matter can be acquired in the 
coannihilation region\footnote{The measured abundance is obtained also 
in the focus-point region in which measured abundance of dark matter can be 
reproduced by the large mixing of the bino and the Higgsino 
components~\cite{Feng:1999mn, Feng:1999zg}. The latest XENON100 dark 
matter search, however, excludes most of parameter spaces of the focus-point 
region~\cite{Aprile:2011hi, Buchmueller:2012hv}}. 
The heavy neutralino dark matter requires large coannihilation rate. The large 
coannihilation rate sufficiently reduces the relic number density of the neutralino, 
and can reproduce the measured abundance of dark matter. The large 
coannihilation rate needs the tight degeneracy in mass between the neutralino 
LSP and the stau NLSP (NLSP: next to the lightest SUSY particle)~\cite{Griest:1990kh, 
Edsjo:1997bg}. 
Indeed in large part of parameter space wherein the mass of the 
Higgs boson is consistent with the reported one, the mass difference between 
the neutralino LSP and the stau NLSP is smaller than the mass of tau 
lepton, $m_\tau$~\cite{Aparicio:2012iw, Citron:2012fg}. 
Such a tight degeneracy makes the stau NLSP to be long-lived charged massive 
particle (CHAMP)~\cite{Profumo:2004qt, Jittoh:2005pq}.

The long-lived CHAMPs can modify the chain of nuclear reactions in a stage of 
big-bang nucleosynthesis (BBN), and hence distort the primordial abundances of 
light elements. 
The success and failure of the nucleosynthesis are quite sensitive to the property 
of the long-lived CHAMPs, e.g., the lifetime, the number density, the electric charge, 
and so on~\cite{Pospelov:2006sc, Kohri:2006cn, Kaplinghat:2006qr, 
Cyburt:2006uv, Hamaguchi:2007mp, Kawasaki:2007xb, Bird:2007ge, 
Jittoh:2007fr, Kawasaki:2008qe, Jittoh:2008eq, Jedamzik:2007cp, 
Pradler:2007is, Pospelov:2008ta, Kamimura:2008fx, Kusakabe:2008kf, 
Bailly:2008yy, Bailly:2009pe, Jittoh:2010wh, Kusakabe:2010cb, Jittoh:2011ni, 
Jedamzik:2004er, Kawasaki:2004yh, Kawasaki:2004qu, Cumberbatch:2007me, 
Kohri:2008cf, Cyburt:2009pg, Pospelov:2010cw, Kawasaki:2010yh, Kohri:2012gc, 
Kusakabe:2012ds, Pospelov:2010hj, Kusakabe:2013tra}. 
The success of the nucleosynthesis means to reproduce the measured abundances 
of light elements. Notice that measured abundance of the lithium-7 ($^7$Li) is 
reported to be inconsistent with the theoretical prediction in the standard BBN; 
the measured one is 
$^7\text{Li/H} = 1.48 \pm 0.41 \times 10^{-10}$~\cite{Monaco:2010mm} 
and the theoretical prediction is $^7\text{Li/H} = 5.24 \times 
10^{-10}$~\cite{Coc:2011az}. This inconsistency is known as the lithium-7 
problem~\cite{Spite:1982dd}, and the success of the nucleosynthesis includes 
also solving the problem. 
It is important to emphasize that the success of the nucleosynthesis constrains and 
predicts the property of the stau NLSP.

The purpose of this work is to give theoretical clues to experimentally identify the 
CMSSM as the new physics in the light of the natures of the Higgs boson, the 
measured abundance of dark matter, and the success of the nucleosynthesis. 
We concentrate on the scenario wherein the mass difference of the neutralino LSP 
and the stau NLSP is smaller than the mass of tau lepton, and the longevity 
of the stau is guaranteed by the tight phase space.

Both the mass and the signal strength of the Higgs boson are correlated with 
parameters of the stop sector. So it sheds light on the sign($\mu$), $m_0$, 
$A_0$, and $\tan\beta$.  The predictability of the derived relations, however, 
is not so strong because of too large degrees of freedom of their parameter space. 
Besides the property of the long-lived stau is predicted/constrained from the 
nucleosynthesis, and is projected mainly on the values of $m_0$, $A_0$, and 
$\tan\beta$. 
It should be noted that, combining the relations from the natures of the Higgs 
boson and the values of sfermion parameters from the BBN, as we will find, 
the linear relation between $m_0$ and $A_0$ are derived. 
Furthermore the degeneracy of the neutralino LSP and the stau NLSP in the 
coannihilation region set a relation between  $M_{1/2}$ and sfermion parameters. 
Thus accumulating all of relations and constraints, we give theoretical clues 
on the parameter space of the CMSSM. Then we calculate the observables of 
terrestrial experiments based on the analysis. 
After the discovery of SUSY signals, by checking those with the calculation we
make it possible to confirm the CMSSM in the near future.

This paper is organized as follows. In next section we recall the framework of 
the CMSSM, keeping an eye on the coannihilation scenario. Then we set the 
constraint on some input parameters from the viewpoint of the report of the 
Higgs boson and the observed abundances of both dark matter and light elements. 
In Sec.~\ref{sec:prediction}, we show the predictions for SUSY mass spectrum, 
the anomalous magnetic moment of muon, branching fractions of the $B$-meson 
rare decays, and direct detection of the neutralino dark matter in the allowed region.  
We show the number of the signals of the long-lived stau and the neutralino 
at the LHC experiment, 
and then we discuss the verification of the scenario in Sec~\ref{sec:LHCexp}.  
Sec.~\ref{sec:sum} is devoted to a summary and a discussion.

\section{Constraints}  \label{sec:parameter region} 

Here we show our constraints to derive the experimentally favored 
parameter space in the CMSSM.

Before we explain the detail of our constraints, we briefly review the CMSSM. 
The CMSSM is described by four parameters and a sign, 
  \begin{equation}
    m _{0}, M _{1/2}, A _{0}, \tan \beta, \mathrm{sign}(\mu),
  \end{equation}
where the first three parameters are the universal scalar mass, the 
universal gaugino mass, and the universal trilinear coupling at the scale 
of grand unification, respectively. 
Here $\tan \beta$ is the ratio of vacuum expectation values of two Higgs bosons, 
and $\mu$ is the supersymmetric Higgsino mass parameter. 
In the CMSSM, we describe all observables by these parameters, 
and hence we can derive the favored parameter space by putting 
the experimental/observational constraints on the calculated observables.

From now on, we explain in more details our constraints and how to 
apply those to our numerical calculations.  The first requirement comes 
from the Higgs boson mass. The latest reports on its mass, $m_h$, are 
\begin{equation}
  m _{h} = 125.8 \pm 0.4 (\text{stat}) \pm 0.4 (\text{syst})~\text{GeV}, 
\label{eq:mhCMS}
\end{equation}
by the CMS collaboration~\cite{CMS:aya}, and 
\begin{equation}
  m _{h} = 125.2 \pm 0.3 (\text{stat}) \pm 0.6 (\text{syst})~\text{GeV}, 
  \label{eq:mhATLAS}
\end{equation}
by the ATLAS collaboration~\cite{ATLAS:2012klq}, respectively.
It is known that the Higgs boson mass calculated by each public codes 
fluctuates by about $\pm 3$GeV~\cite{Allanach:2001hm, Djouadi:2002nh, 
Allanach:2003jw, Allanach:2004rh}. Taking into account these uncertainties, 
we apply a more conservative constraint as 
\begin{equation}
  m _{h} = 125.0 \pm 3.0~\text{GeV}.
  \label{eq:higgs-mass-bound}
\end{equation}

The second constraint comes from the observation for the relic 
abundance of the dark matter. The WMAP satellite reported the value 
at the 3 sigma level~\cite{Hinshaw:2012fq},
\begin{equation}
     0.089 
    \leq \Omega _{\text{DM}} h ^2 
    \leq 0.136.
    \label{eq:DMabundance}
\end{equation}
In most of the CMSSM parameter space, the relic abundance of the neutralino 
LSP is over-abundant against the measured value. The correct dark 
matter abundance requires the unique parameter space where the 
bino-like neutralino LSP and the stau NLSP are degenerate in mass 
so that the coannihilation mechanism works well~\cite{Griest:1990kh, Edsjo:1997bg}.

The third and fourth constraints are required from solving the lithium-7 problem. 
The stau NLSP is long-lived if the mass difference is smaller than the mass of 
tau lepton~\cite{Profumo:2004qt, Jittoh:2005pq}. In the case that the stau 
survives until the BBN epoch, the lithium-7 density can be reduced to the 
measured value with the exotic nuclear reactions induced by the stau. 
As the third condition, we impose the mass difference, $\delta m$, to be 
\begin{align} 
\delta m \equiv m_{\tilde \tau_1} - m_{\tilde \chi_1^0} \leq 0.1\text{GeV},
\label{eq:deltam}
\end{align}
where $m_{\tilde \tau_1}$ and $m_{\tilde \chi_1^0}$ are the masses of the 
stau NLSP and the neutralino LSP, respectively. With this mass difference, 
the stau NLSP is sufficiently long-lived so that it can survive until the BBN 
era~\cite{Jittoh:2007fr, Jittoh:2010wh, Jittoh:2011ni}.
In our numerical calculation, we use the pole mass of top quark as an input 
which involves an uncertainty of $\mathcal{O}(1)$GeV.  It is expected that 
the SUSY spectrum also includes the same order of uncertainties.
So in the numerical results in the next section, we show also the case of 
\begin{equation} 
 \delta m \leq 1~\text{GeV},
  \label{eq:deltam2}
\end{equation}
to make more conservative prediction on the CMSSM parameters\footnote{In 
a precise sense, a too small mass difference makes the stau too long lived.  
If the stau lives long sufficient to form a bound state with a helium, the stau 
converts it another nuclei, a deuteron and a triton.  Those reactions make 
number densities of converted nuclei too large compared with those of 
observations~\cite{Jittoh:2011ni}.  Number densities of these nuclei 
therefore become inconsistent with observed values through those reactions 
and too small mass difference is not allowed. This is, however, considerably 
relaxed by introducing a tiny lepton flavor violation~\cite{Kohri:2012gc}.}.

The forth constraint is the upper bound on the mass of the neutralino LSP, 
\begin{equation}
  m _{\tilde \chi_1^0} \leq 450\text{GeV}.
  \label{eq:neutralinomass}
\end{equation}
This upper bound is derived from the requirement of the abundances of both 
the dark matter and light elements.  
The yield value of the negatively charged stau at the BBN epoch is required to 
be larger than $Y_{\tilde \tau _{1}}^\text{BBN} \gtrsim 1.0 \times 10 ^{-13}$, 
where $Y_{\tilde \tau _{1}} = n_{\tilde \tau _{1}} / s$ ($s$ is the entropy 
density), to sufficiently reduce the lithium-7~\cite{Jittoh:2011ni}. 
Assuming the enough longevity of the stau, $Y_{\tilde \tau _{1}}^\text{BBN}$ 
is fixed at when the ratio of number densities of the stau NLSP and the neutralino 
LSP are frozen out. The yield value $Y_{\tilde \tau _{1}}^\text{BBN}$ is 
expressed with the relic density of the neutralino dark matter 
$Y_{\tilde \chi _{1} ^{0}}^\text{relic}$
\begin{equation}
\begin{split}
   Y_{\tilde \tau _{1}}^\text{BBN} = 
   \frac{Y_{\tilde \chi _{1} ^{0}}^\text{relic}}
   {2 (1 + e^{\delta m/T_{f}})}. 
\label{Ystau_YDM}   
\end{split} 
\end{equation}
Here $T_{f}$ is the freeze-out temperature of the ratio, $T_{f} \simeq (m_\tau - 
\delta m)/25$. 
The relic abundance of the dark matter is bounded (see Eq.~(\ref{eq:DMabundance})), 
\begin{equation}
  \Omega_\text{DM}h^2 
  \equiv 
  \frac{Y_{\tilde \chi _{1} ^{0}}^\text{relic} s_0 m_\text{DM} h^2}{\rho_c} \leq 0.136.  
\label{eq:DMabundance}
\end{equation}
Here $s_0$ is the today's entropy density, $h$ is the scale factor of the 
Hubble expansion rate, and $\rho_c$ is the critical density. Combining 
these facts, we obtain the upper bound on the mass of the neutralino 
dark matter, 
\begin{equation}
\begin{split}
   m_{\tilde \chi _{1} ^{0}} \lesssim 
   \frac{\rho_c}{2 s_0 h^2 (1 + e^{\delta m/T_{f}})} 
   \hspace{0.5mm}
   \frac{0.136}{1.0 \times 10^{-13}}. 
\label{m_bound}   
\end{split} 
\end{equation}
Hence the upper bound on the neutralino mass is $m_{\tilde \chi_1^0} 
\leq 450$GeV.

\section{Allowed region in the CMSSM}  

In this section, we show the allowed region in the CMSSM parameter space 
by imposing the constraints explained in Sec.~\ref{sec:parameter region}. 
We calculate the SUSY spectrum, the DM abundance and 
other observables using micrOMEGAs~\cite{Belanger:2013oya} implementing SPheno~\cite{Porod:2003um, Porod:2011nf}, and the 
lightest Higgs mass using FeynHiggs~\cite{Heinemeyer:1998yj, Heinemeyer:1998np, Degrassi:2002fi, Frank:2006yh}.

\subsection{$A _{0}$ - $m _{0}$ plane} \label{sec:A0m0}   

In Fig.~\ref{fig:A0-m0plane}, we show the allowed region in 
$m _{0}$-$A _{0}$ plane for $\tan \beta = 10,~20,$ and $30$ 
from top to bottom and $\delta m \leq 1$ and $0.1$GeV from left 
to right panels, respectively. Color represents $M _{1/2}$.

Notably, one can see that $A _{0}$ and $m _{0}$ have an almost linear 
relation for fixed $M _{1/2}$. The relations can be parameterized as 
\begin{equation}
\begin{split}
 m _{0} &= - 5.5 \times 10 ^{-3} A _{0} \tan \beta + b, \\
  b &\simeq 
  \begin{cases}
  (122, 190)~\text{GeV} \quad \text{for} \tan \beta = 10, \\
  (165, 228)~\text{GeV} \quad \text{for} \tan \beta = 20, \\
  (225, 283)~\text{GeV} \quad \text{for} \tan \beta = 30. 
  \end{cases}  
\label{eq:linearrelation}
\end{split}
\end{equation}
These linear relations come from the tight degeneracy in mass between the 
stau and the neutralino.  
We explain the linear relation as follows.  
Firstly we note that signs of $\mu$ and $A _{0}$ is opposite each other 
due to obtaining the large Higgs boson mass as we explain latter.  
In this scenario we choose $\mu > 0$ and $A _{0} < 0$, respectively.  
For a fixed $m_{\tilde \chi _{1} ^{0}}$, in response to an increasing 
of $m_0$, the mass of the lighter stau is also increased. In order to keep 
the mass difference to be smaller than 1GeV(0.1GeV), $|A_0|$ also has 
to be increased. Increasing of $|A_0|$ makes a non-diagonal element 
of the stau mass matrix to be large, and hence the mass eigenvalue of 
the lighter stau is decreased after diagonarization to be within the small 
mass difference. As a result, the linear behavior in the $A_0$-$m_0$ 
plane is found.

One can see in Fig.~\ref{fig:A0-m0plane} that $m _0$ in the allowed region 
increases as $\tan \beta$ increases.  
This is because the soft masses of the staus are more decreased for larger 
$\tan\beta$ in RG running due to the tau Yukawa couplings. 
The terms with tau Yukawa coupling in RGE decrease the soft masses in the 
running~\cite{Martin:1997ns}. These become significant when $\tan\beta$ is 
large because the tau Yukawa coupling is proportional to $1/\cos\beta$. Thus 
larger $m_0$ is required for larger $\tan\beta$ to obtain the stau mass 
satisfying with Eq.\eqref{eq:deltam} or \eqref{eq:deltam2}.

The upper and the lower edges of the allowed region in all panels of the 
Fig.~\ref{fig:A0-m0plane} are determined by the constraints of the correct 
abundances both of the dark matter and the light elements.
This can be understood as follows.
Since the dark matter abundance is depend on the neutralino LSP mass 
as shown in Eq.~(\ref{eq:DMabundance}), 
the dark matter abundance gives the 
lower bound on the neutralino mass.  
On the other hand, the light elements abundance gives the upper bound as 
shown in Eq.~(\ref{eq:neutralinomass}).  The bounds for the lighter stau 
mass is nearly same as that for the neutralino mass due to the tight degeneracy.  
Therefore once $A _{0}$ is fixed, $m _{0}$ can vary in range satisfying the 
bound for the lighter stau mass. Thus, the upper and the lower edges are 
determined by the constraints.

On the other hand, the right side of the allowed region is determined 
by the lower bound on the Higgs boson mass.  We can see from the the left 
panel of Fig.~\ref{fig:A0-m0plane_mh_Xt} that the Higgs boson mass reaches 
to $122$GeV. 
The mass square of the Higgs boson with one-loop corrections is given by 
\begin{align}
  m _{h} ^2 & = 
   m _{Z} ^2 \cos ^2 2 \beta \nonumber \\
    &~~+ \frac{3 m _{t} ^4}{16 \pi ^2 v ^2}
   \left [ 
   \log \left ( \frac{m _{\tilde t} ^2}{m _{t} ^2} \right ) 
 + \frac{X _{t} ^2}{m _{\tilde t} ^2} 
   \left ( 1 - \frac{X _{t} ^2}{12 m _{\tilde t} ^2} \right )  
   \right ], 
   \label{eq:higgsmass}  \\
  &(X _{t} = A _{t} - \mu \text{cot} \beta ,\ \  
   m _{\tilde t} = \sqrt{m _{\tilde t _{1}} m _{\tilde t _{2}} } ), \nonumber 
\end{align}
where the first line represents the tree-level mass square and the second 
one is the one-loop corrections~\cite{Haber:1990aw, Ellis:1990nz, 
Okada:1990vk, Casas:1994us}. The tree-level contribution is simply given by 
the $Z$ boson mass $m _{Z}$ while the radiative corrections are given by the 
masses of the top $m_t$, the lighter/heavier stop $m _{\tilde t_1/ \tilde t_2}$, 
the stop mixing parameter $X_t$, and the vacuum expectation values of 
the Higgs bosons, $v = \sqrt{v _{u} ^2 + v _{d} ^2}$, here $v _{u}$ and $v _{d}$ 
are vacuum expectation values of up-type and down-type Higgs boson, respectively.   
The radiative corrections are sensitive to the stop mixing parameter. 
As is well known, the second term in the bracket decreases quadratically from 
its maximum at 
\begin{equation}
  |X _{t}| = \sqrt{6} m _{\tilde t}.  
\label{eq:maximal}
\end{equation}
To obtain maximum value of $|X _{t}|$, signs of $\mu$ and $A _{0}$ have 
to be opposite each other, and we choose $\mu > 0$ and $A _{0} < 0$ in 
this paper.  
The right panel of Fig.~\ref{fig:A0-m0plane_mh_Xt} shows $|X_t|/\sqrt{6} 
m_{\tilde t}$ in the allowed region for $\tan\beta=20$ and $\delta m \leq 
1$GeV. It can be seen that $|X _{t}|$ is smaller than $\sqrt{6} m_{\tilde t}$ 
at the right side edge, and hence the one-loop corrections are not so large that  
$m _{h}$ is pushed up to the lower bound. 
Similarly, the left side edge is also determined by the Higgs boson mass bound. 
The value of $|X_t|/\sqrt{6} m_{\tilde t}$ gradually becomes large as $|A_0|$ 
increases.  At a large $(|A_0|, m_0)$ point, the value of $|X_t|/\sqrt{6} 
m_{\tilde t}$ is equal to $1$, and therefore $m _{h}$ receives the maximal 
loop correction. 
In the region where $|X_t|/\sqrt{6} m_{\tilde t} > 1$, the loop corrections 
to $m_h$ are smaller and $m_h$ decreases from the maximal value. 
Therefore $m_h$ is smaller than $122$GeV again at the left side edge as 
can be seen in the right panel of Fig.~\ref{fig:A0-m0plane_mh_Xt}.

\clearpage

\begin{widetext}
\begin{center}
\begin{figure}[h]
\begin{tabular}{l}
\hspace{-2mm}
\begin{minipage}{80mm}
\begin{center}
 \includegraphics[width=8.4cm,clip]{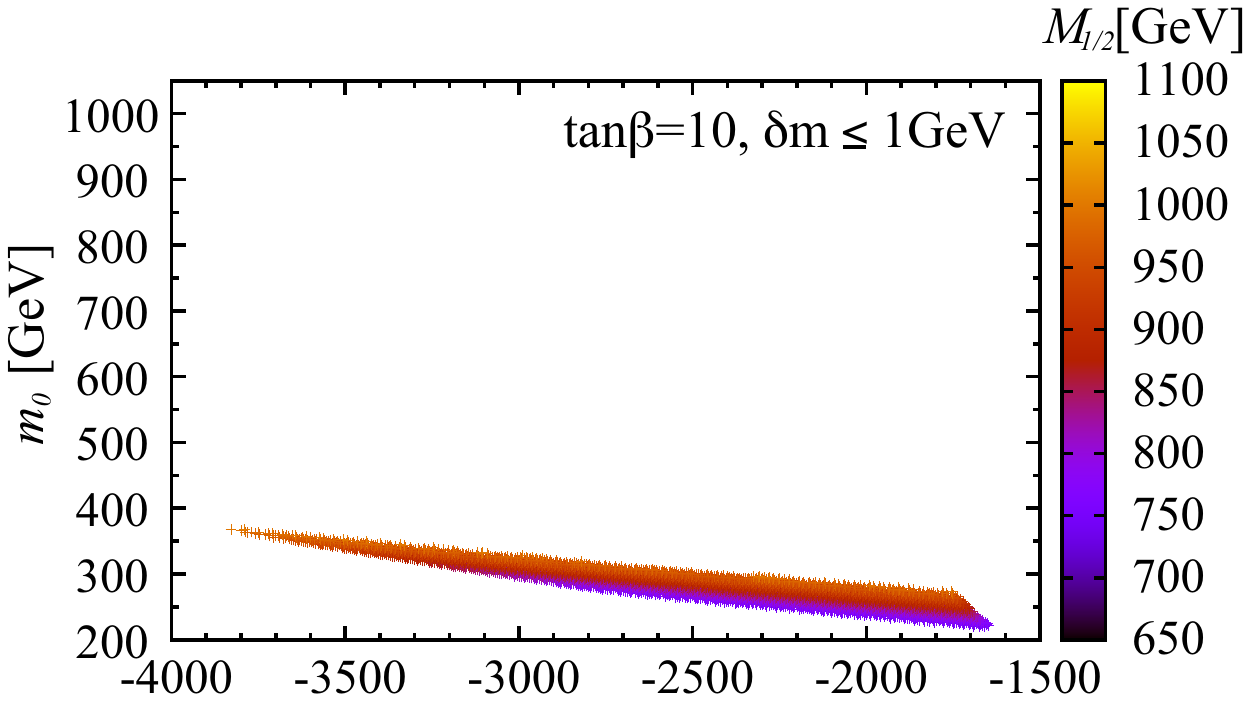}
 \label{fig:tb10_cut_A0m0m12_dm1}
\end{center}
\end{minipage}
\hspace{4mm}
\begin{minipage}{80mm}
\begin{center}
 \includegraphics[width=8.02cm,clip]{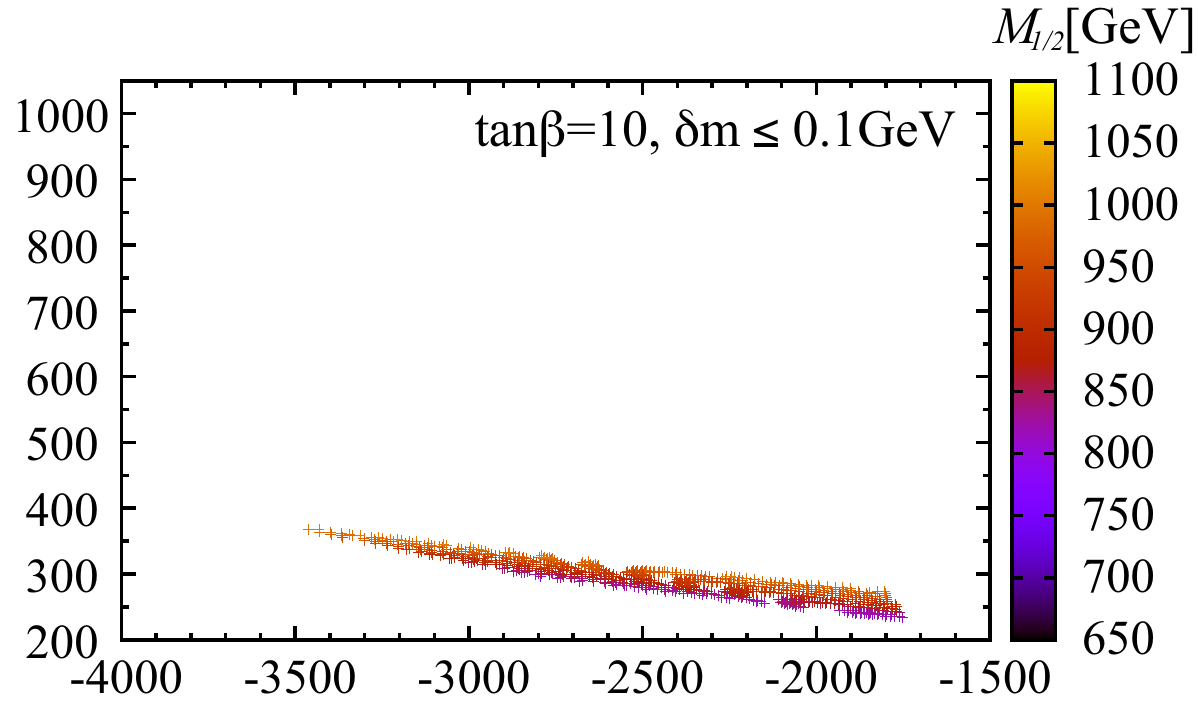}
 \label{fig:tb10_cut_A0m0m12_dm01}
\end{center}
\end{minipage}
\\[-3mm]
\hspace{-2mm}

\begin{minipage}{79mm}\vspace{-0.8mm}
\begin{center}
  \includegraphics[width=8.3cm,clip]{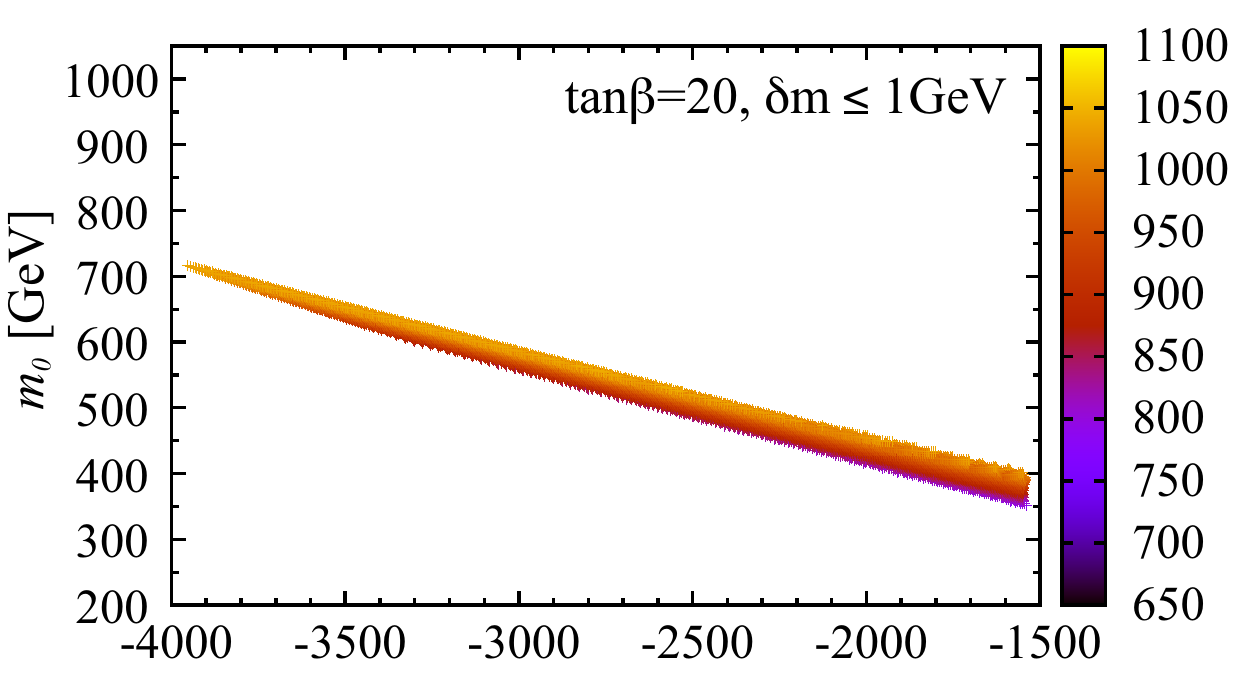}
  \label{fig:tab20_cut_A0m0m12_dm1}
\end{center}
\end{minipage}
\hspace{4.54mm}
\begin{minipage}{79.8mm}\vspace{0.5mm}
\begin{center}
 \includegraphics[width=7.92cm,clip]{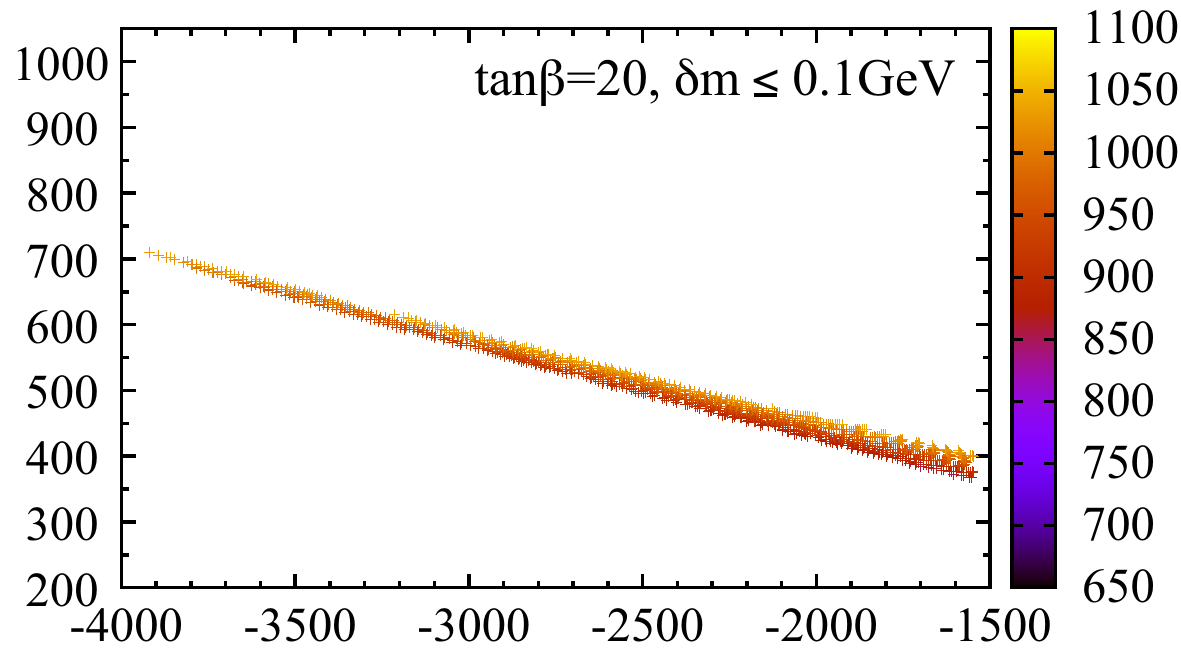}
 \label{fig:tb20_cut_A0m0m12_dm01}
\end{center}
\end{minipage}
\\[5mm]
\hspace{-2mm}
\begin{minipage}{79mm}\vspace{-4.1mm}
\begin{center}
 \includegraphics[width=8.28cm,clip]{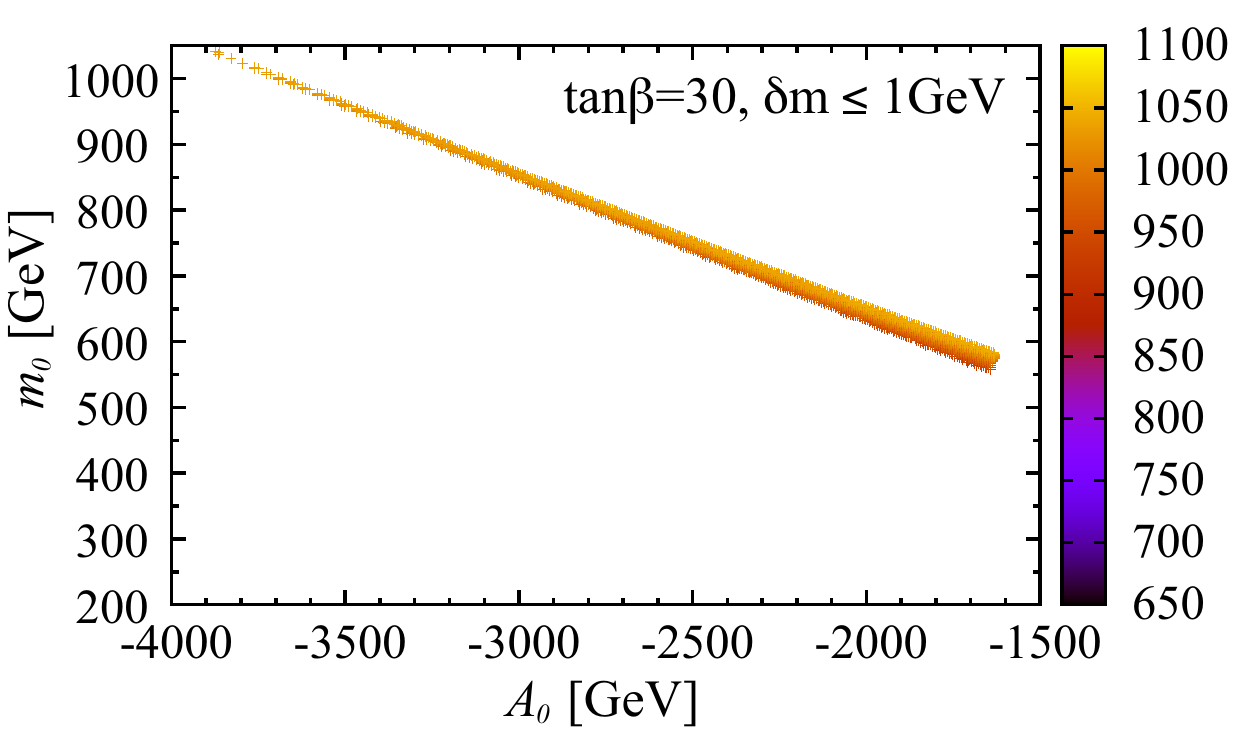}
 \label{fig:tb30_cut_A0m0m12_dm1}
\end{center}
\end{minipage}
\hspace{4.5mm}
\begin{minipage}{79.8mm}\vspace{-2.5mm}
\begin{center}
 \includegraphics[width=7.92cm,clip]{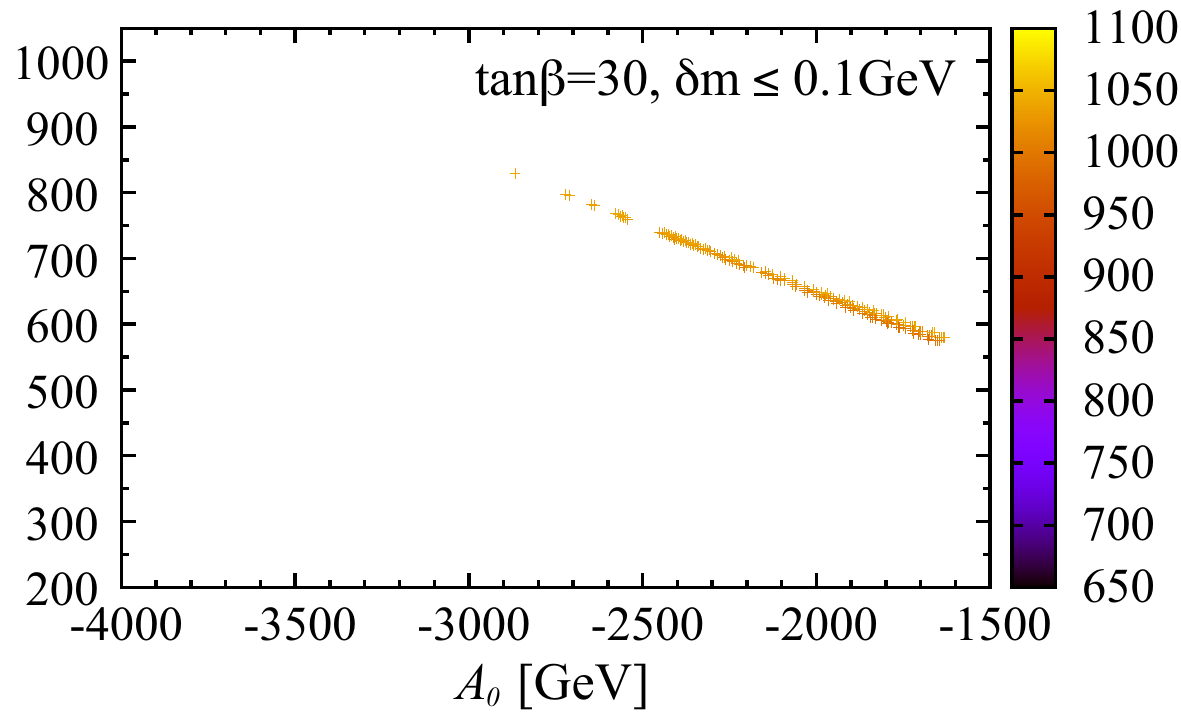}
 \label{fig:tb30_cut_A0m0m12_dm01}
\end{center}
\end{minipage}
\end{tabular}
\caption{Allowed region in $A _{0}$-$m _{0}$ plane.  
We fix $\tan \beta $ to $10,~20$ and$~30$ from top to bottom 
and $\delta m \leq 1~$ and $0.1~\text{GeV}$ from left to right.  
A gradation of colors represents $M _{1/2}$.  
Light color indicates large value, and dark color indicates small value.}
\label{fig:A0-m0plane}
\begin{tabular}{l}
\begin{minipage}{83mm}\vspace{10mm}
\begin{center}
 \includegraphics[width=8.1cm,clip]{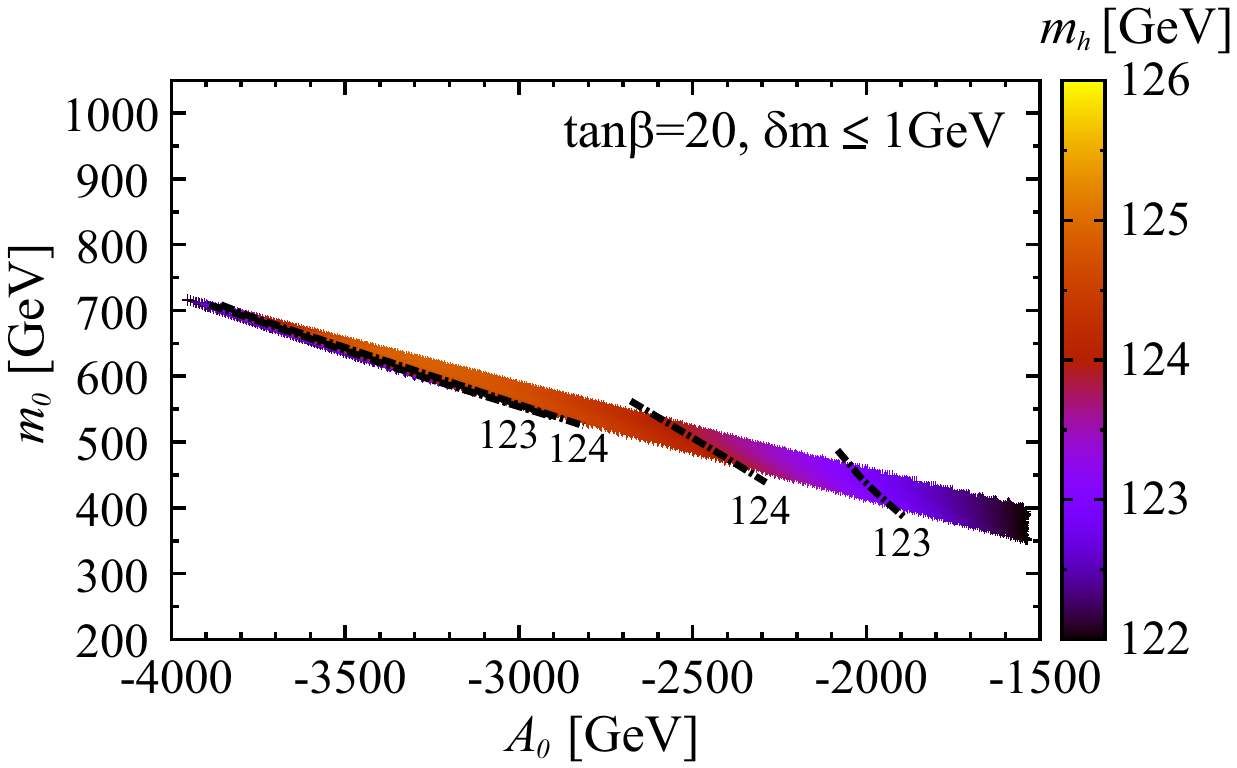}
 \label{fig:tb20_cut_A0m0mh_dm1}
\end{center}
\end{minipage}
\hspace{-3mm}
\begin{minipage}{83mm}\vspace{10mm}
\begin{center}
 \includegraphics[width=8.02cm,clip]{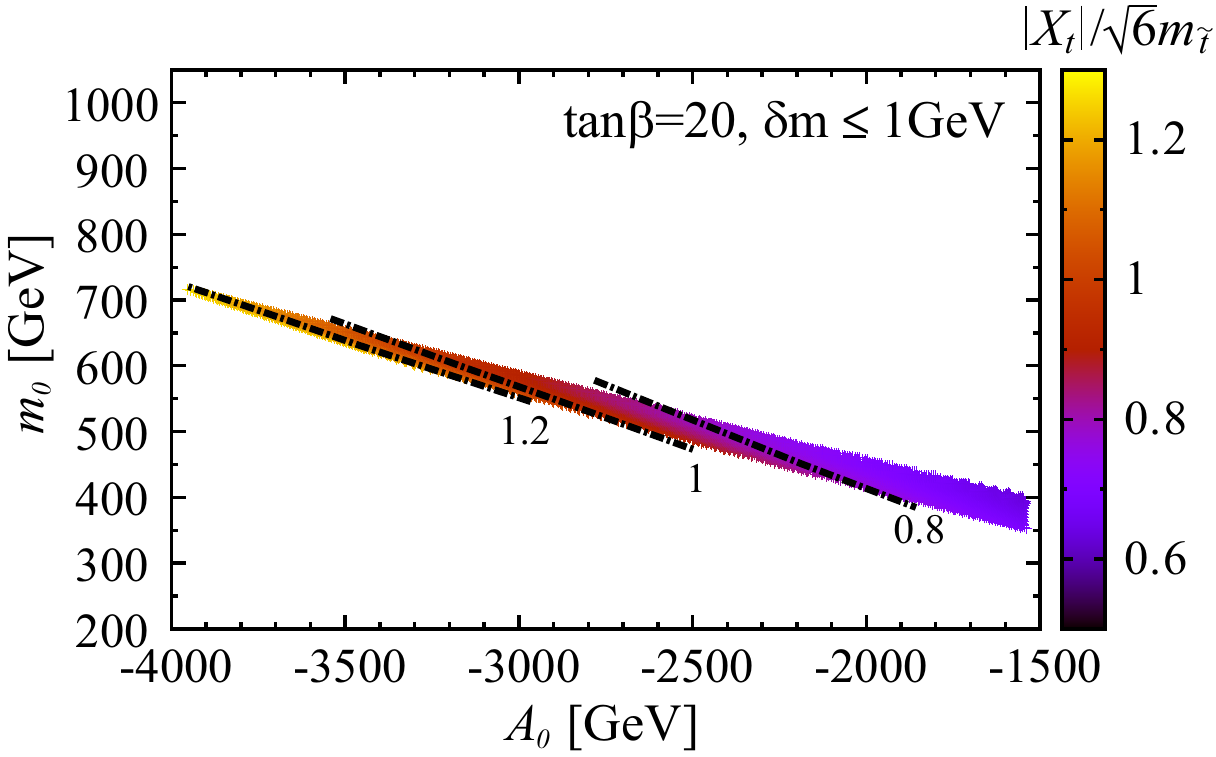}
 \label{fig:tb20_cut_A0m0Xt_dm1}
\end{center}
\end{minipage}
\end{tabular}
\caption{Left panel:the value of the Higgs boson mass.  
Right panel: the ratio of $|X _{t}|$ to $\sqrt{6} m _{\tilde t}$.
A gradation of colors represents each value.   
Light color indicates large value, and dark color indicates small value.
We fix $\tan \beta $ to 20 and $\delta m \leq 1~\text{GeV}$ in each figure.
}
\label{fig:A0-m0plane_mh_Xt}
\end{figure}
\end{center}
\clearpage

\end{widetext}

\subsection{$m _{0}$ - $M _{1/2}$ plane} \label{sec:m0M12} 

\begin{center}
\begin{figure*}[h]
\begin{tabular}{l}
\hspace{-2mm}
\begin{minipage}{85mm}
\begin{center}
 \includegraphics[width=8.3cm,clip]{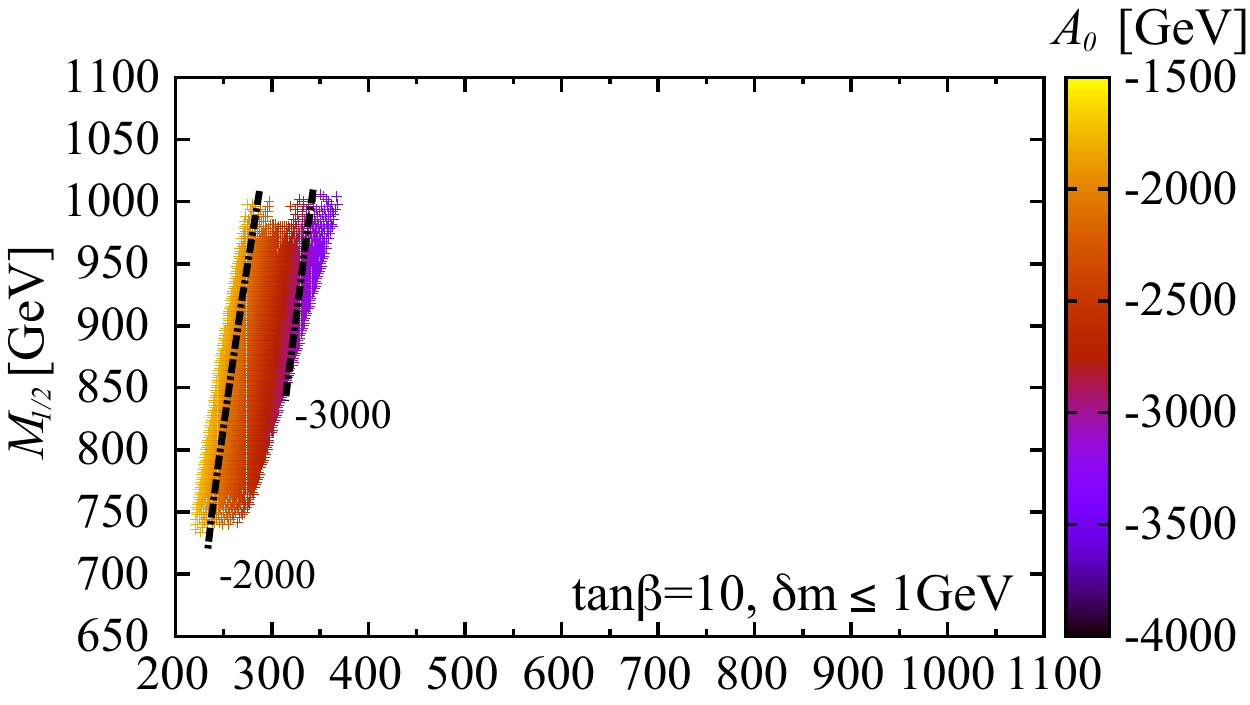}
 \label{fig:tb10_cut_m0m12A0_dm1}
\end{center}
\end{minipage}
\hspace{-3mm}
\begin{minipage}{85mm}
\begin{center}
 \includegraphics[width=8cm,clip]{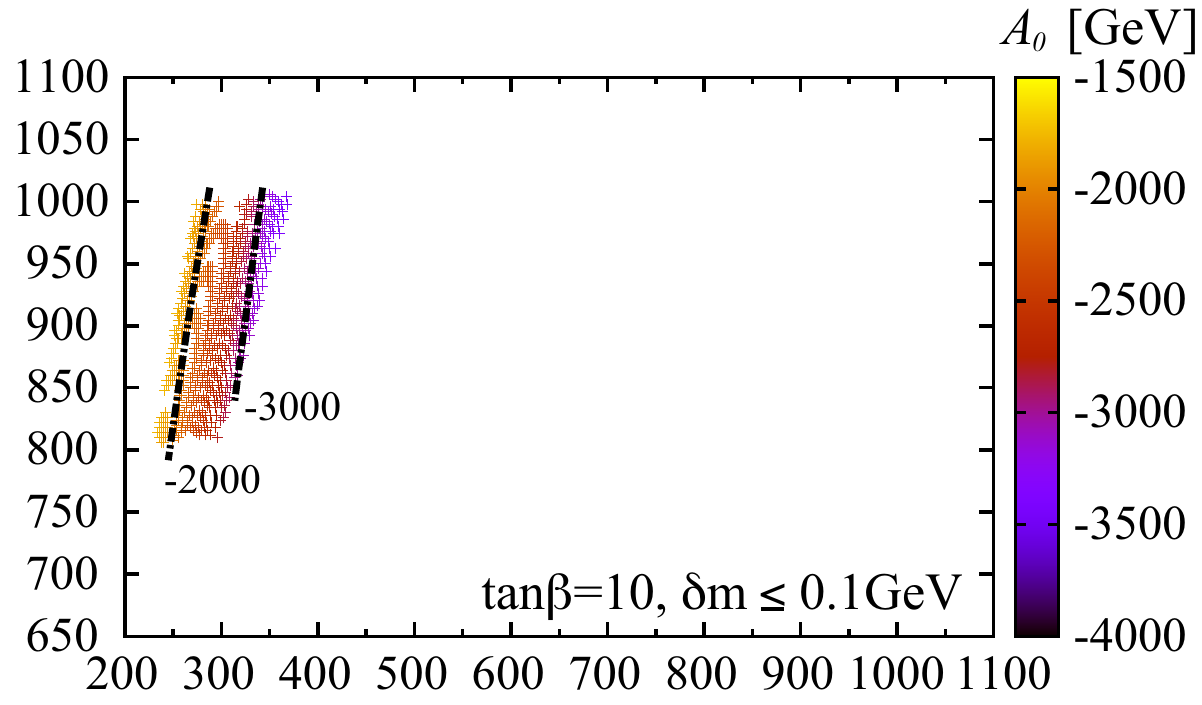}
 \label{fig:tb10_cut_m0m12A0_dm01}
\end{center}
\end{minipage}
\\[-3mm]
\hspace{-2mm}
\begin{minipage}{84.3mm}\vspace{4.6mm}
\begin{center}
 \includegraphics[width=8.23cm,clip]{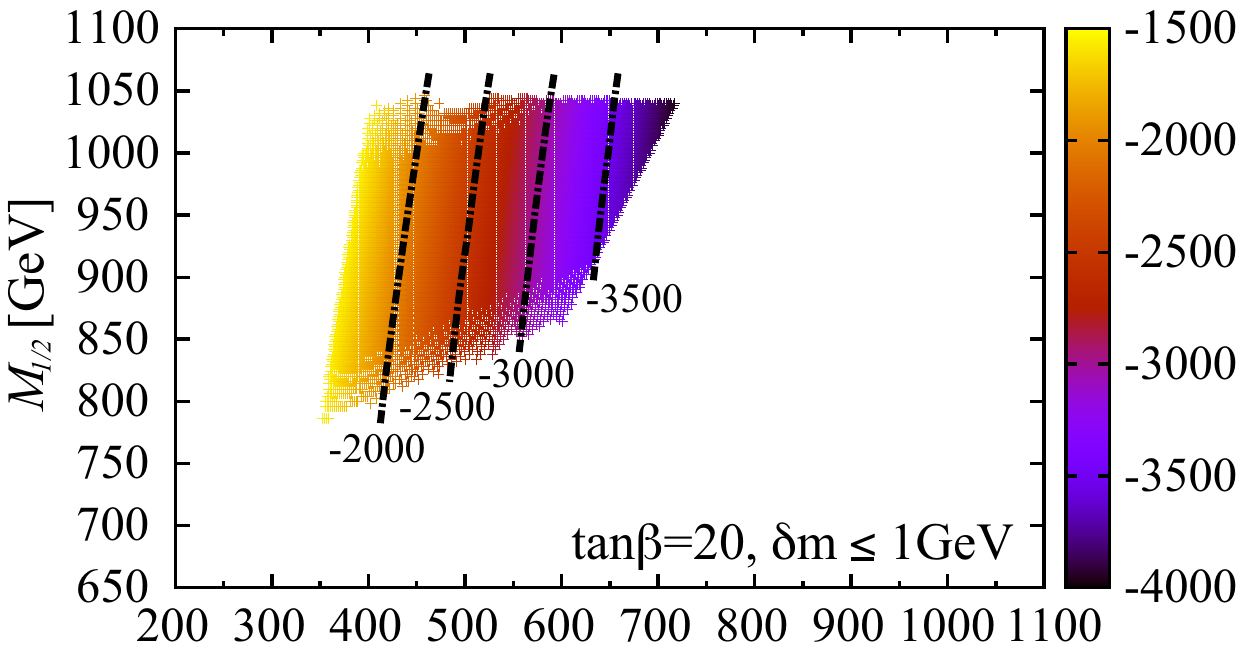}
 \label{fig:tb20_cut_m0m12A0_dm1}
\end{center}
\end{minipage}
\hspace{-3mm}
\begin{minipage}{85.4mm}\vspace{4mm}
\begin{center}
 \includegraphics[width=7.91cm,clip]{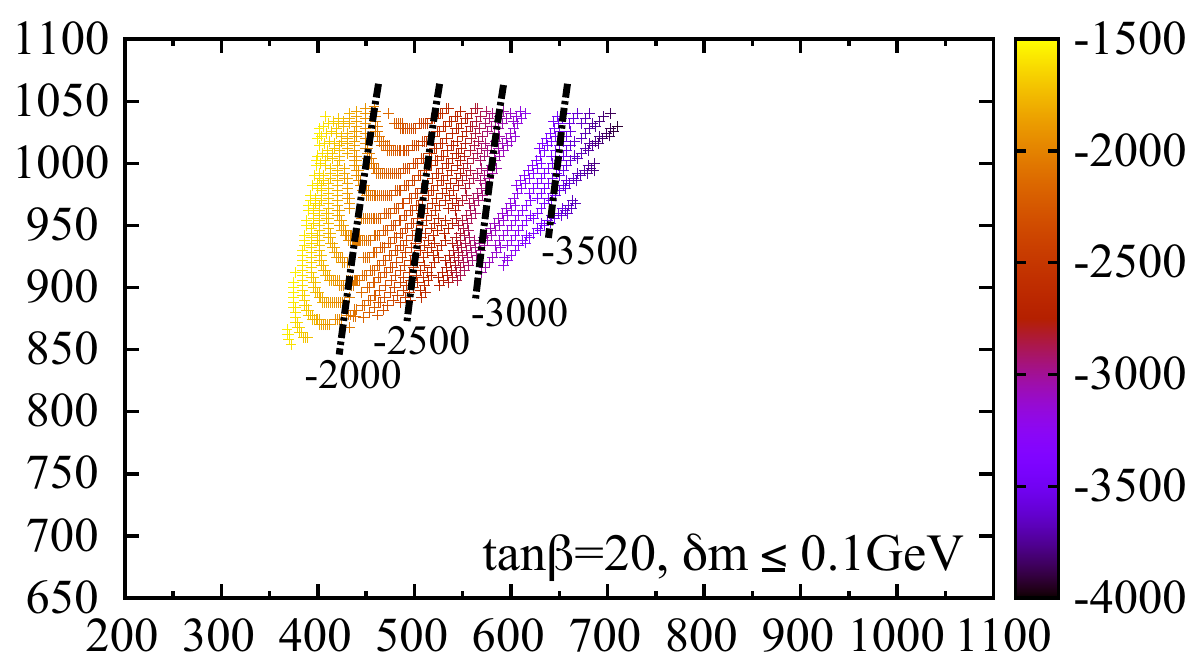}
 \label{fig:tb20_cut_m0m12A0_dm01}
\end{center}
\end{minipage}

\\[5mm]
\hspace{-2mm}
\begin{minipage}{84.3mm}\vspace{1.2mm}
\begin{center}
 \includegraphics[width=8.23cm,clip]{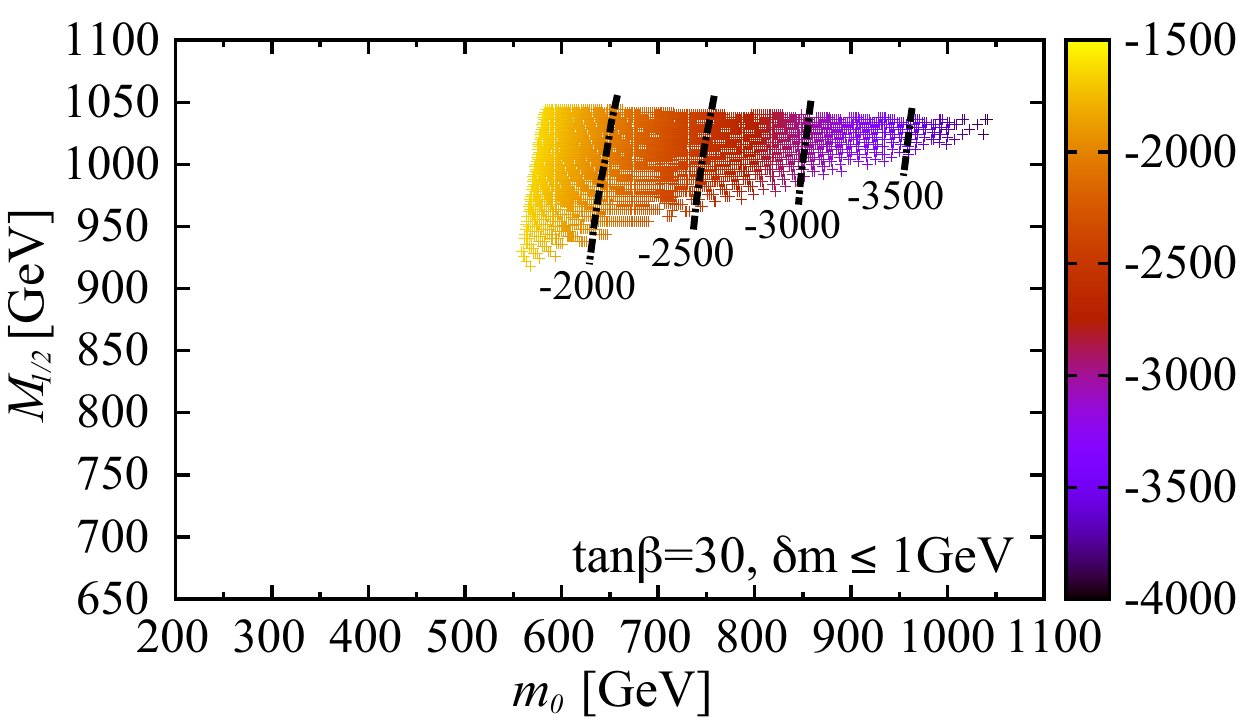}
 \label{fig:tb30_cut_m0m12A0_dm1}
\end{center}
\end{minipage}
\hspace{-3mm}
\begin{minipage}{85.61mm}\vspace{1.1mm}
\begin{center}
 \includegraphics[width=7.89cm,clip]{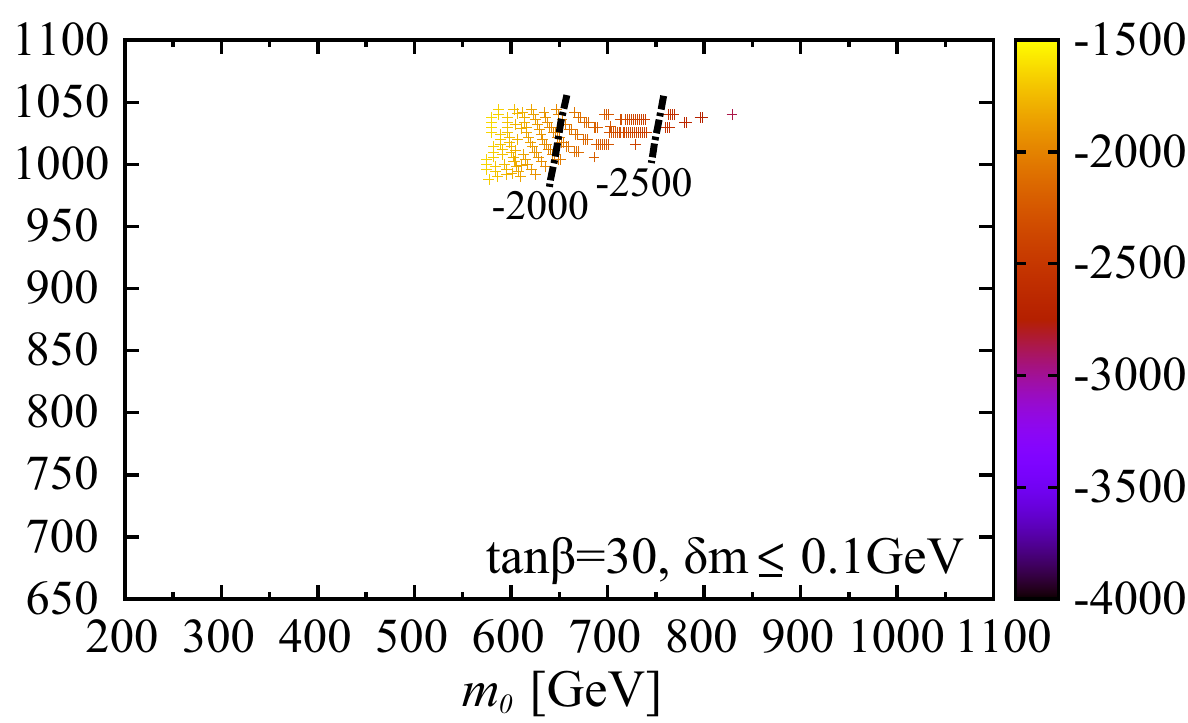}
 \label{fig:tb30_cut_m0m12A0_dm01}
\end{center}
\end{minipage}
\end{tabular}
\caption{Allowed parameter region in $m _{0}$-$M _{1/2}$ plane. 
We fix $\tan \beta $ to $10,~20$ and $30$ from top to bottom 
and $\delta m \leq 1~$and~$0.1~\text{GeV}$ from left to right, respectively.  
A gradation of colors represents $A _{0}$. 
Light color indicates large value, and dark color indicates small value. }
\label{fig:m0-M12plane}
\begin{tabular}{l}
\hspace{-2mm}
\begin{minipage}{85mm}\vspace{10mm}
\begin{center}
 \includegraphics[width=8.3cm,clip]{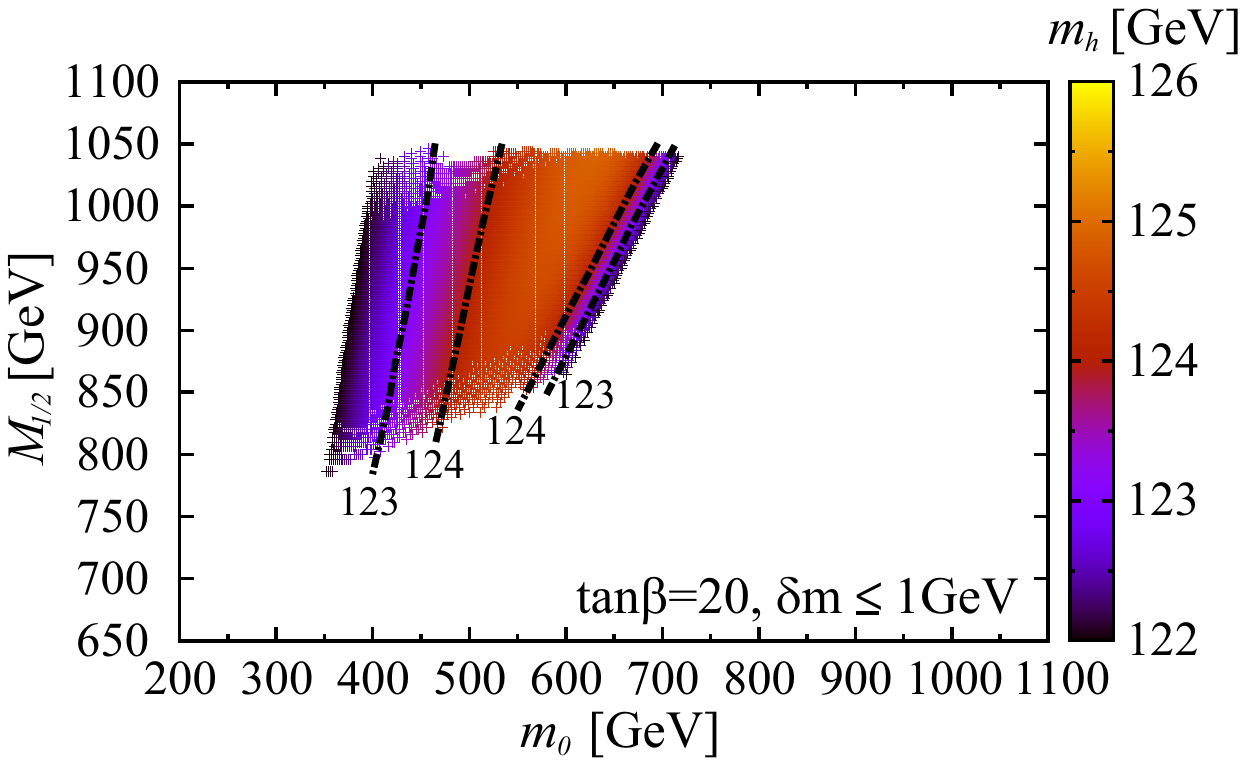}
 \label{fig:tb20_cut_m0m12mh_dm1}
\end{center}
\end{minipage}
\hspace{-5mm}
\begin{minipage}{85mm}\vspace{10mm}
\begin{center}
 \includegraphics[width=8.3cm,clip]{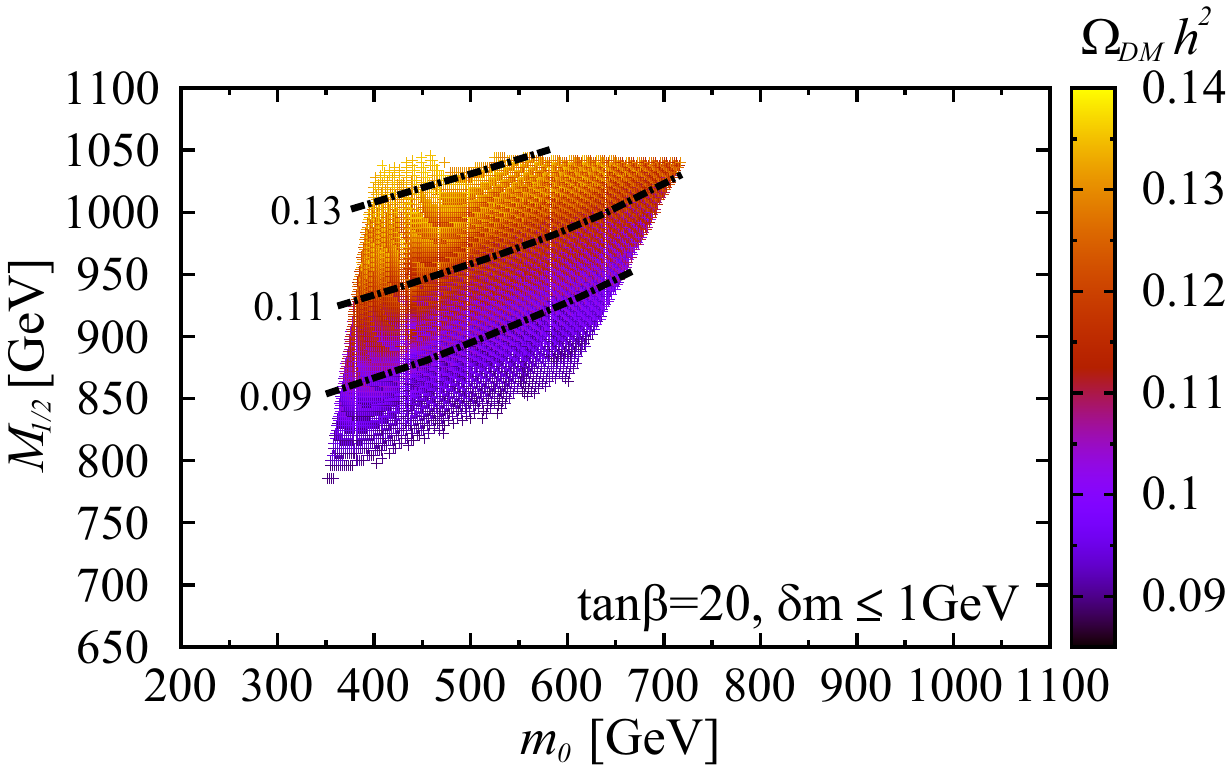}
 \label{fig:tb20_cut_m0m12omega_dm1}
\end{center}
\end{minipage}
\end{tabular}
\caption{Left panel:the value of the Higgs boson mass.  
Right panel: the value of the relic abundance of the dark matter.
A gradation of colors represents each value.  
Light color indicates large value, and dark color indicates small value.
We fix $\tan \beta $ to 20 and $\delta m \leq 1~\text{GeV}$ in each figure.}
\label{fig:m0-M12plane_mh_DM}
\end{figure*}
\end{center}

We show in Fig.~\ref{fig:m0-M12plane} the allowed region in $m_0$-$M_{1/2}$ 
plane for $\tan \beta = 10,~20$ and $30$ from top to bottom and $\delta m 
\leq 1$ and $0.1$GeV from left to right panels, respectively. Color and the 
contours represents $A_0$.  
Figure~\ref{fig:m0-M12plane_mh_DM} shows the Higgs boson mass in the 
left panel and the relic abundance of the dark matter in the right panel 
by color on the allowed region.  We fix $\tan \beta$ to 20 and $\delta m 
\leq 1$GeV in the panels.

In Fig.~\ref{fig:m0-M12plane}, it can be seen that $|A_0|$ is larger 
as $m_0$ is larger while it is slightly dependent on $M_{1/2}$.  
This is because the lighter stau mass is mainly determined 
by $m_0$ and $A_0$ as is explained in Sec.~\ref{sec:A0m0}. 
In the figure, we obtain $m _{h} \simeq 126$GeV,
at $(m _{0}, M _{1/2}) = (350, 1000) $GeV for $\tan \beta = 10$,  
$(640, 1050) $GeV for $\tan \beta = 20$, and 
$(800, 1050) $GeV for $\tan \beta = 30$, respectively.
These points are located at the middle of $m_0$ and the upper edge of 
$M_{1/2}$ in the allowed region. The value of $|X _{t}| $ is equal to 
$\sqrt{6} m _{\tilde t} $ at these points, and hence $m _{h}$ receives 
the large loop corrections.
Furthermore the logarithmic term in Eq.~(\ref{eq:higgsmass}) becomes 
large as $M _{1/2}$ increases since the stops become heavy as we show 
later in Fig.~\ref{fig:mstop}. Thus, the Higgs boson mass is pushed up 
to 126 GeV. The Higgs boson mass decreases as the parameters deviate 
from these points.
In the left panel of Fig.~\ref{fig:m0-M12plane_mh_DM}, it is clearly seen 
that the lower bound on the Higgs boson mass determines the left and the 
right edges in the allowed region.
Since $|A _{0}|$ is small at the left edge, the value of $|X _{t}|$ is 
smaller than $\sqrt{6} m _{\tilde t}$. Therefore $m _{h}$ receives a 
small one-loop correction.  
On the contrary, $|A _{0}|$ is large at the right edge, and $|X _{t}|$ is 
larger than $\sqrt{6} m _{\tilde t}$.  It means that $m _{h}$ receives 
a small one-loop correction as we explained in the previous subsection.

Note that the minimum value of $M _{1/2}$ in the allowed region are 
different in each panel. The minimum value is determined by the lower 
bound on the relic abundance of the dark matter.  
In the right panel of Fig.~\ref{fig:m0-M12plane_mh_DM}, we can see 
that the neutralino relic abundance reaches to the lower bound at the 
bottom edge of the allowed region\footnote{In the panel, we see moderate 
stripes on the allowed region. 
These stripes result from the low accuracy of the numerical calculation. 
We are not able to collect all data with the mass difference within $0.1$~GeV.  
If we collect all data, these stripes are not on the allowed region.}.  
The minimum value of $M_{1/2}$ becomes large as $\tan \beta$ 
increases and/or the mass difference becomes small.

The $\tan \beta$ dependence on the relic abundance is understood as 
follows. 
In the favored parameter region, the stau NLSP consists mostly of the 
right-handed stau. Then the dominant contribution to the total coannihilation 
rate of the dark matter is the stau-stau annihilation into tau leptons 
via Higgsino exchange for large $\tan\beta$~\cite{Pradler:2008qc}. 
The interactions of the stau-tau lepton-Higgsino is proportional to 
$\tan\beta$, and so the large $\tan\beta$ leads the large coannihilation 
rate of the dark matter. Larger $\tan\beta$ therefore makes the relic 
abundance to be smaller.

The dependence of the relic number density on the mass difference is 
understood in terms of the ratio of number density of the stau to the 
neutralino. 
At the freeze out of total number of all of SUSY particles, the ratio is 
propotional to $e^{-\delta m/T_f^\text{total}}$. Since the freeze out 
temperature $T_f^\text{total}$ is almost the same for $\delta m=0.1$ 
and $1$GeV as long as $m_{\tilde \chi_1^0}$ is  fixed, the stau number 
density is relatively large for $\delta m = 0.1$GeV. 
This leads that the total coannhilation rate become enlarged and hence 
the relic number density of the neutralino is reduced. Such reduction of 
the number density can be compensated by increasing the neutralino 
mass. Thus, the minimum value of $M _{1/2}$ should be larger for larger 
$\tan\beta$ and smaller $\delta m$ to meet the lower bound for the 
dark matter abundance.

On the other hand, the maximum value of $M _{1/2}$ is fixed from the upper 
bound in Eq.~(\ref{eq:neutralinomass}). Note that the upper bound is derived 
by the fact that the large number density of the stau is required to reduce the 
lithium-7 to be the measured abundance. 
This requirement forbids too large mass of the neutralino.  
As there is a relation between the neutralino mass and $M_{1/2}$, i.e., 
$m_{\tilde \chi_1^0} \simeq 0.43 M_{1/2}$, the upper bound of $M_{1/2}$ 
is estimated by the neutralino mass, and is fixed to be $1$TeV in the present 
paper.

We emphasize here that the maximum value of $M _{1/2}$ in the 
allowed region is unique in the parameter space. If we do not take into 
account the lithium-7 problem, heavier neutralino or $M_{1/2}$ is 
allowed, and hence the resultant range of $M_{1/2}$ is wider. 
\clearpage
%
%
%

\section{SUSY spectrum, $(g - 2) _{\mu}$, 
 $B$ meson rare decays and dark matter detection} 
\label{sec:prediction} 

In this section, we show our predictions on the SUSY spectrum, the muon anomalous 
magnetic moment, rare decays of $B$ mesons, and the dark matter direct detections. 
We calculate mass spectrum of SUSY particles by using SPheno~\cite{Porod:2003um, 
Porod:2011nf} and that of Higgs boson by using FeynHiggs~\cite{Heinemeyer:1998yj, 
Heinemeyer:1998np, Degrassi:2002fi, Frank:2006yh}.

\subsection{Spectra of SUSY particles with current limits} \label{sec:spectrum} 

We show the mass spectra of SUSY particles in the allowed region. 
Figure~\ref{fig:msquarksmgluino} shows the masses of the gluino, the first 
and the second generations squarks with respect to the lightest neutralino 
mass, respectively. From top to bottom, $\tan \beta $ is varied with 
$10,~20$ and $30$, and in the left and the right panels $\delta m \leq 1$ and 
$0.1$GeV.  
Similarly the masses of the stop and sbottom are shown in 
Fig.~\ref{fig:mstop}, those of the neutralino, the slepton and  the heavier 
Higgs in Fig.~\ref{fig:mneutralino}, Fig.~\ref{fig:mslepton} and in 
Fig.~\ref{fig:mheavyhiggs}, respectively.
Note that in all of figures we have excluded the region with $m_{\tilde \chi_1^0} 
(\simeq m_{\tilde \tau_1}) \lsim 339$GeV, which is the direct bound 
on the long-lived CHAMP at the LHC~\cite{Chatrchyan:2013oca}.
We explain behaviors of these figures in following subsections.

\subsubsection{Gluino, neutralinos and heavy Higgs boson masses} 

In the CMSSM, the gluino mass parameter $M _{3}$ is related to 
the bino and the wino mass parameter $M _{1}$ and $M _{2}$ at 
one-loop RGE as follows, 
\begin{equation}
  M _{3} 
  = \frac{\alpha _{s}}{\alpha} \sin ^2 \theta _{W} M _{2} 
  = \frac{3}{5} \frac{\alpha _{s}}{\alpha} \cos ^2 \theta _{W} M _{1}.  
\end{equation}
From this relations, we can obtain the ratio 
\begin{equation}
   M_3 : M_2 : M_1 \simeq 6 : 2 : 1, 
\end{equation}
around the TeV scale. We can see that $m _{\tilde g}$ is nearly 6 times larger 
than $m _{\tilde \chi _{1} ^{0}}$ in Fig.~\ref{fig:msquarksmgluino}, 
and $m _{\tilde \chi _{2} ^{0}}$ is about twice as large as that in 
Fig.~\ref{fig:mneutralino}.  
This means that the second lightest neutralino is almost the neutral wino. 
On the other hand,  $m_{\tilde \chi_3^0}$ and $m_{\tilde \chi_4^0}$ 
extend above $1$ TeV because these consist of the neutral Higgsinos. 
Their masses are given by $\mu$ parameter that is sensitive to $m_0$ for fixed 
$m_{\tilde \chi_1^0}$.  
The modulus $|\mu|$ is determined by the EWSB conditions.  
At tree level the corresponding formula for the correct EWSB leads as  
\begin{equation}
   |\mu| ^2 = \frac{1}{2} \left[ \tan 2 \beta 
   \left( M_{H_u}^2 \tan \beta - M_{H_d}^2 \cot \beta \right) 
   - m_Z^2  \right].  
\label{eq:mu1}
\end{equation}
where $M_{H_d}$ and $M _{H_u}$ are the down-type and the up-type 
Higgs soft SUSY breaking masses.  For $\tan \beta \gg 1$, 
Eq.~\eqref{eq:mu1} is approximated as follows  
\begin{equation}
  |\mu|^2 \simeq - M_{H_u}^2.  
\label{eq:mu2}
\end{equation}
The soft mass $M _{H _{u}}^2$ is sensitive to $m _{0}$.  The 
approximate solution of the one-loop RGE for $M _{H _{u}}^2$ is given by
\begin{equation}
\begin{split}
  m^2_{H_u}
  &\simeq
  - 3.5 \times 10^3 \cot^2 \beta m_0^{\prime 2} \\
  &+ 87 \cot \beta  M_{1/2} m_0^{\prime} - 2.8 M_{1/2}^2, 
\end{split}
\end{equation}
where $m_0^{\prime} \equiv m_0 - b$, and $b$ is defined in 
Eq.~(\ref{eq:linearrelation}). Therefore $m_{\tilde \chi_3^0}$ and 
$m_{\tilde \chi_4^0}$ become large with increasing $m_0$.

Meanwhile, the mass of the CP-odd Higgs boson, $m _{A}$, is given by, 
\begin{equation}
   m_A^2 \simeq |\mu|^2.  
\end{equation}
Thus, the masses of the heavy Higgs boson and the CP-odd Higgs boson are 
determined by $|\mu|$, and these are close to $m _{\tilde \chi _{3} ^{0}}$ and 
$m _{\tilde \chi _{4} ^{0}}$ as is shown in Figs.~\ref{fig:mneutralino} and 
\ref{fig:mheavyhiggs}.

\subsubsection{First and Second generation squarks, sleptons masses}

\begin{center}
\begin{figure*}[h]
\begin{tabular}{ll}
\begin{minipage}{84mm}
\begin{center}
 \includegraphics[width=8.2cm,clip]{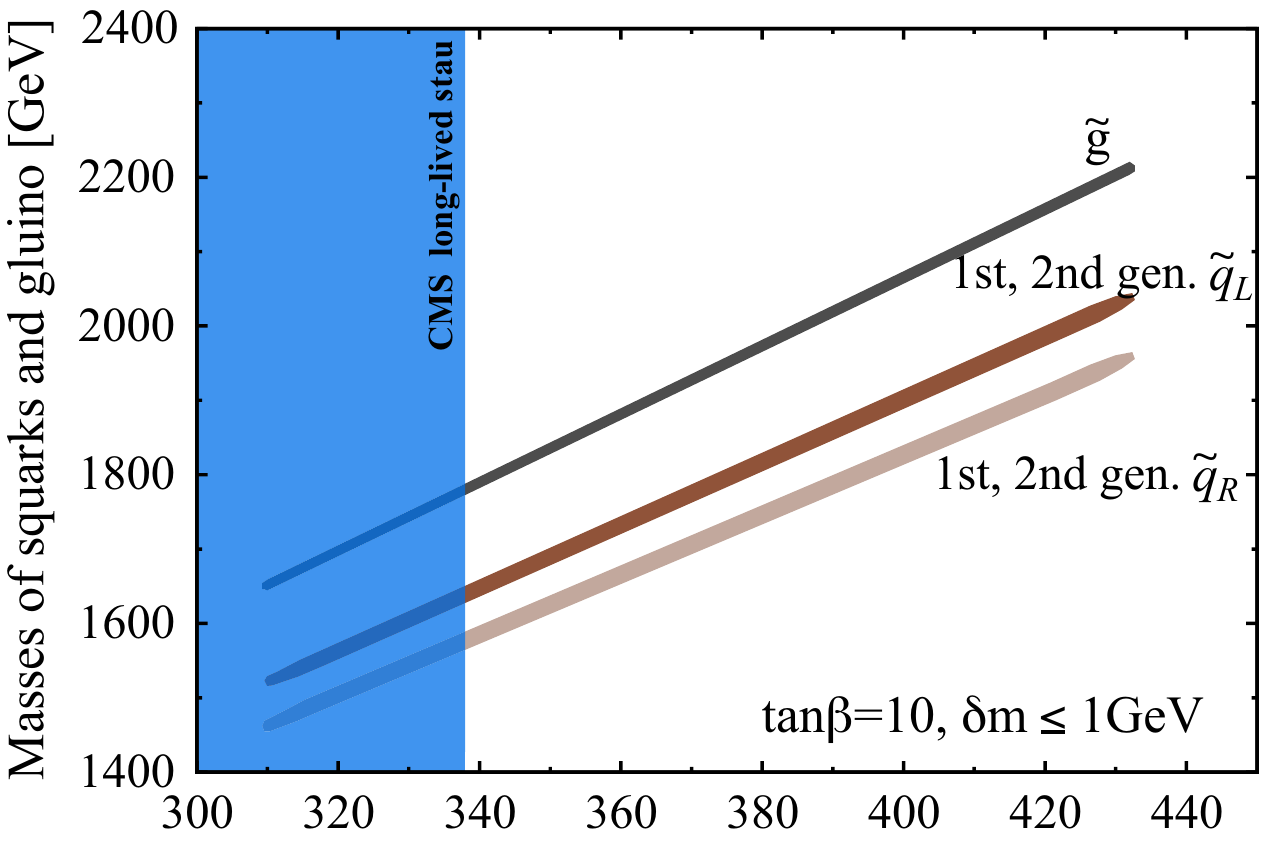}
\end{center}
\end{minipage}
&
\hspace{-4.5mm}
\begin{minipage}{84mm}
\begin{center}
 \includegraphics[width=7.7cm,clip]{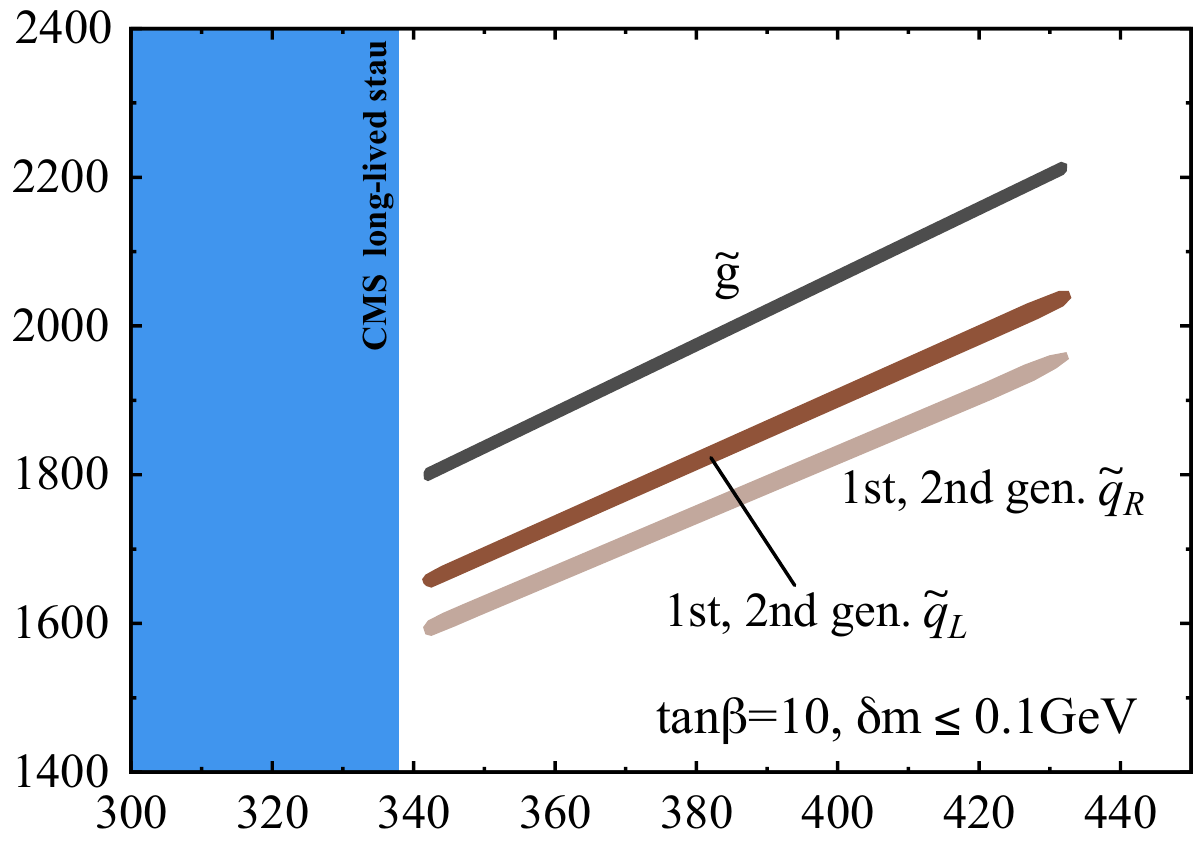}
\end{center}
\end{minipage}
\\[5mm]
\hspace{-1.5mm}
\begin{minipage}{84mm}\vspace{2.4mm}
\begin{center}
 \includegraphics[width=8.1cm,clip]{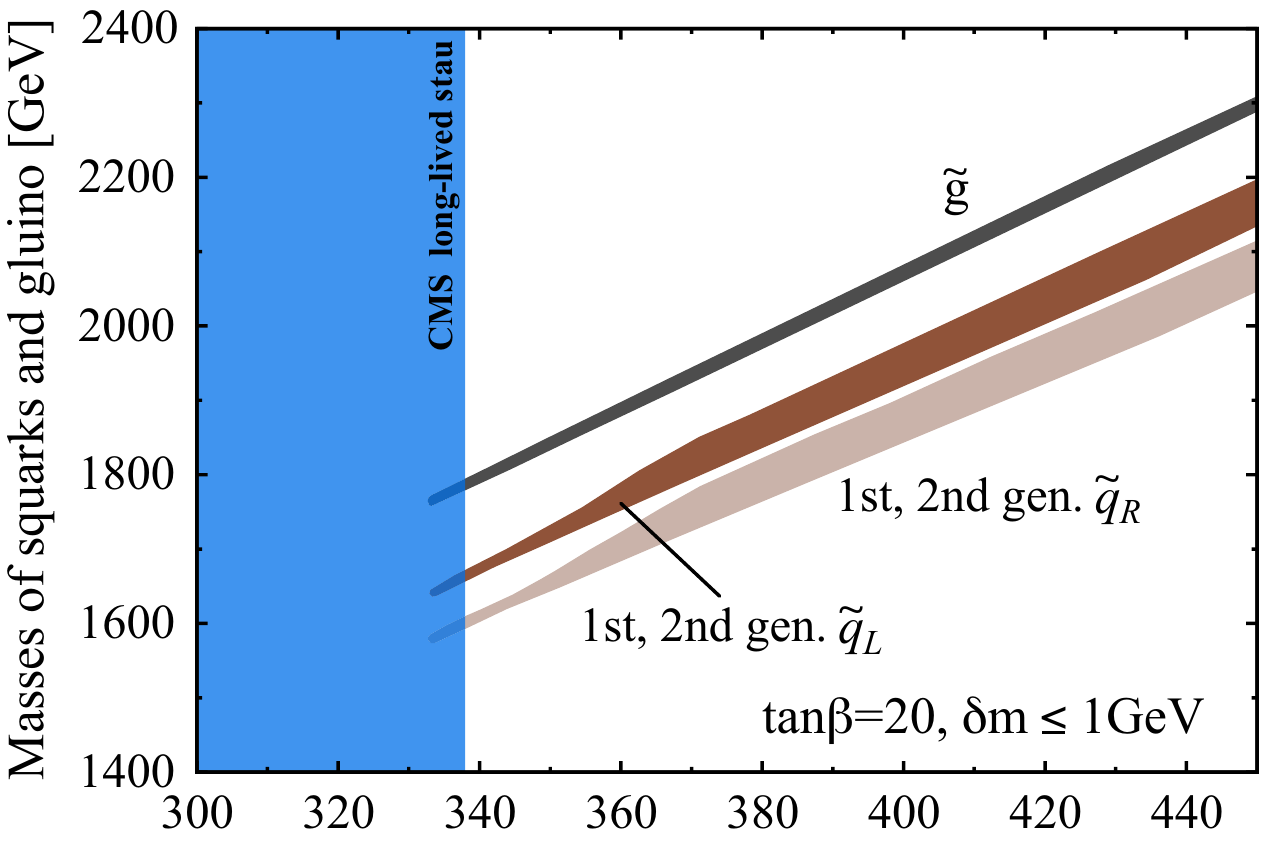}
\end{center}
\end{minipage}
&
\hspace{-4.5mm}
\begin{minipage}{84mm}\vspace{2.3mm}
\begin{center}
 \includegraphics[width=7.7cm,clip]{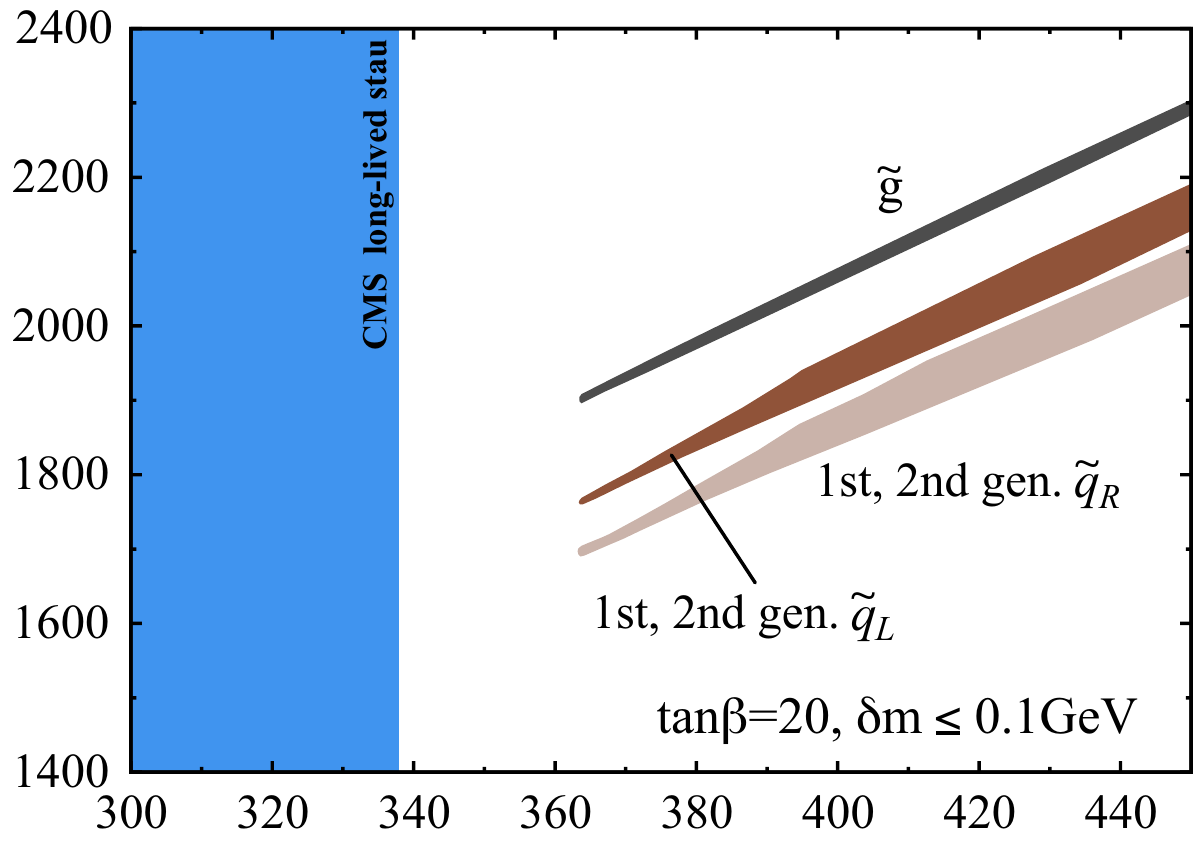}
\end{center}
\end{minipage}
\\[5mm]
\hspace{-1.5mm}
\begin{minipage}{84mm}\vspace{2.5mm}
\begin{center}
 \includegraphics[width=8.1cm,clip]{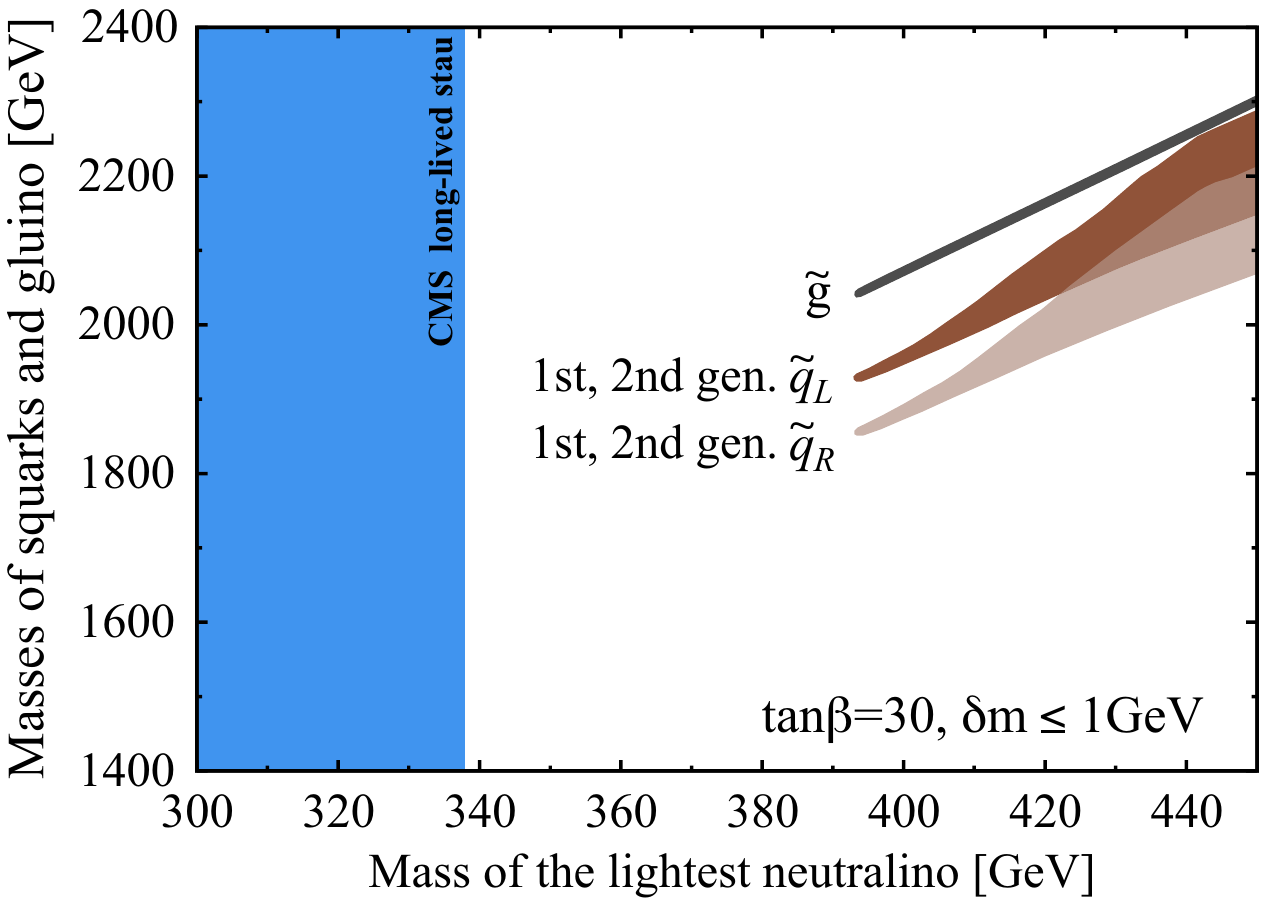}
\end{center}
\end{minipage}
&
\hspace{-4.5mm}
\begin{minipage}{84mm}\vspace{2.3mm}
\begin{center}
 \includegraphics[width=7.7cm,clip]{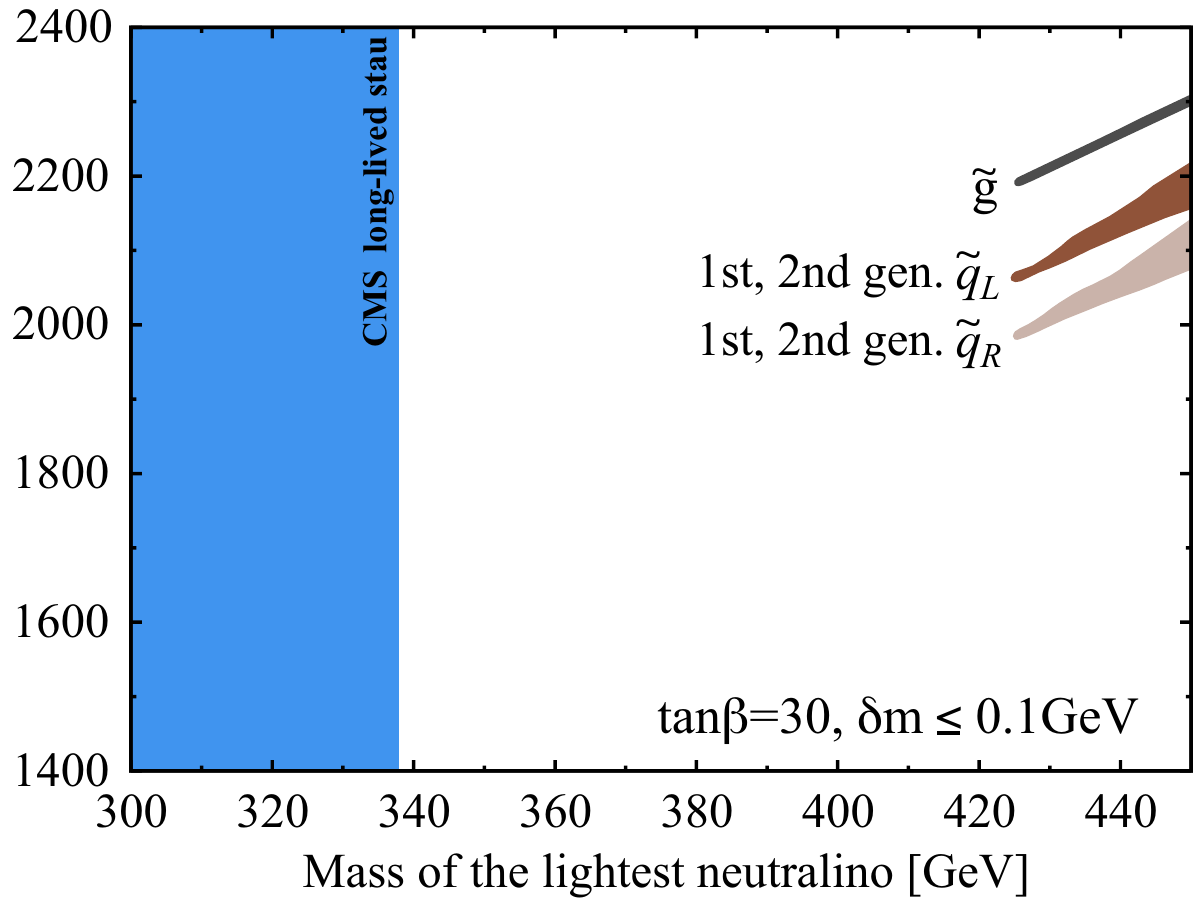}
\end{center}
\end{minipage}
\end{tabular}
\begin{center}
\caption{Mass spectra of the gluino and the first and second generations squarks.  
The horizontal axis represents the mass of the LSP neutralino.   
We fix $\tan \beta$ to $10,~20$ and $30$ from top to bottom 
and $\delta m \leq 1$~and~$0.1~\text{GeV}$ from left to right, respectively.}
\label{fig:msquarksmgluino}
\end{center}
\end{figure*}
\end{center}

\begin{center}
\begin{figure*}[h]
\begin{tabular}{ll}
\begin{minipage}{84mm}
\begin{center}
\includegraphics[width=8.2cm,clip]{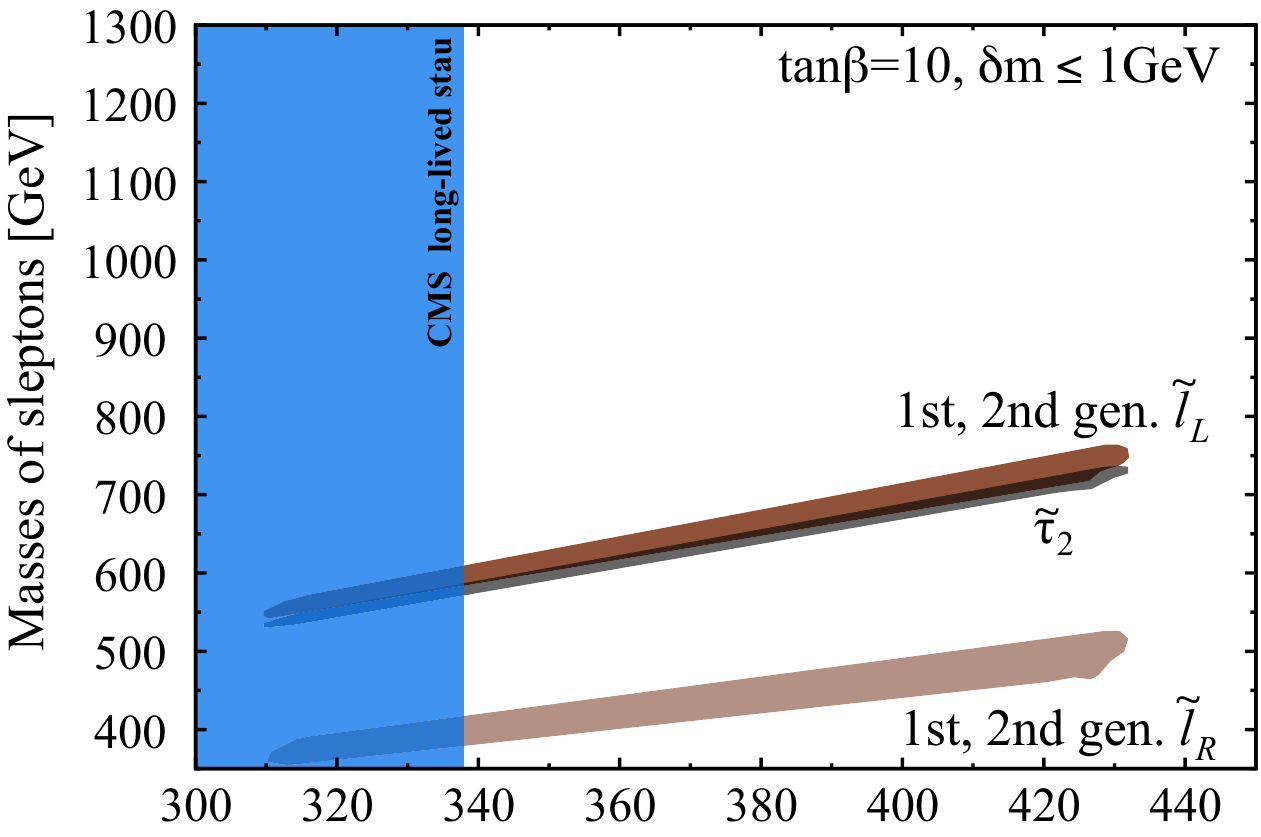}
\end{center}
\end{minipage}
&
\hspace{-3.5mm}
\begin{minipage}{84mm}\vspace{2.6mm}
\begin{center}
 \includegraphics[width=7.76cm,clip]{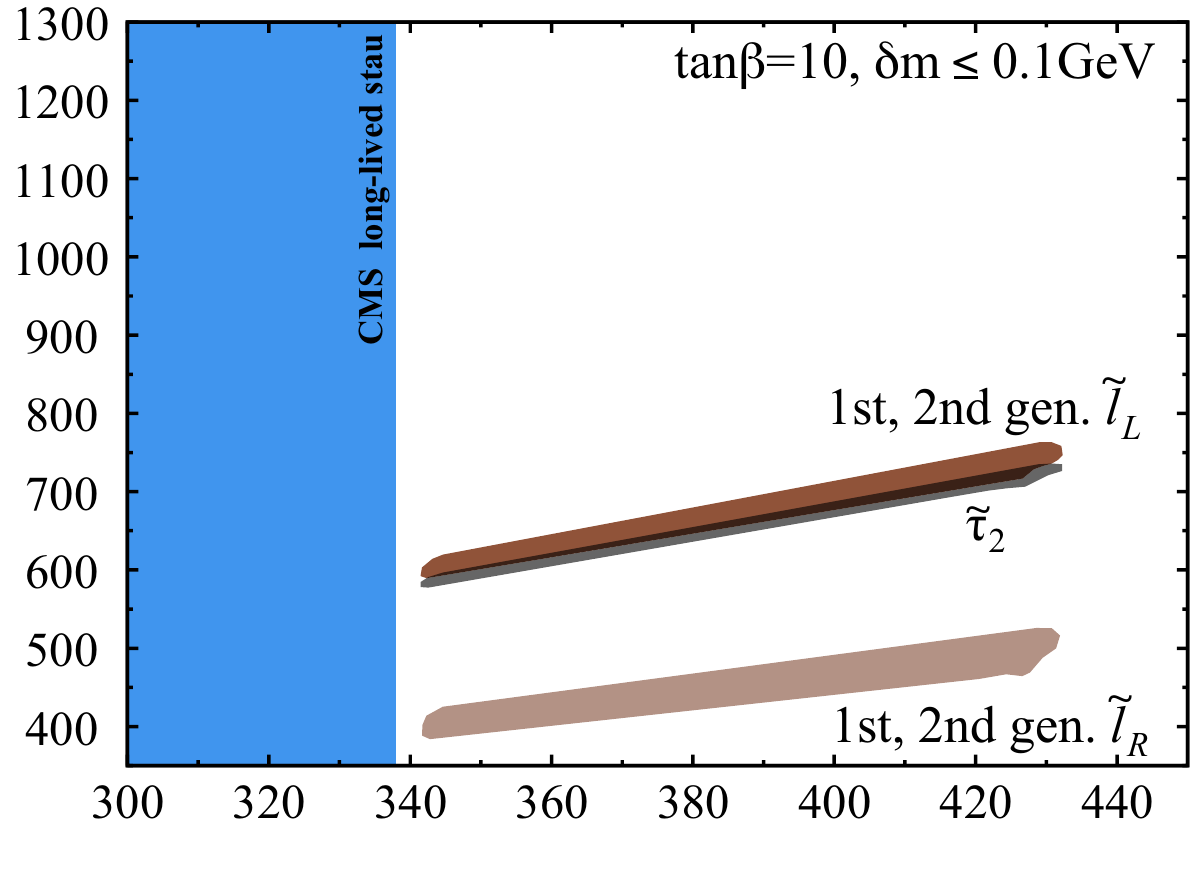}
\end{center}
\end{minipage}
\\[5mm]
\begin{minipage}{84mm}\vspace{-0.6mm}
\begin{center}
 \includegraphics[width=8.2cm,clip]{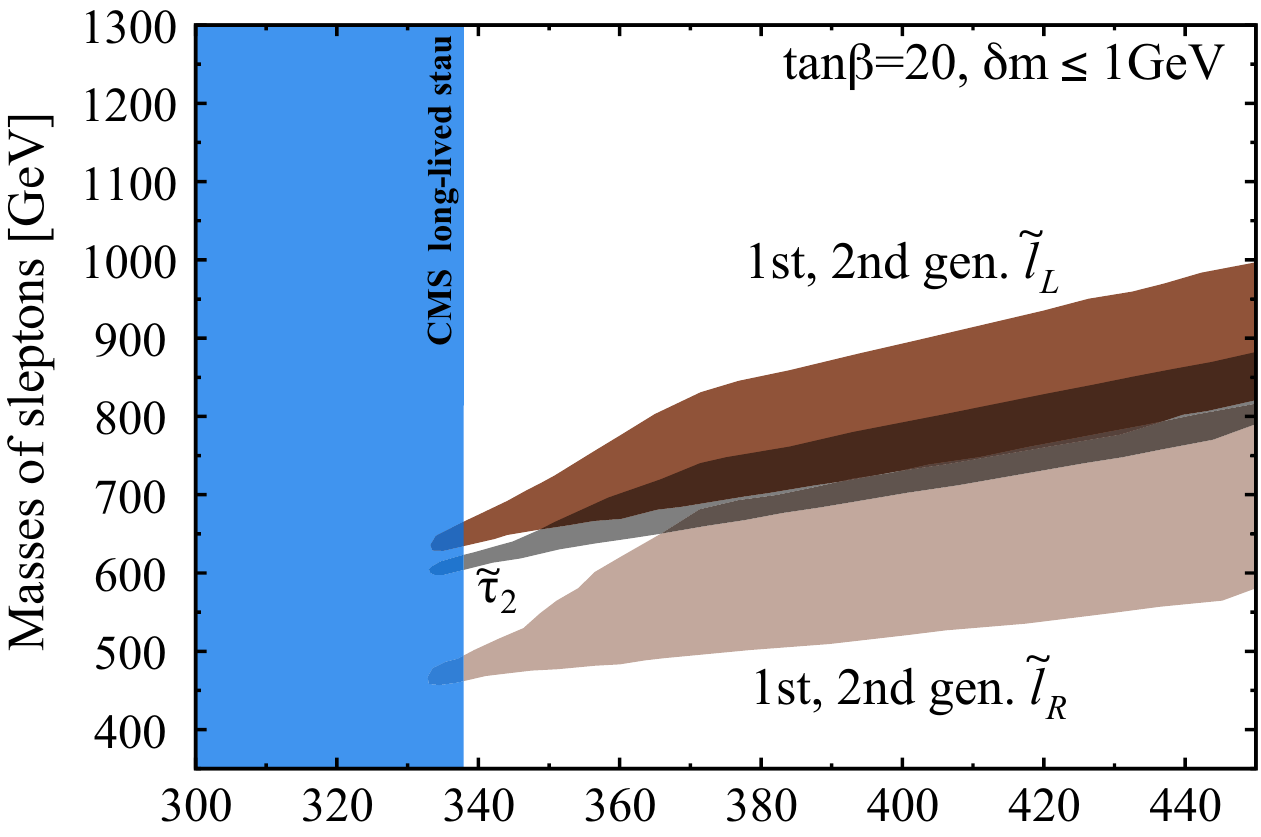}
\end{center}
\end{minipage}
&
\hspace{-3.5mm}
\begin{minipage}{84mm}\vspace{-0.6mm}
\begin{center}
 \includegraphics[width=7.76cm,clip]{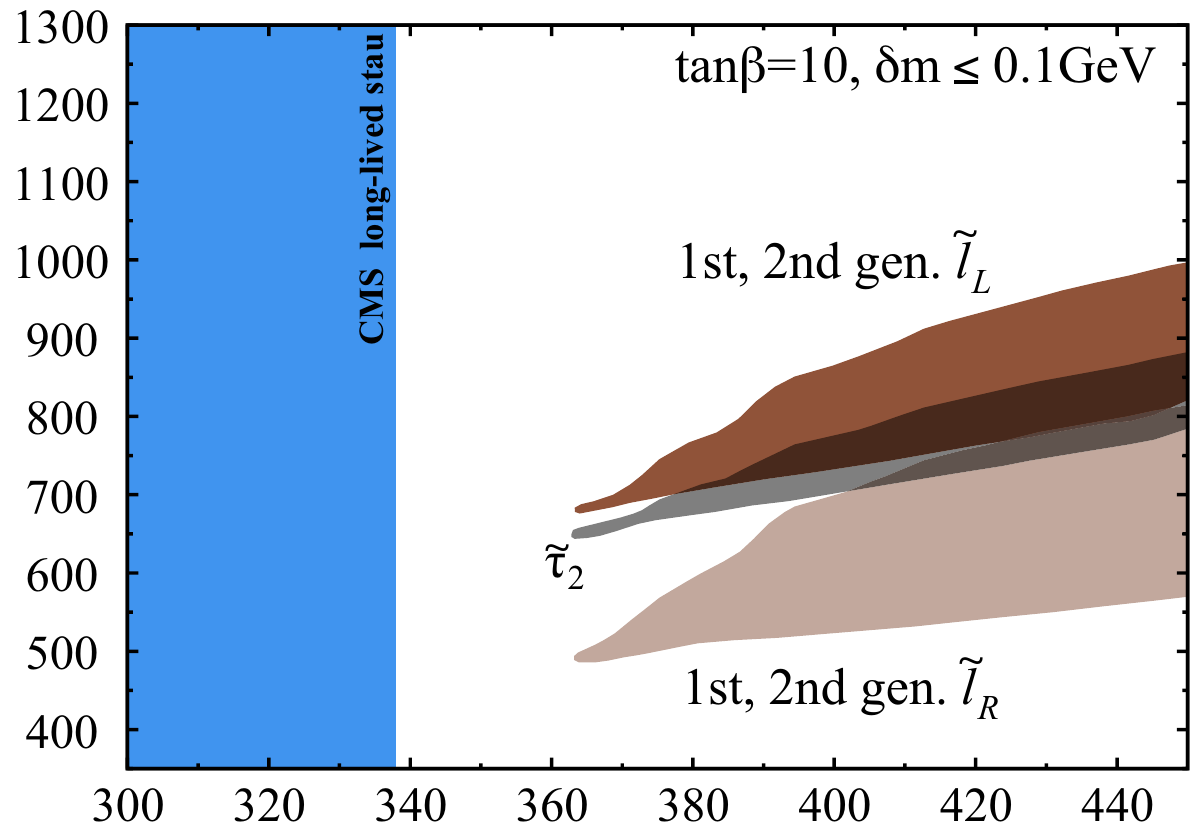}
\end{center}
\end{minipage}
\\[5mm]
\begin{minipage}{84mm}\vspace{2mm}
\begin{center}
\includegraphics[width=8.2cm,clip]{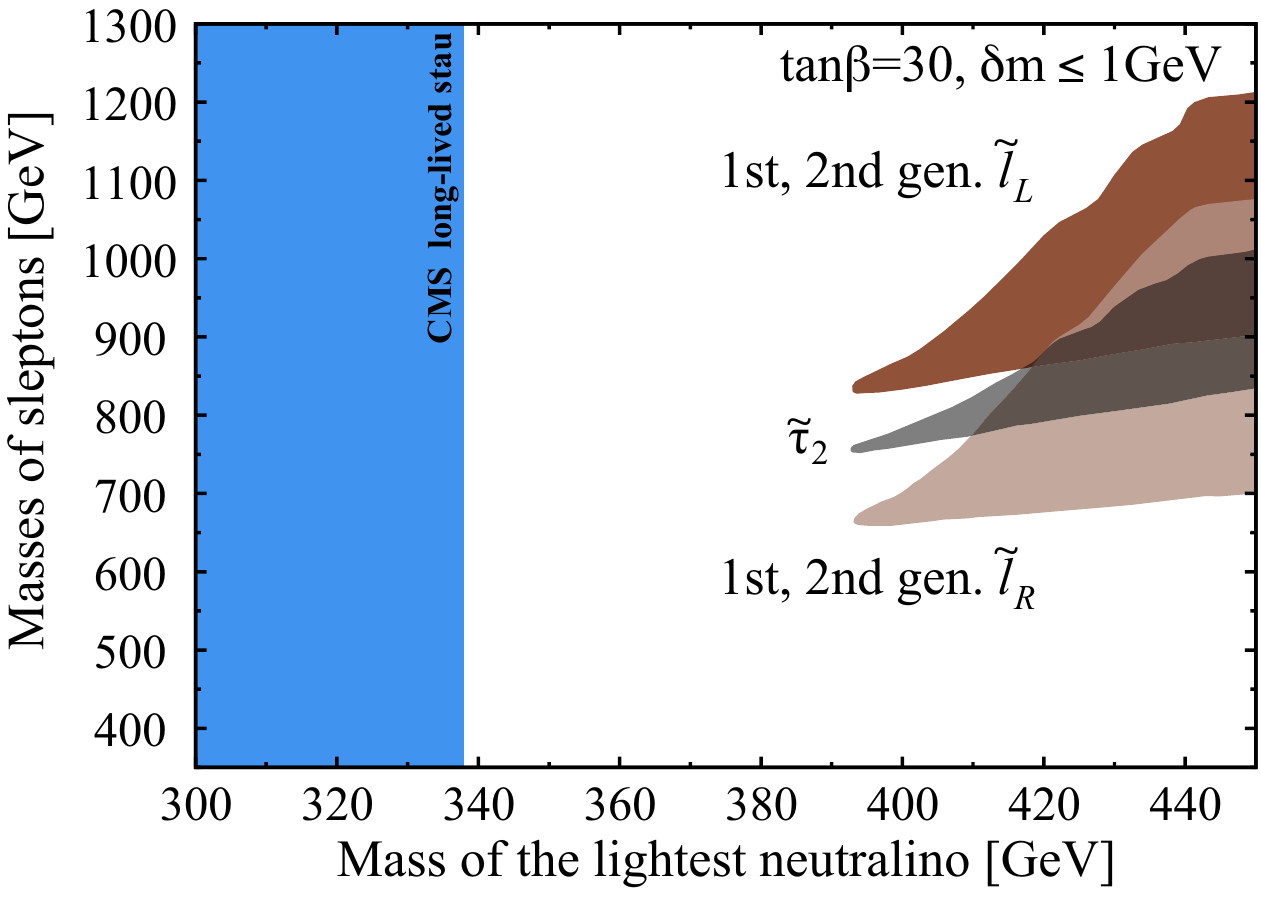}
\end{center}
\end{minipage}
&
\hspace{-3.5mm}
\begin{minipage}{84mm}\vspace{2mm}
\begin{center}
  \includegraphics[width=7.76cm,clip]{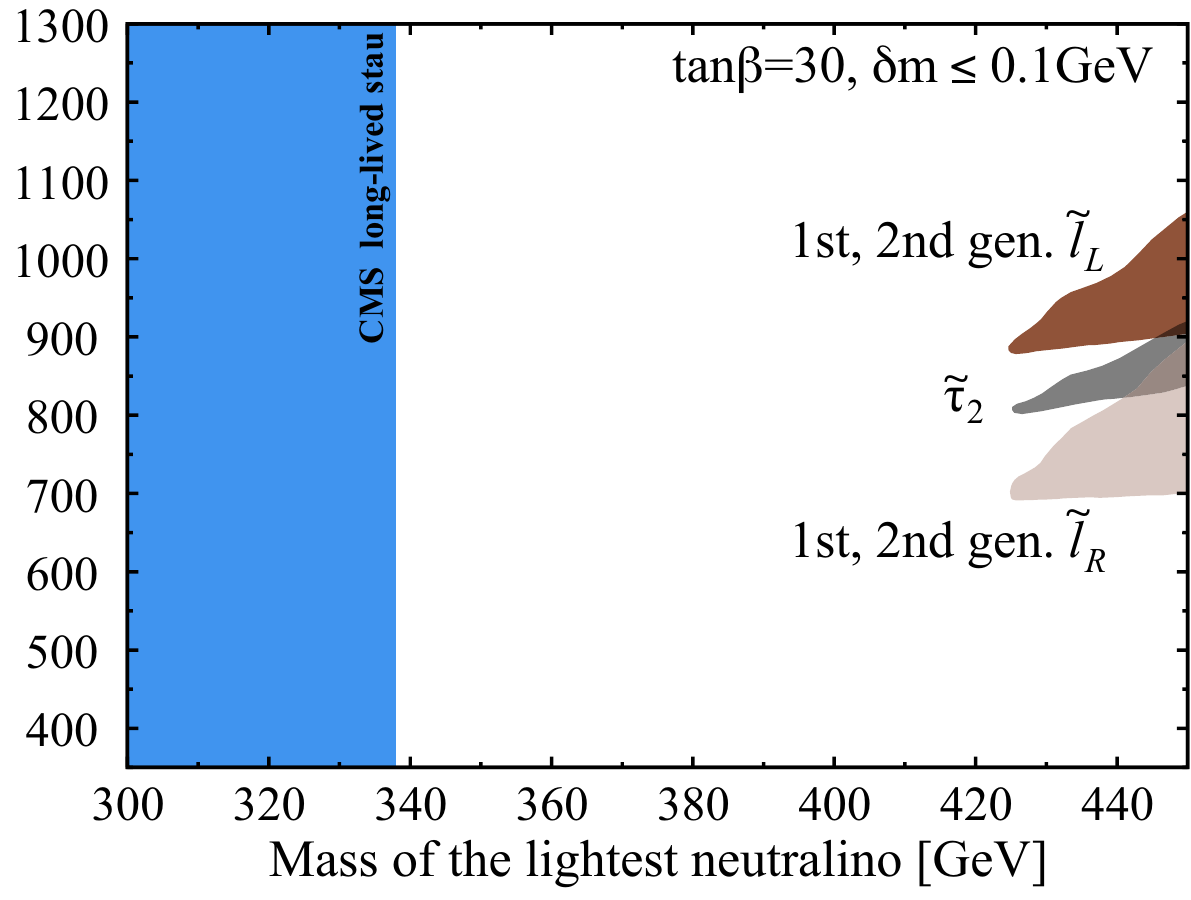}
\end{center}
\end{minipage}
\end{tabular}
\caption{Mass spectra of the slepton.  
The horizontal axis represents the mass of the LSP neutralino.   
We fix $\tan \beta$ to $10,~20$ and $30$ from top to bottom 
and $\delta m \leq 1$~and~$0.1~\text{GeV}$ from left to right, respectively.}
\label{fig:mslepton}
\end{figure*}
\end{center}

For the first and the second generation squarks and sleptons, the effects of 
the corresponding Yukawa couplings are negligible in RG evolutions of their 
soft masses. 
The soft SUSY breaking masses are parameterized up to the one-loop order 
as~\cite{Martin:1997ns},  
\begin{subequations}
\begin{align}
  m _{\tilde q _{L}} ^{2} 
  &\simeq m _{0} ^2 + 4.7 M _{1/2} ^2, 
  \label{eq:qlmass} \\
  m _{\tilde q _{R}} ^{2} 
  &\simeq m _{0} ^2 + 4.3 M _{1/2} ^2, 
  \label{eq:qrmass} \\
  m _{\tilde e _{L}} ^{2}
  &\simeq m _{0} ^2 + 0.5 M _{1/2} ^2, 
  \label{eq:elmass} \\
  m _{\tilde e _{R}} ^{2}
  &\simeq m _{0} ^2 + 0.1 M _{1/2} ^2.
  \label{eq:ermass}
\end{align}
\end{subequations}
Note that the slepton masses are sensitive to $m _{0}$
in Fig.~\ref{fig:mslepton} 
while the squark masses are insensitive to $m _{0}$ in Fig.~\ref{fig:msquarksmgluino}.  
In Eqs.~(\ref{eq:qlmass}) and (\ref{eq:qrmass}) the second term is 
dominant, and hence we approximate these equations more roughly 
as follows:
\begin{equation}
\begin{split}
    m _{\tilde q _{L}} 
  &\simeq 2.2 M _{1/2}, \\
  m _{\tilde q _{R}} 
  &\simeq 2.1 M _{1/2}. 
\end{split}
\end{equation}
Meanwhile, in Eqs.~(\ref{eq:elmass}) and (\ref{eq:ermass}) the 
contributions to soft mass from the first term is comparable to that of 
the second term. Therefore, the slepton masses are sensitive to 
$m _{0}$.

\subsubsection{Stop mass spectra} 

\begin{center}
\begin{figure*}[h]
\begin{tabular}{ll}
\begin{minipage}{84mm}
\begin{center}
 \includegraphics[width=8.2cm,clip]{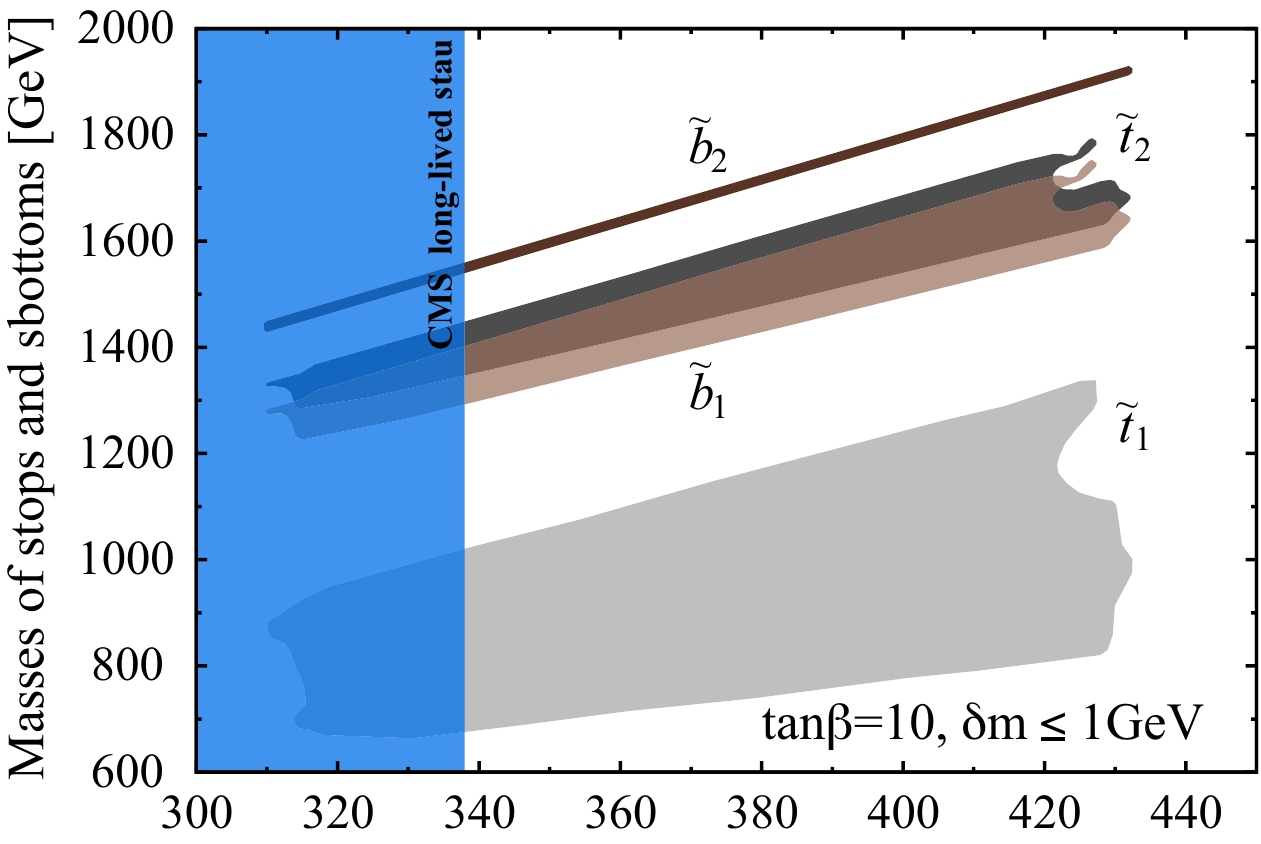}
\end{center}
\end{minipage}
&
\hspace{-3.5mm}
\begin{minipage}{84mm}
\begin{center}
\includegraphics[width=7.78cm,clip]{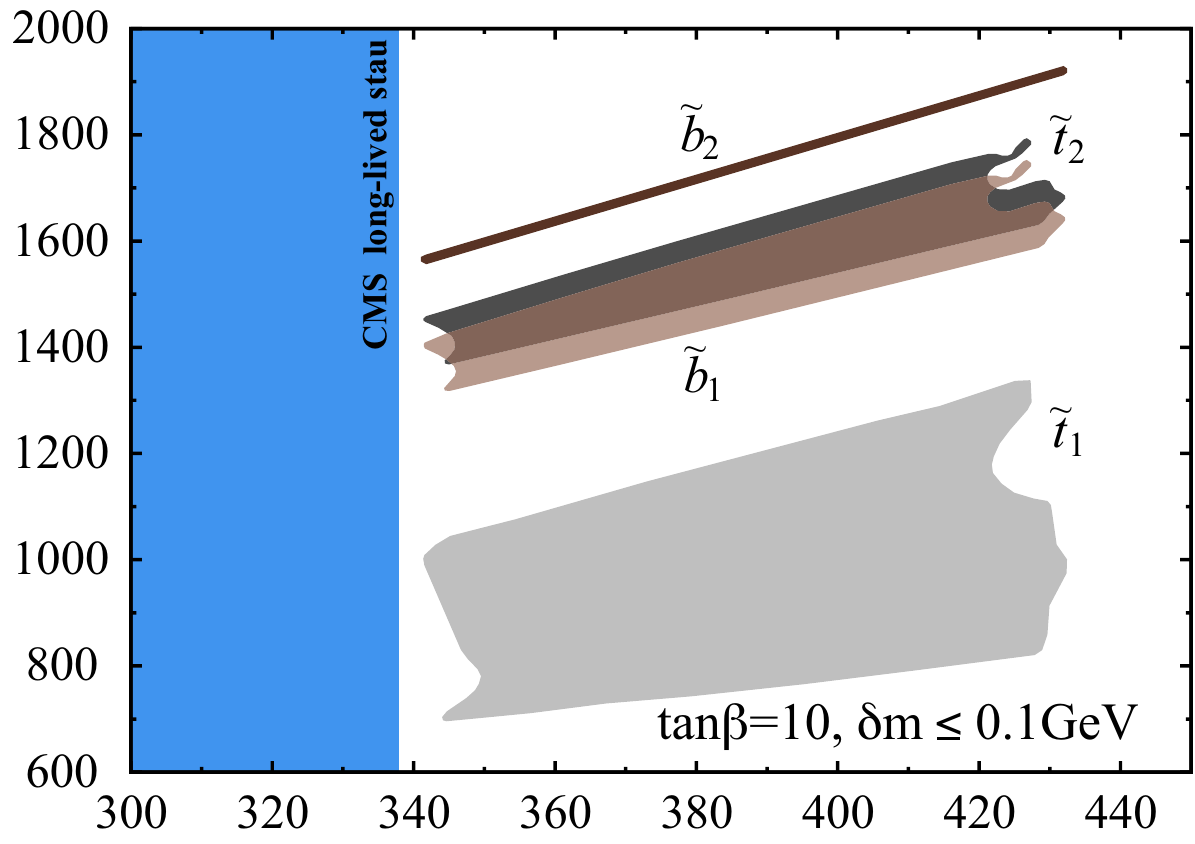}
\end{center}
\end{minipage}
\\[5mm]
\begin{minipage}{84mm}\vspace{2.4mm}
\begin{center}
 \includegraphics[width=8.2cm,clip]{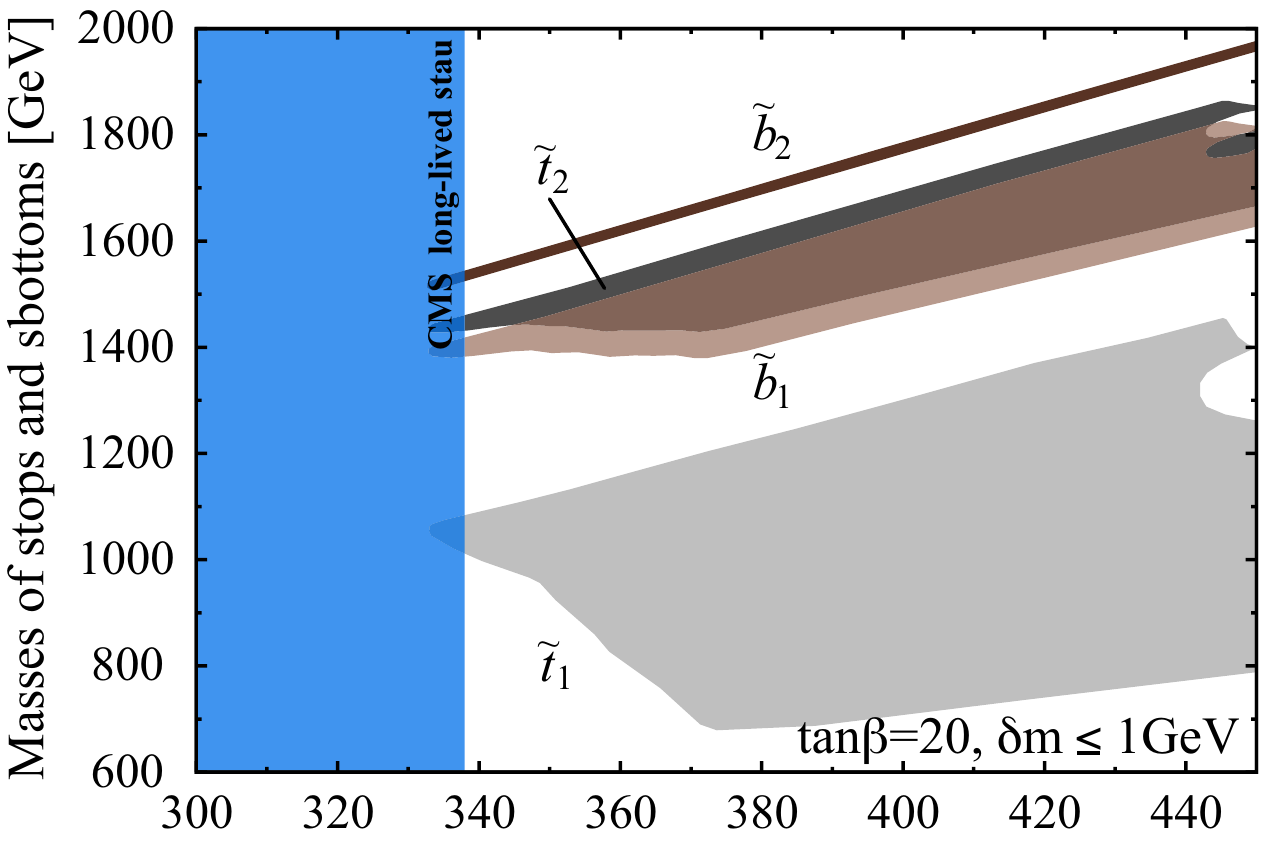}
\end{center}
\end{minipage}
&
\hspace{-3.5mm}
\begin{minipage}{84mm}\vspace{2.4mm}
\begin{center}
 \includegraphics[width=7.78cm,clip]{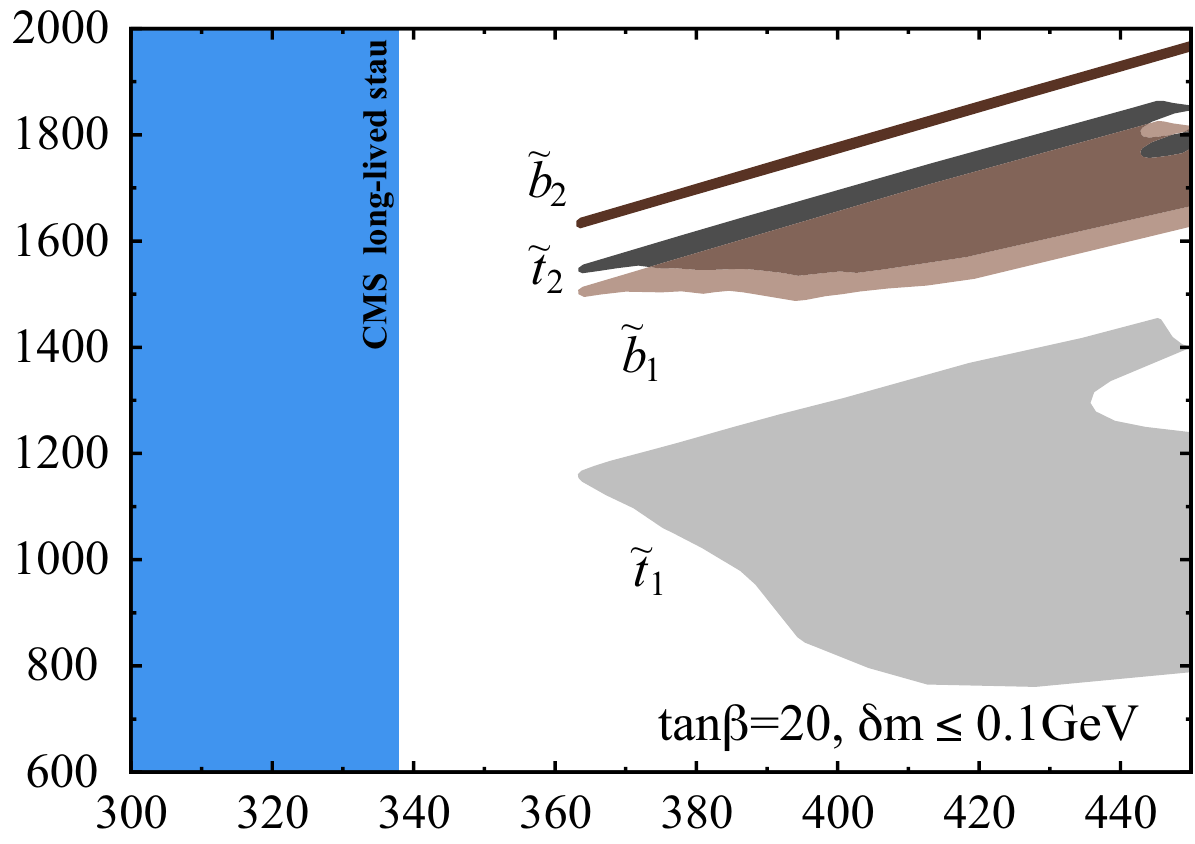}
\end{center}
\end{minipage}
\\[5mm]
\begin{minipage}{84mm}\vspace{2.4mm}
\begin{center}
  \includegraphics[width=8.2cm,clip]{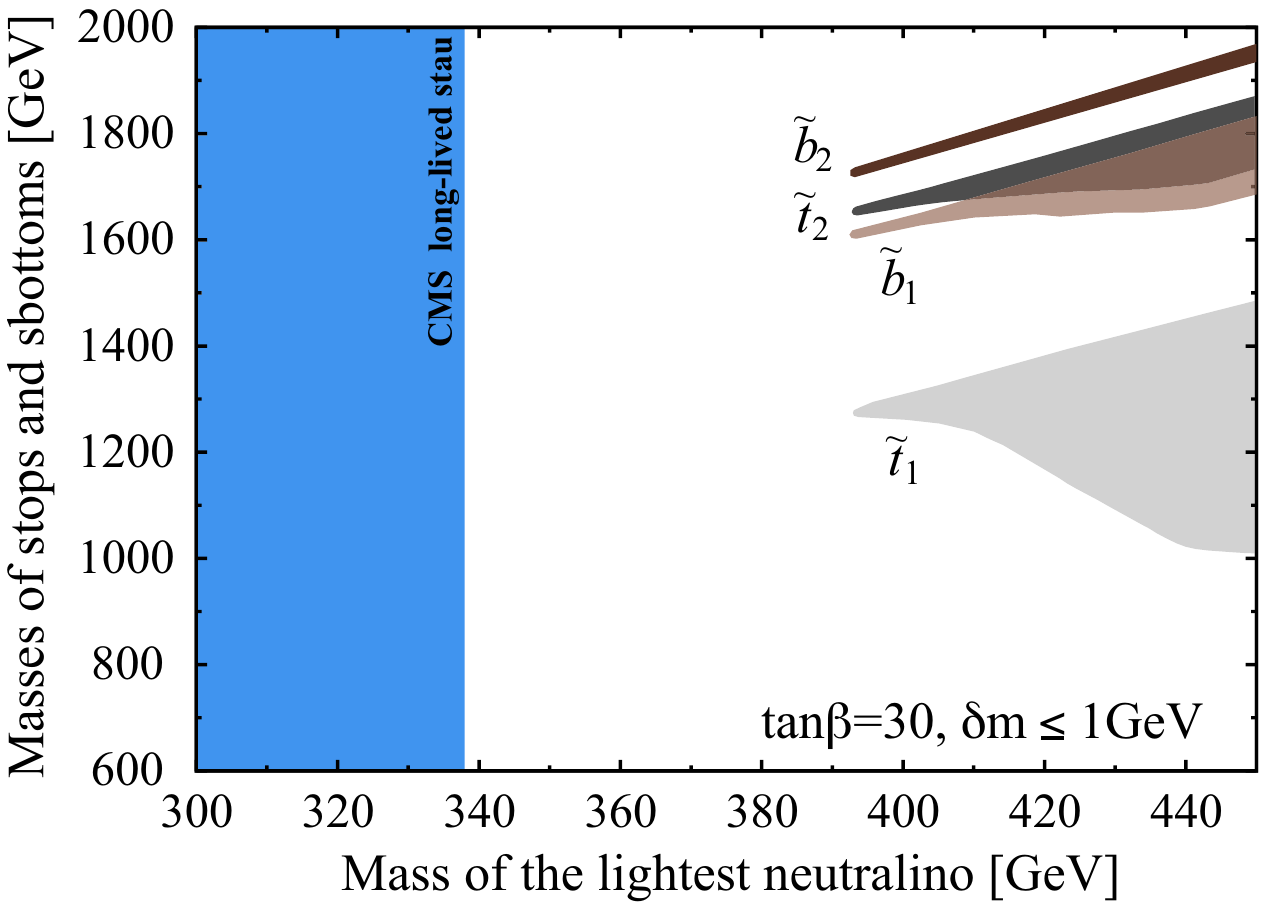}
\end{center}
\end{minipage}
&
\hspace{-3.5mm}
\begin{minipage}{84mm}\vspace{2.4mm}
\begin{center}
\includegraphics[width=7.78cm,clip]{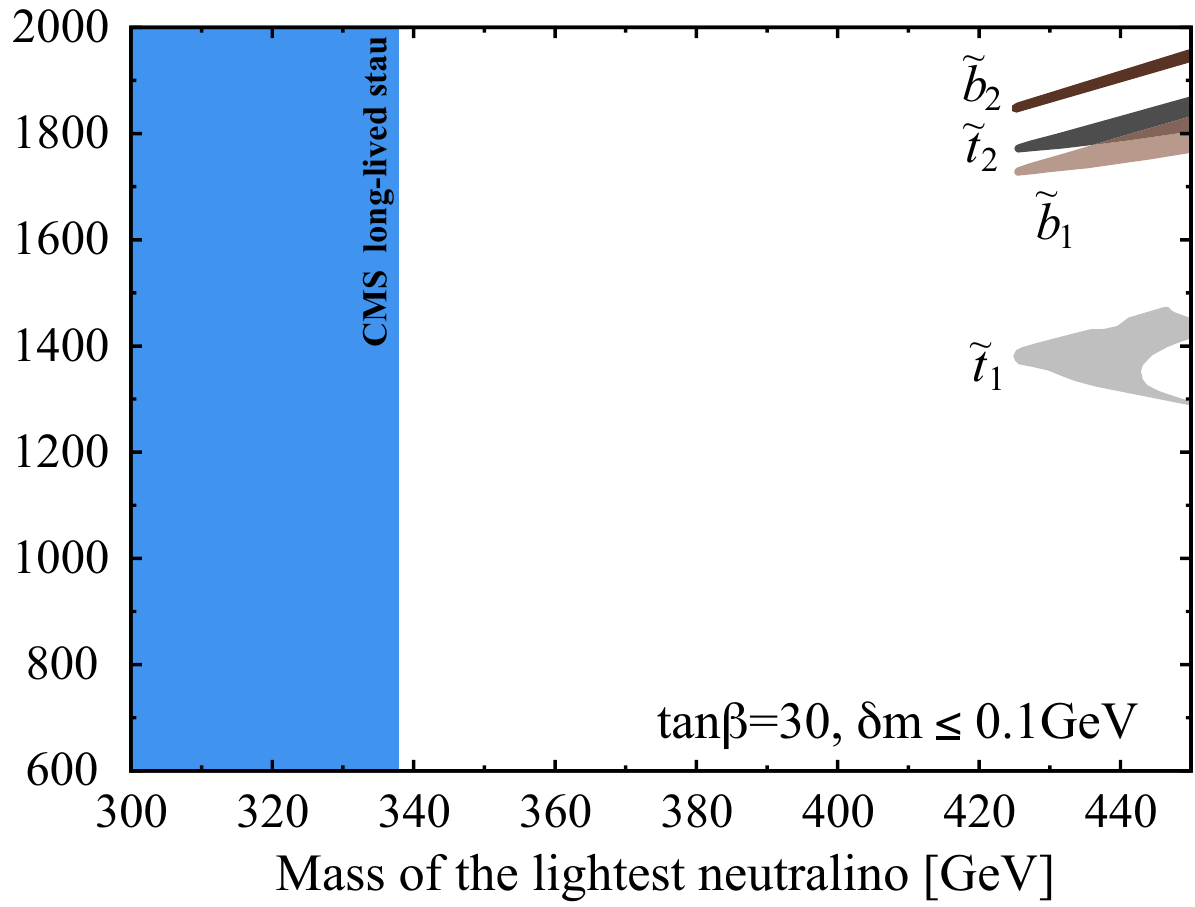}
\end{center}
\end{minipage}
\end{tabular}
\caption{Mass spectra of the stop and the sbottom.  
The horizontal axis expresses the mass of the LSP neutralino.   
We fix $\tan \beta$ to $10,~20$ and $30$ from top to bottom 
and $\delta m \leq 1$~and~$0.1~\text{GeV}$ from left to right, respectively.}
\label{fig:mstop}
\end{figure*}
\end{center}

Figure~\ref{fig:mstop} shows the masses of the stops. Unlike in the 
case of the other sfermions, the distributions of stop masses are in 
nonlinear relation. The spread of the distributions in the lightest 
neutralino mass-stop mass plane is understood as follows.

The masses of the stops in mass eigenstate are given by
\begin{subequations}
\begin{align}
  & 
  m^2_{\tilde t_1, \tilde t_2} \simeq \frac{1}{2} 
  \left(m^2_{Q_3} + m^2_{U_3} \right) \notag 
  \\& \hspace{1cm} \mp 
  \frac{1}{2} \sqrt{(m^2_{Q_3} - m^2_{U_3})^2
  + 4 (m^2_{\tilde t_{LR}})^2},
  \label{eq:stopmassA} 
  \\& 
  m^2_{\tilde t_{LR}}
  = m_t (A_t - \mu \cot \beta ), 
  \label{eq:stopmassB}
\end{align}
\end{subequations}
where $m_{Q_3}$ and $m_{U_3}$ are the soft SUSY breaking masses, and 
$A_t$ is the stop trilinear coupling.  In Eq.~(\ref{eq:stopmassA}) the first 
term is dominant, and it is a decreasing function of $m_0^2$ because of the 
top and bottom Yukawa couplings. 
Meanwhile, the second term in Eq.~(\ref{eq:stopmassA}) is an 
increasing function of $m_0^2$. This is because the dominant term in 
the square root is $m_{\tilde t_{LR}}^2$ involving $A_t$, and the 
coupling $A_t$ is proportional to $m_0$ due to the relation between 
$A_0$ and $m_0$ in Eq.~(\ref{eq:linearrelation}).  
Hence, for a fixed lightest neutralino mass, stop masses are not 
simply determined as a function of the lightest neutralino mass, and 
rather spread depending on the $m_0^2$.

The ATLAS collaboration gives the bound for the stop mass 
from the direct stop search at 8 TeV LHC run~\cite{ATLAS:2013pla}.  
Note that our results of the stop mass are safely above the bound. 
\begin{center}
\begin{figure*}[h]
\begin{tabular}{ll}
\begin{minipage}{84mm}
\begin{center}
 \includegraphics[width=8.2cm,clip]{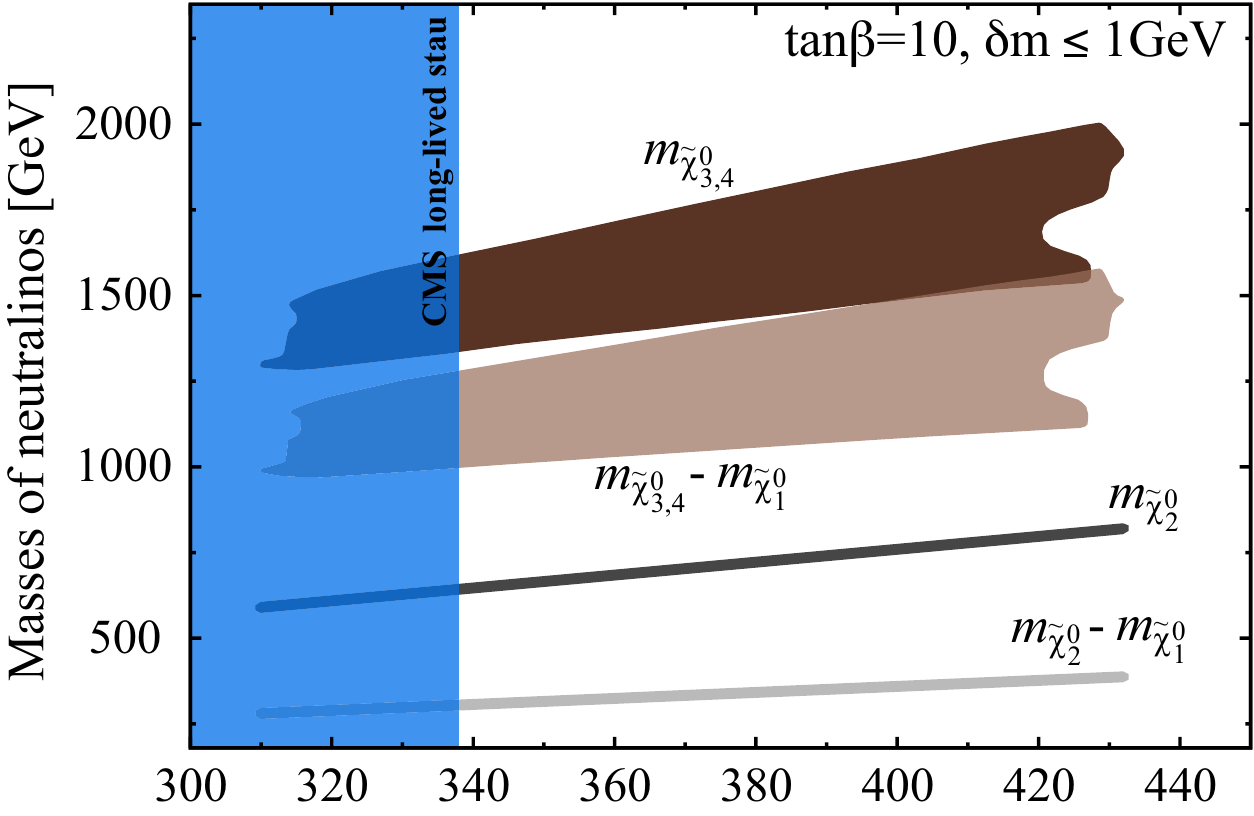}
\end{center}
\end{minipage}
&
\hspace{-3.5mm}
\begin{minipage}{84mm}
\begin{center}
 \includegraphics[width=7.78cm,clip]{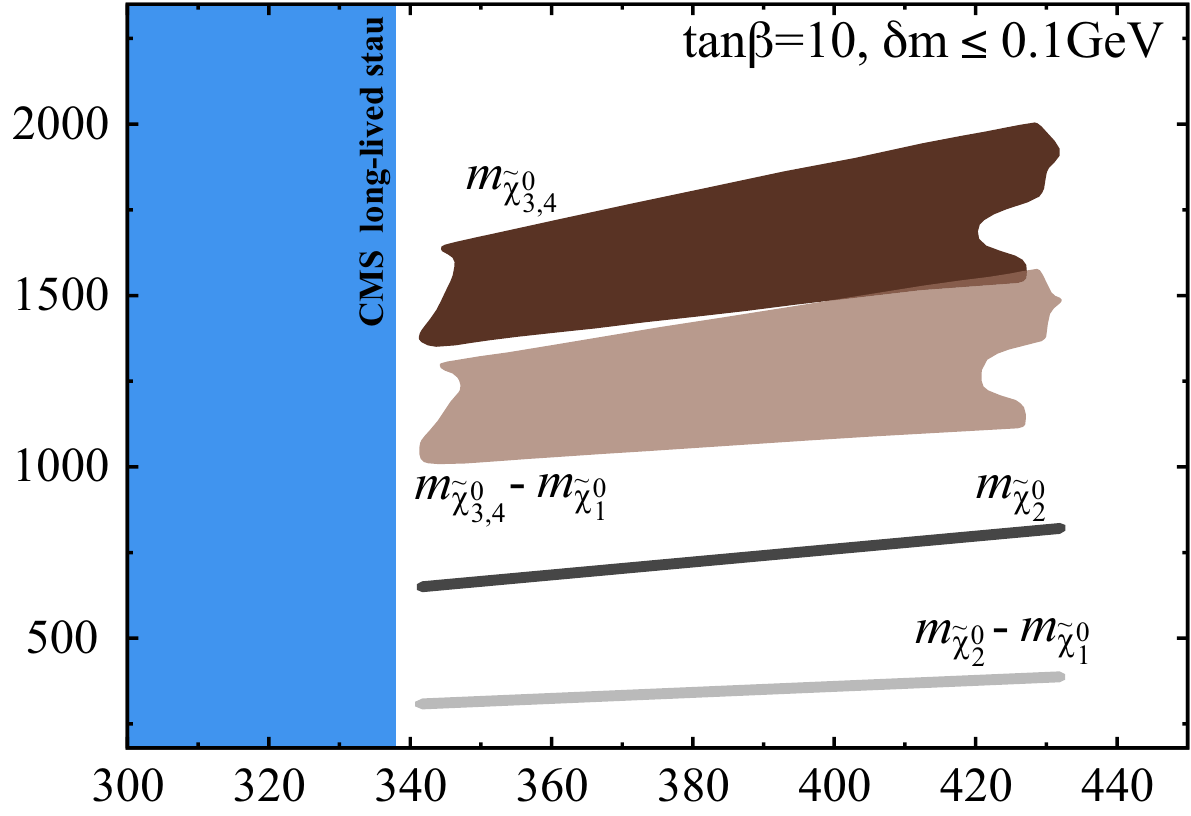}
\end{center}
\end{minipage}
\\[5mm]
\begin{minipage}{84mm}\vspace{3.5mm}
\begin{center}
 \includegraphics[width=8.2cm,clip]{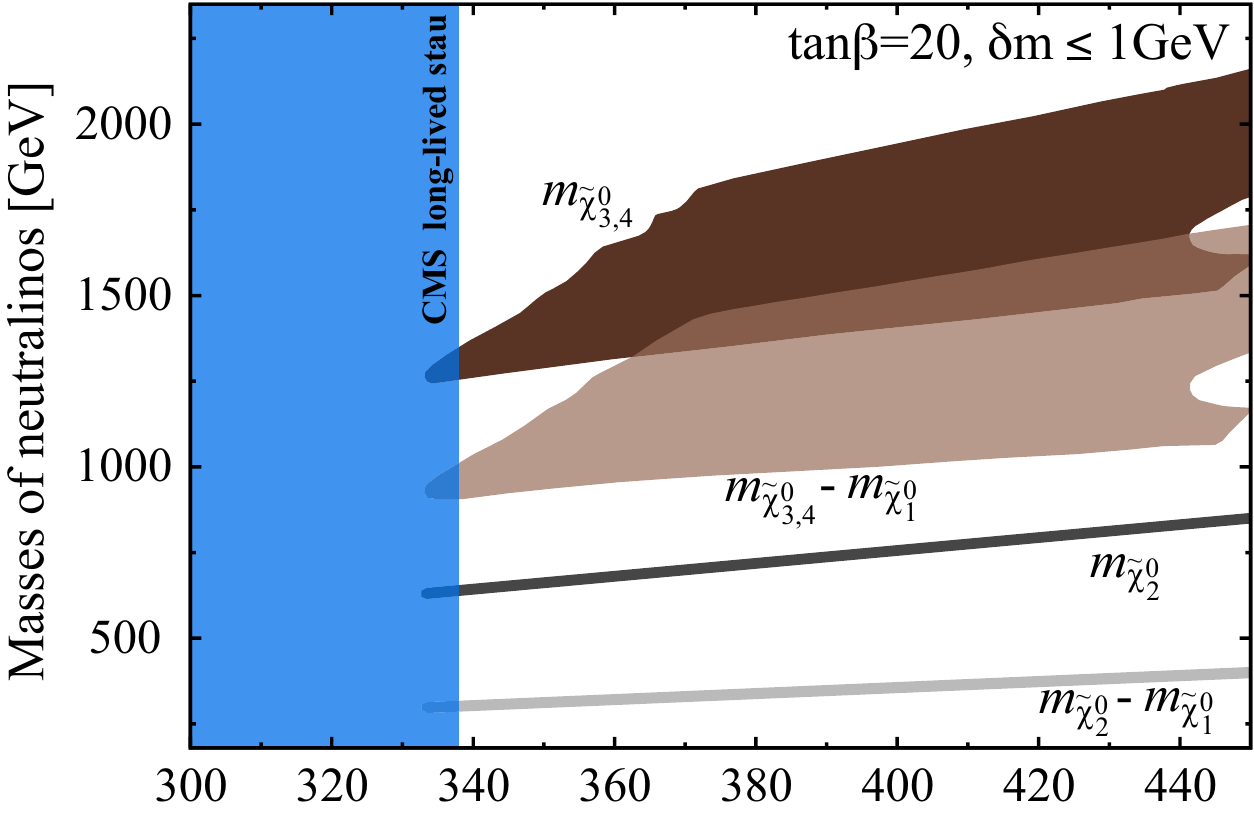}
\end{center}
\end{minipage}
&
\hspace{-3.5mm}
\begin{minipage}{84mm}\vspace{3.5mm}
\begin{center}
 \includegraphics[width=7.78cm,clip]{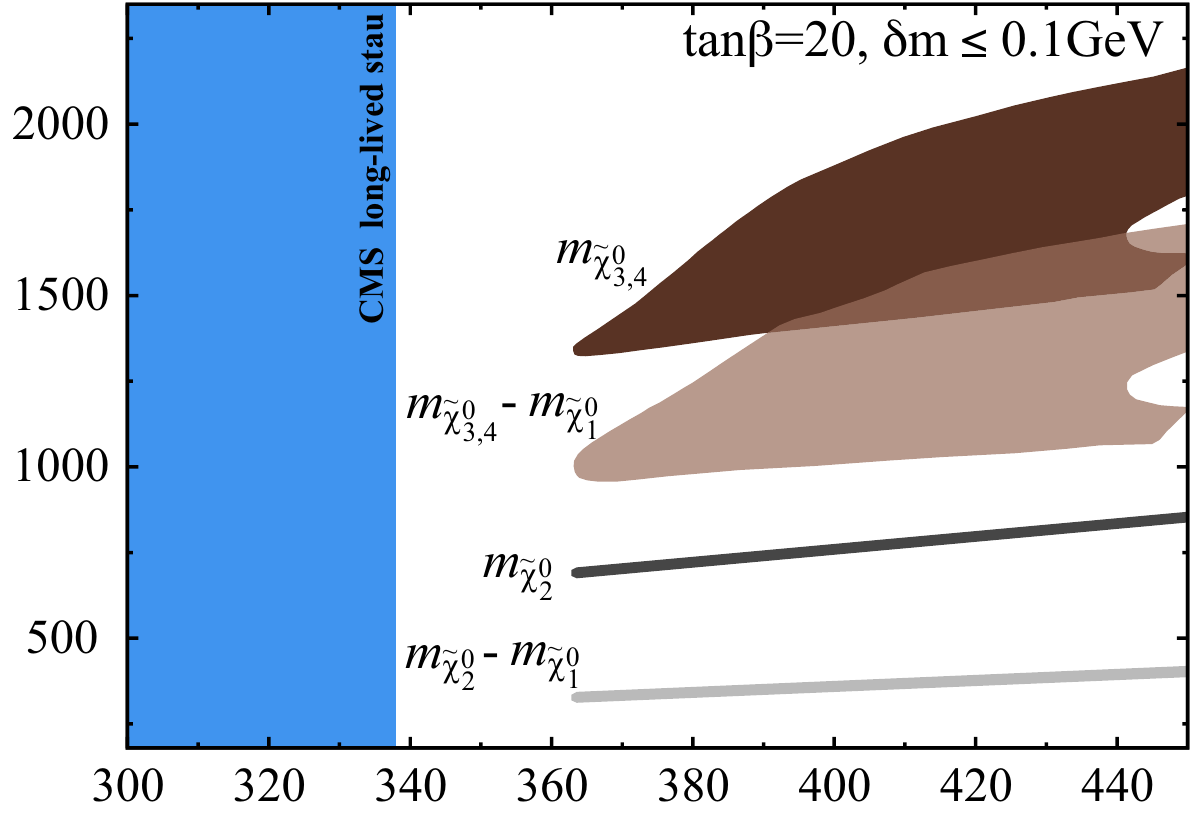}
\end{center}
\end{minipage}
\\[5mm]
\begin{minipage}{84mm}\vspace{3.5mm}
\begin{center}
 \includegraphics[width=8.2cm,clip]{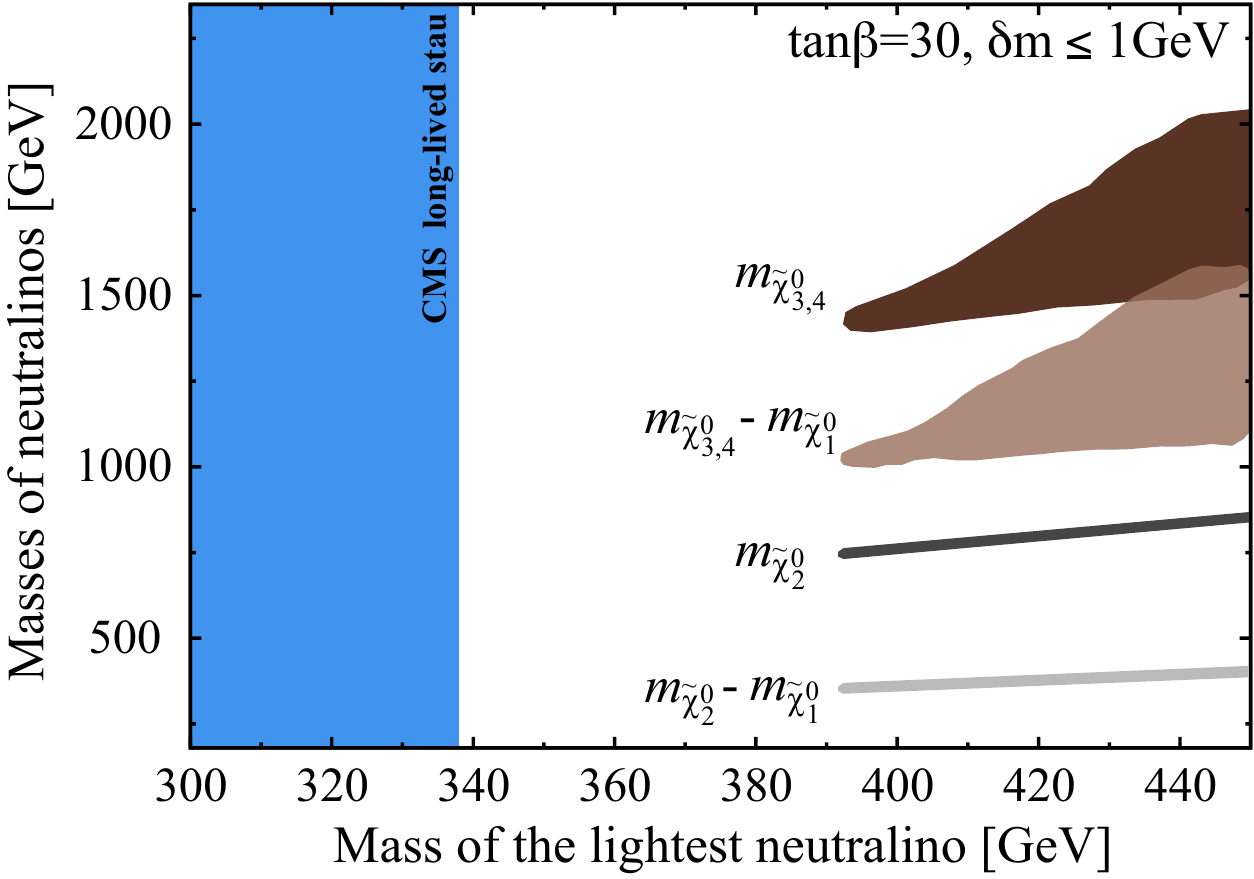}
\end{center}
\end{minipage}
&
\hspace{-3.5mm}
\begin{minipage}{84mm}\vspace{3.5mm}
\begin{center}
  \includegraphics[width=7.78cm,clip]{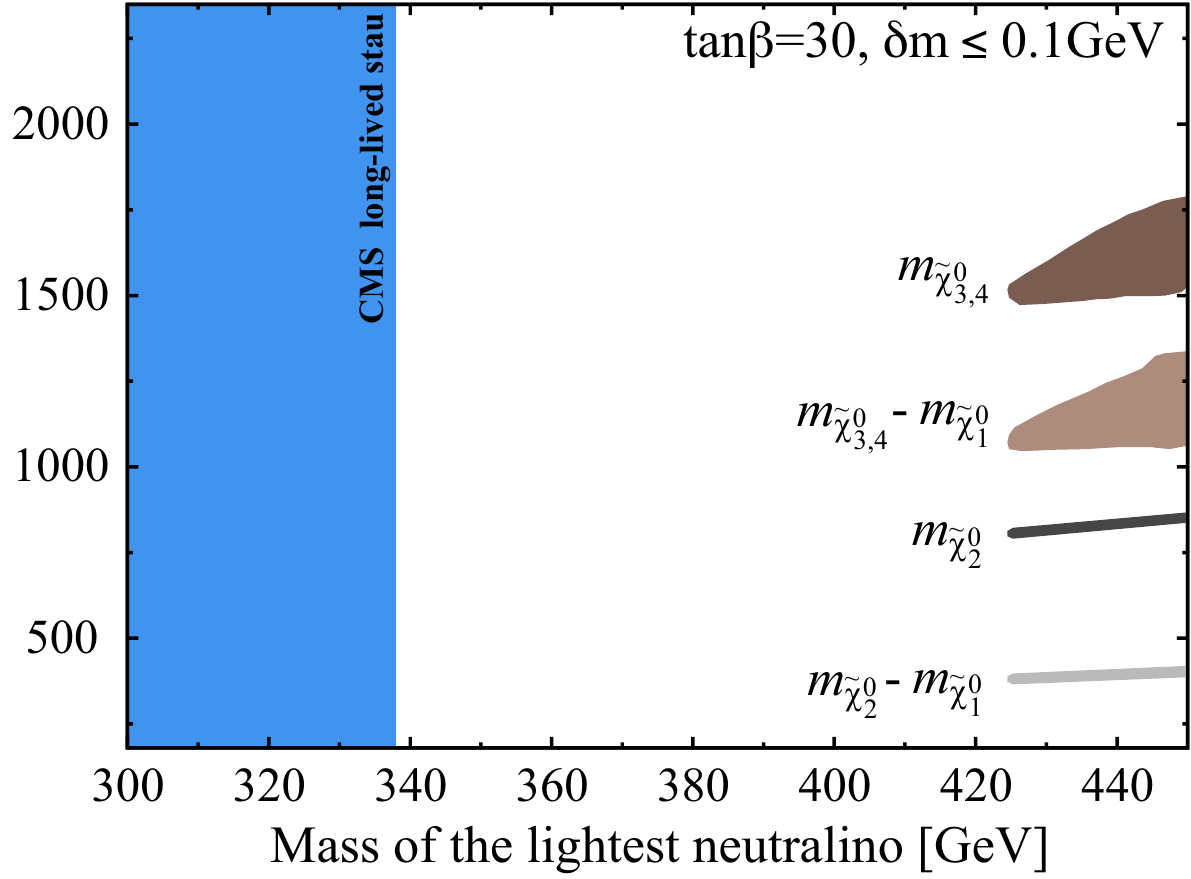}
\end{center}
\end{minipage}
\end{tabular}
\caption{Mass spectra of the neutralino.  
The horizontal axis represents the mass of the LSP neutralino.   
We fix $\tan \beta$ to $10,~20$ and $30$ from top to bottom 
and $\delta m \leq 1$~and~$0.1~\text{GeV}$ from left to right, respectively.}
\label{fig:mneutralino} 
\end{figure*}
\end{center}
\begin{center}
\begin{figure*}[h]
\begin{tabular}{ll}
\begin{minipage}{84mm}
\begin{center}
 \includegraphics[width=8.2cm,clip]{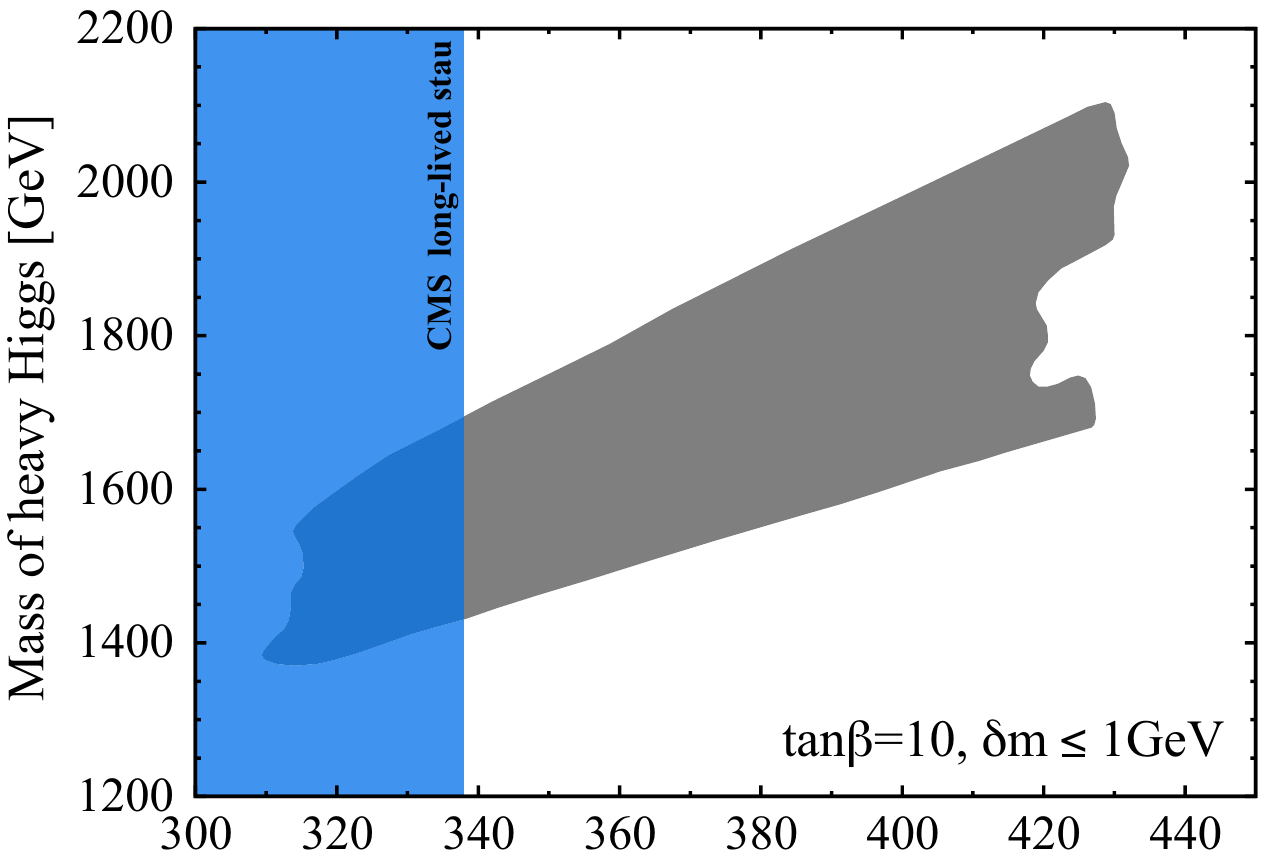}
\end{center}
\end{minipage}
&
\hspace{-3.5mm}
\begin{minipage}{84mm}
\begin{center}
 \includegraphics[width=7.76cm,clip]{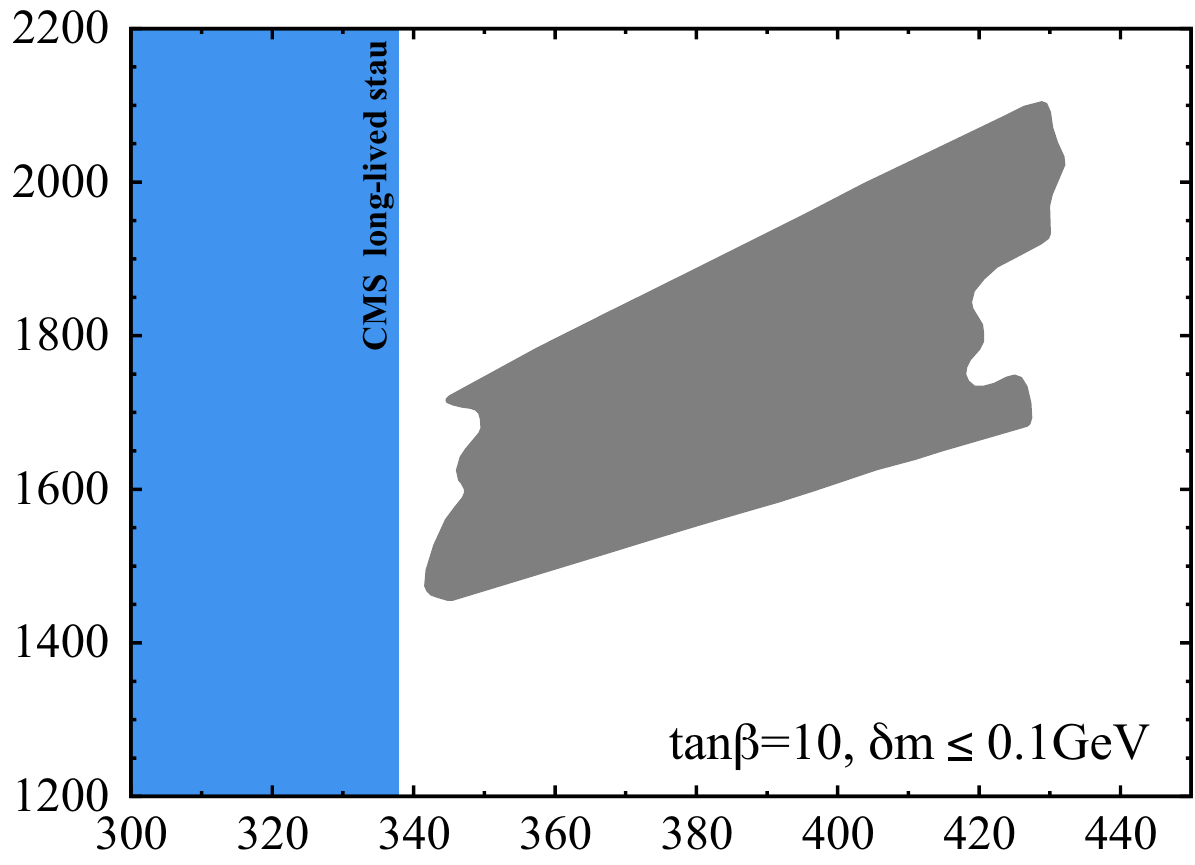}
\end{center}
\end{minipage}
\\[5mm]
\begin{minipage}{84mm}\vspace{2mm}
\begin{center}
 \includegraphics[width=8.2cm,clip]{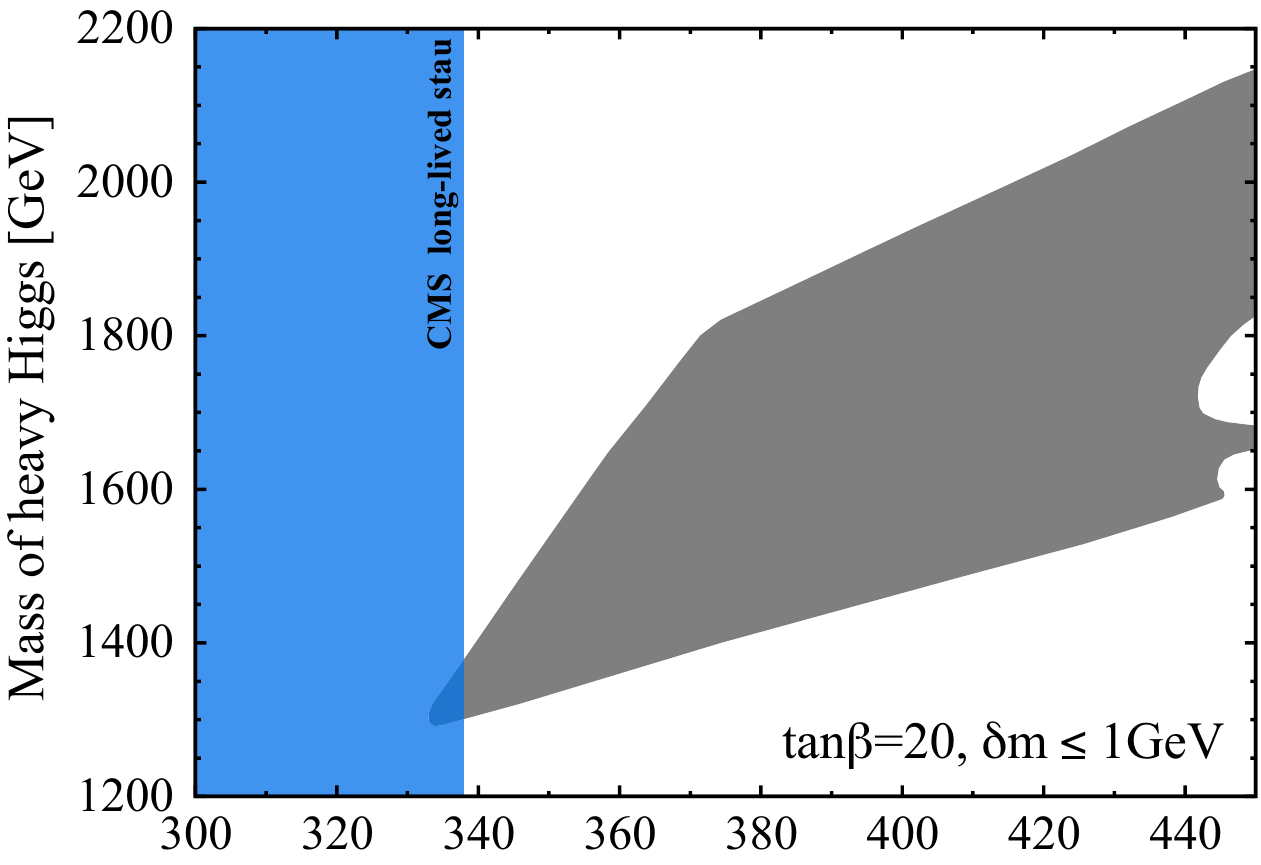}
\end{center}
\end{minipage}
&
\hspace{-3.5mm}
\begin{minipage}{84mm}\vspace{2mm}
\begin{center}
 \includegraphics[width=7.76cm,clip]{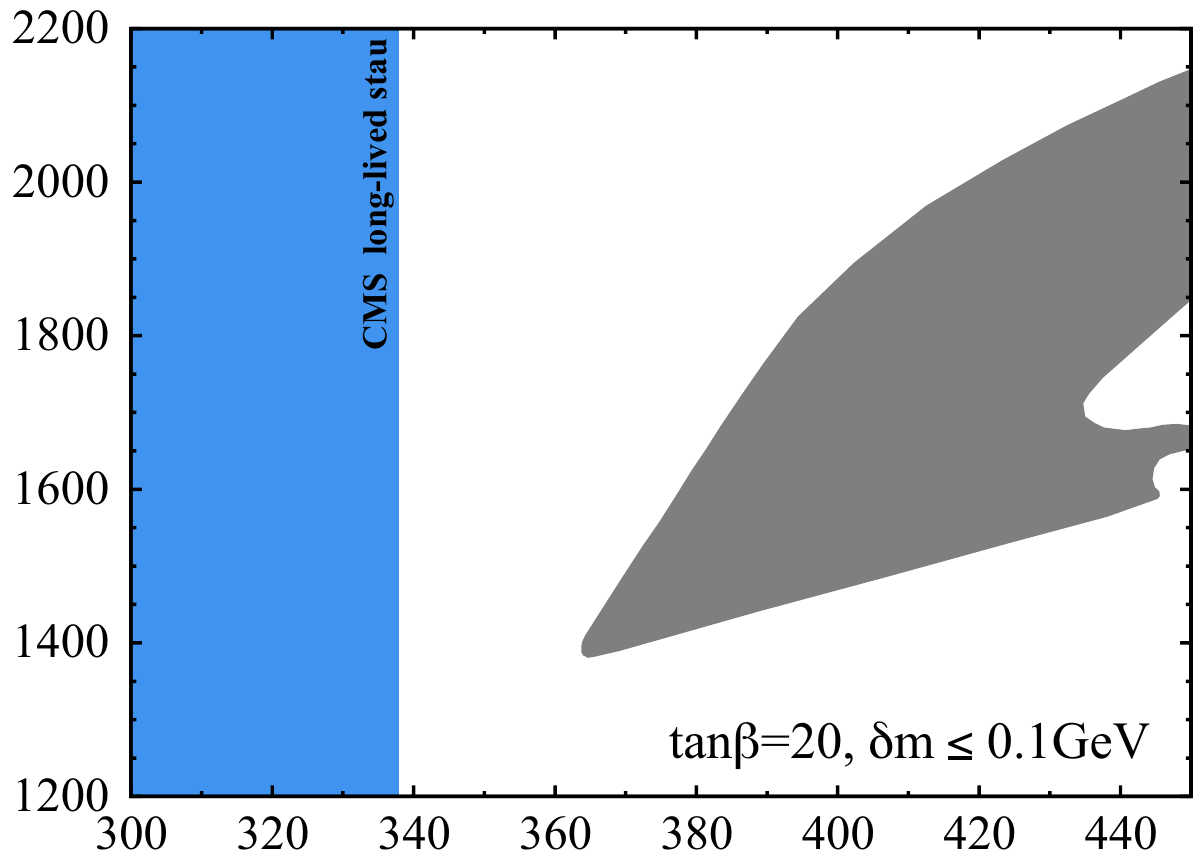}
\end{center}
\end{minipage}
\\[5mm]
\begin{minipage}{84mm}\vspace{2mm}
\begin{center}
 \includegraphics[width=8.2cm,clip]{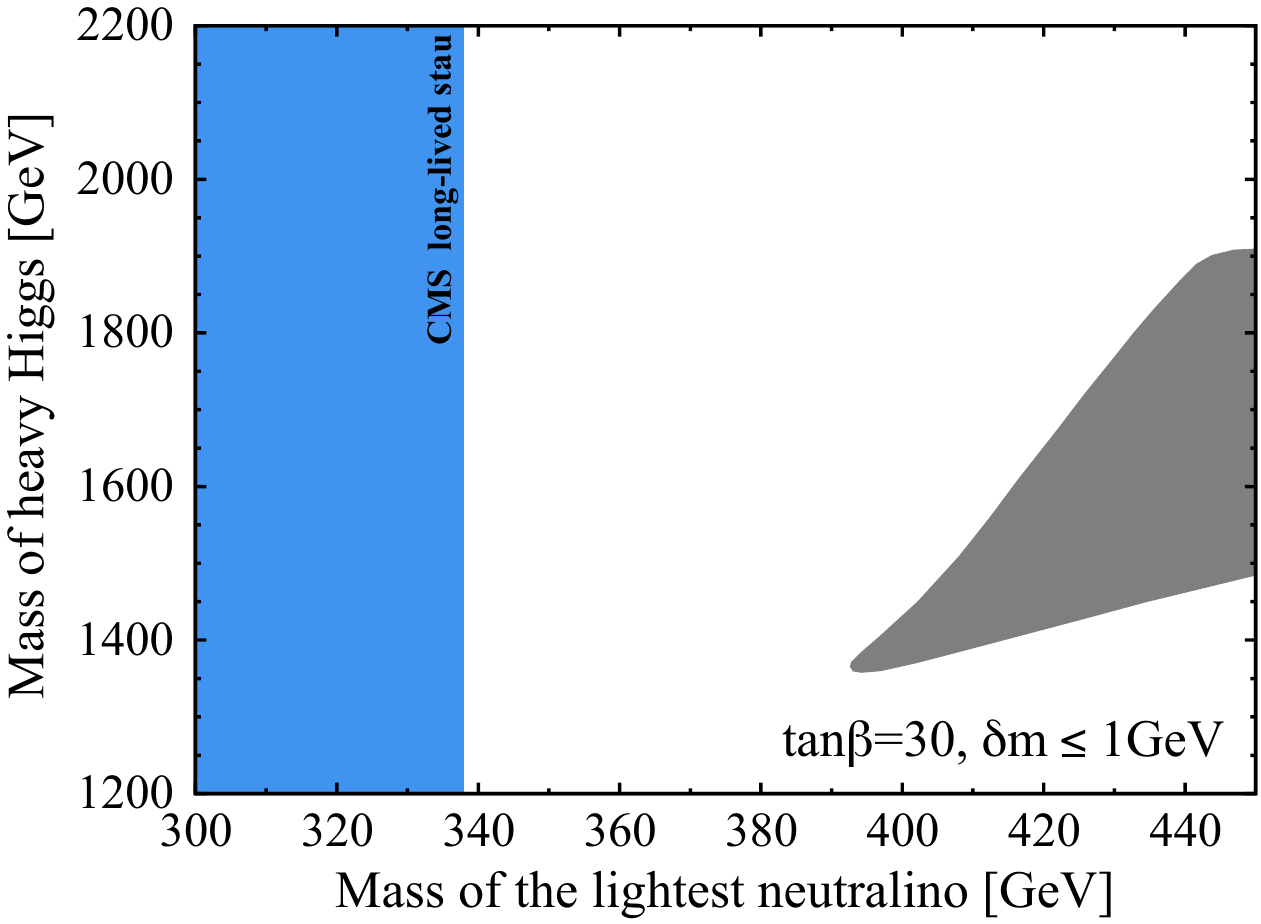}
\end{center}
\end{minipage}
&
\hspace{-3.5mm}
\begin{minipage}{84mm}\vspace{2mm}
\begin{center}
 \includegraphics[width=7.76cm,clip]{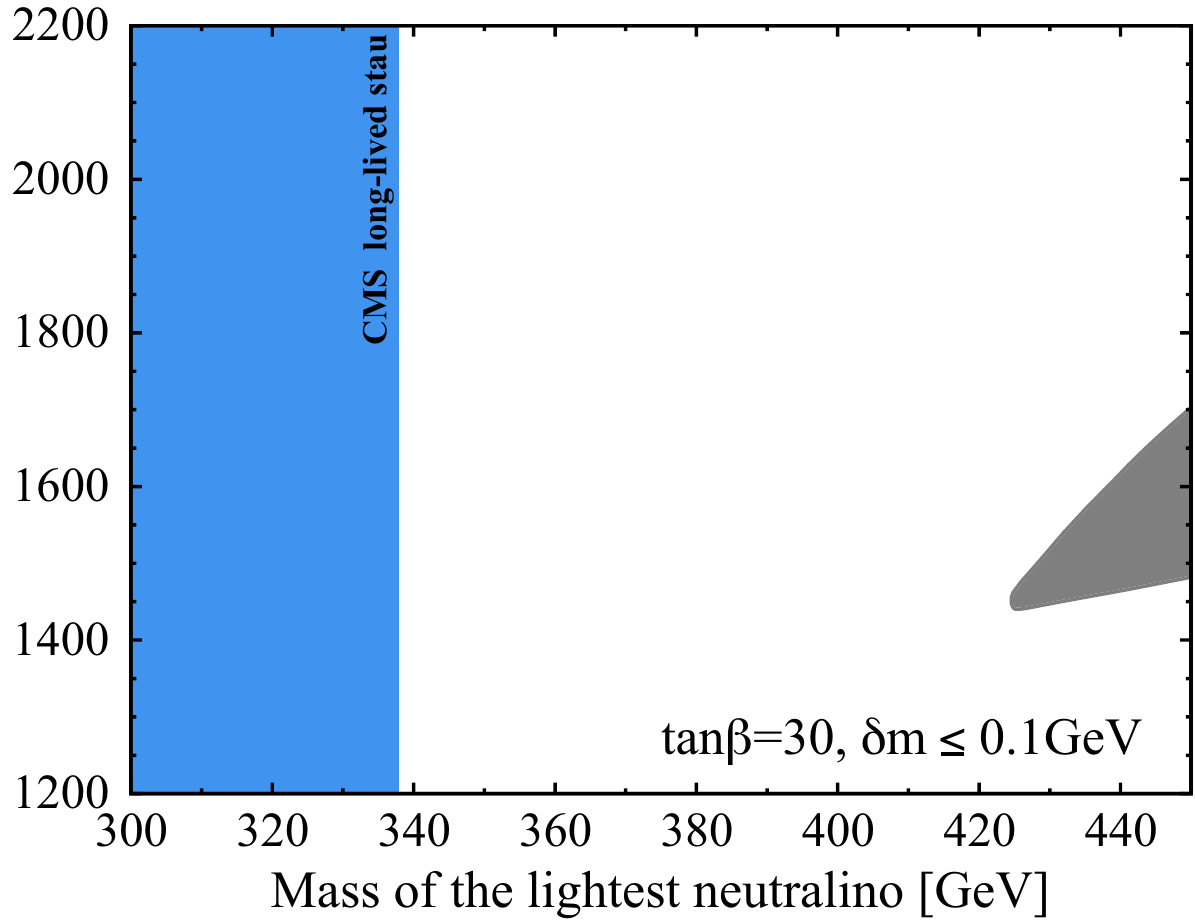}
\end{center}
\end{minipage}
\end{tabular}
\begin{center}
\caption{Mass spectra of the heavy Higgs boson.  
The horizontal axis represents the mass of the LSP neutralino.   
We fix $\tan \beta$ to $10,~20$ and $30$ from top to bottom 
and $\delta m \leq 1$~and~$0.1~\text{GeV}$ from left to right, respectively.}
\label{fig:mheavyhiggs}
\end{center}
\end{figure*}
\end{center}
%
%
%
%
%
%
\clearpage

\begin{figure}[htbp]
\begin{center}
 \includegraphics[width=0.933\linewidth]{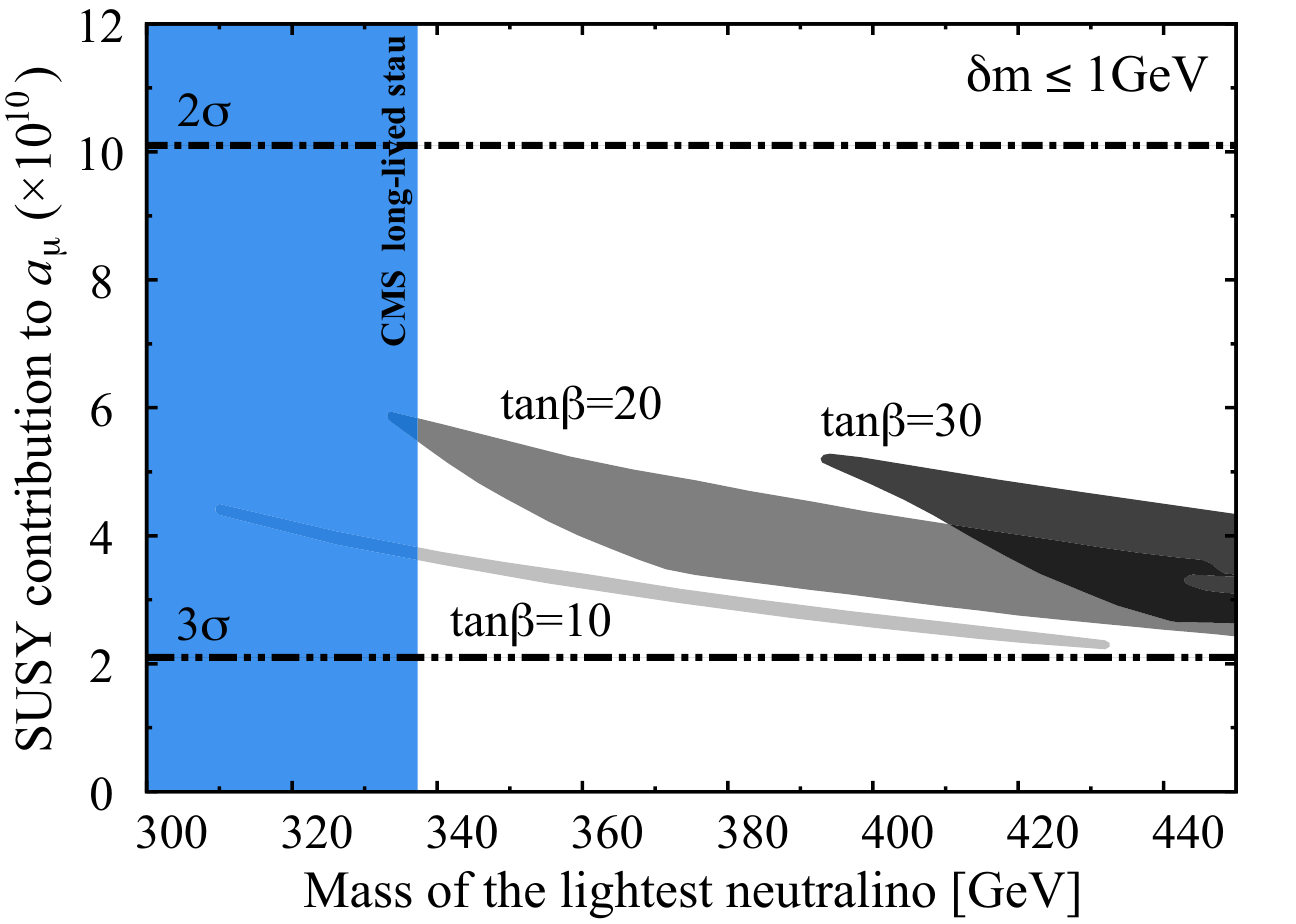} 
 \includegraphics[width=0.933\linewidth]{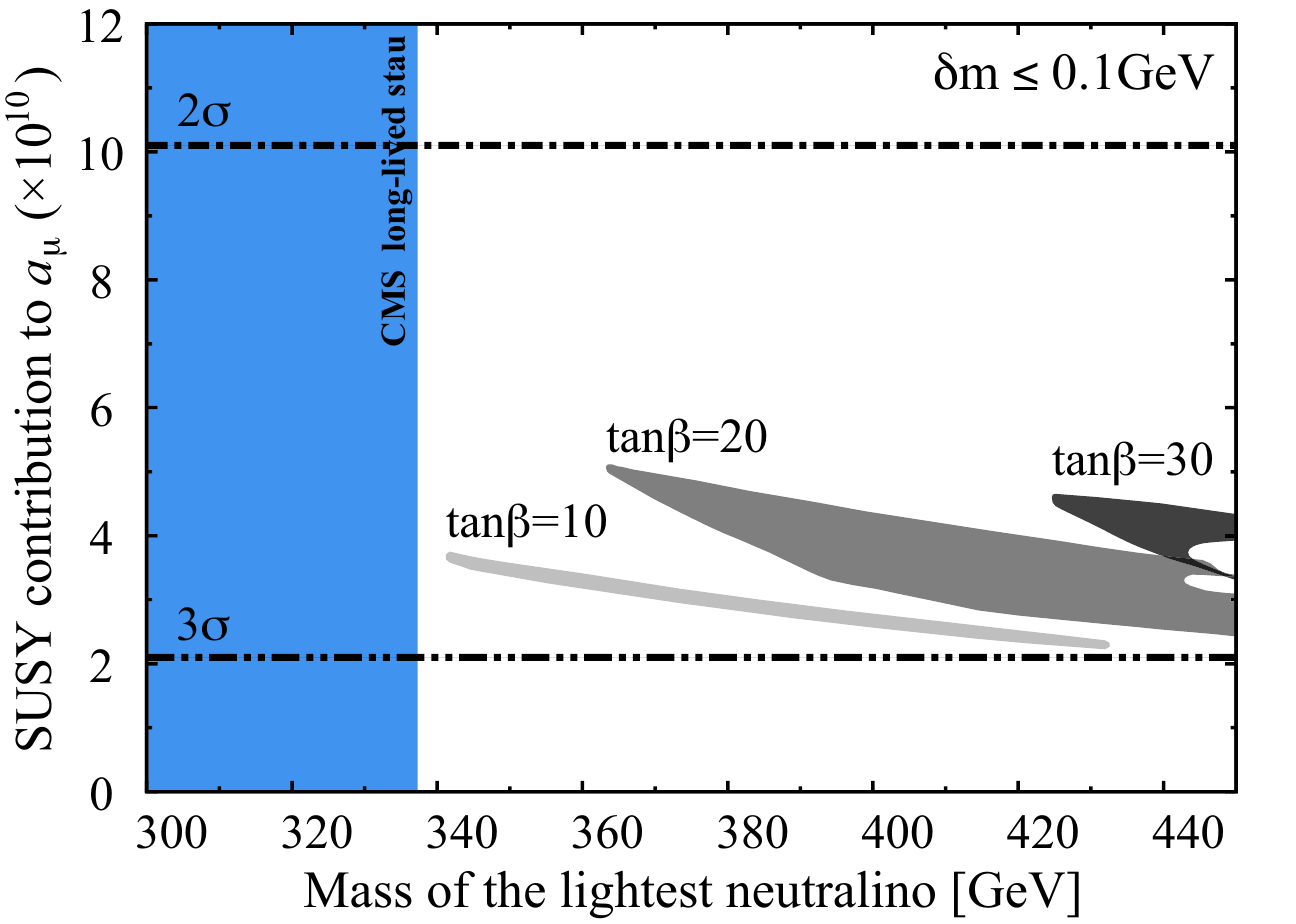} 
\end{center}
\caption{Allowed values of $\delta a _{\mu}$ for neutralino mass.
The upper and lower figures show the results corresponding to the cases of 
$\delta m \leq 1$~GeV and $\delta m \leq 0.1$~GeV, respectively.
}
\label{fig:g-2}
\end{figure}

\subsection{Muon $g-2$}  

We show the muon anomalous magnetic moment, $(g-2)_\mu$, in the 
allowed region. The latest results on $(g-2)_\mu$ have reported that there 
is a $3.3\sigma$ deviation between the SM prediction and the experimental 
data~\cite{Bennett:2006fi, Hagiwara:2011af}:
\begin{equation}
 \delta a_\mu = a_\mu^{\mathrm{exp}} - a_\mu^{\mathrm{SM}}
 = (26.1 \pm 8.0) \times 10^{-10},
\end{equation}
where $a_\mu \equiv (g-2)_\mu/2$.

For a fixed $\tan \beta$, lighter SUSY particles yield larger contributions. 
The masses are light as $m_0$ and $M_{1/2}$ are small. The small $m_0$ 
is, however, excluded by the constraint on the Higgs boson mass as shown 
in~\ref{sec:m0M12}. 
The sizable contributions to $\delta a_\mu$ are obtained in the regions of small 
$M_{1/2}$. When we take into account $\delta m \leq 1$ and $0.1$GeV, 
the dominant SUSY contributions come from the smuon-bino like neutraino loop 
and the muon sneutrino-charged higgsino loop. Since we took $\mu > 0$, both 
contributions are positive. 
In Fig.~\ref{fig:g-2}, the SUSY contributions in the allowed region are shown 
for $\delta m \leq 1$ and $0.1$GeV in the top and the bottom panel, respectively. 
In the panels, $\tan\beta$ is taken to be $10$, $20$, and $30$. 
It should be noted that the muon anomalous magnetic moment is consistent with 
current measurements within $3 \sigma$ levels in this scenario.

\begin{figure}[tp]
\begin{center}
 \includegraphics[width=0.9\linewidth]{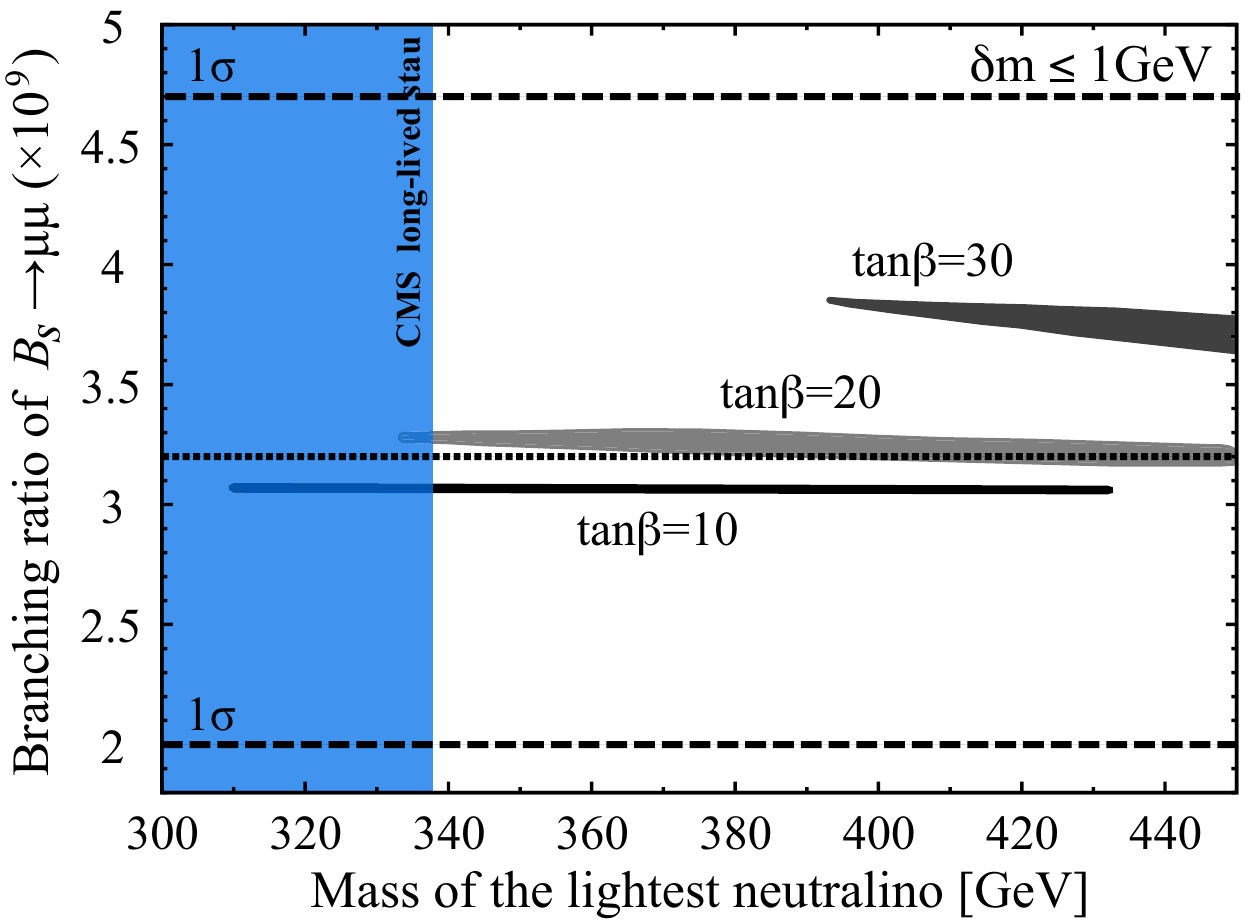} 
 \includegraphics[width=0.9\linewidth]{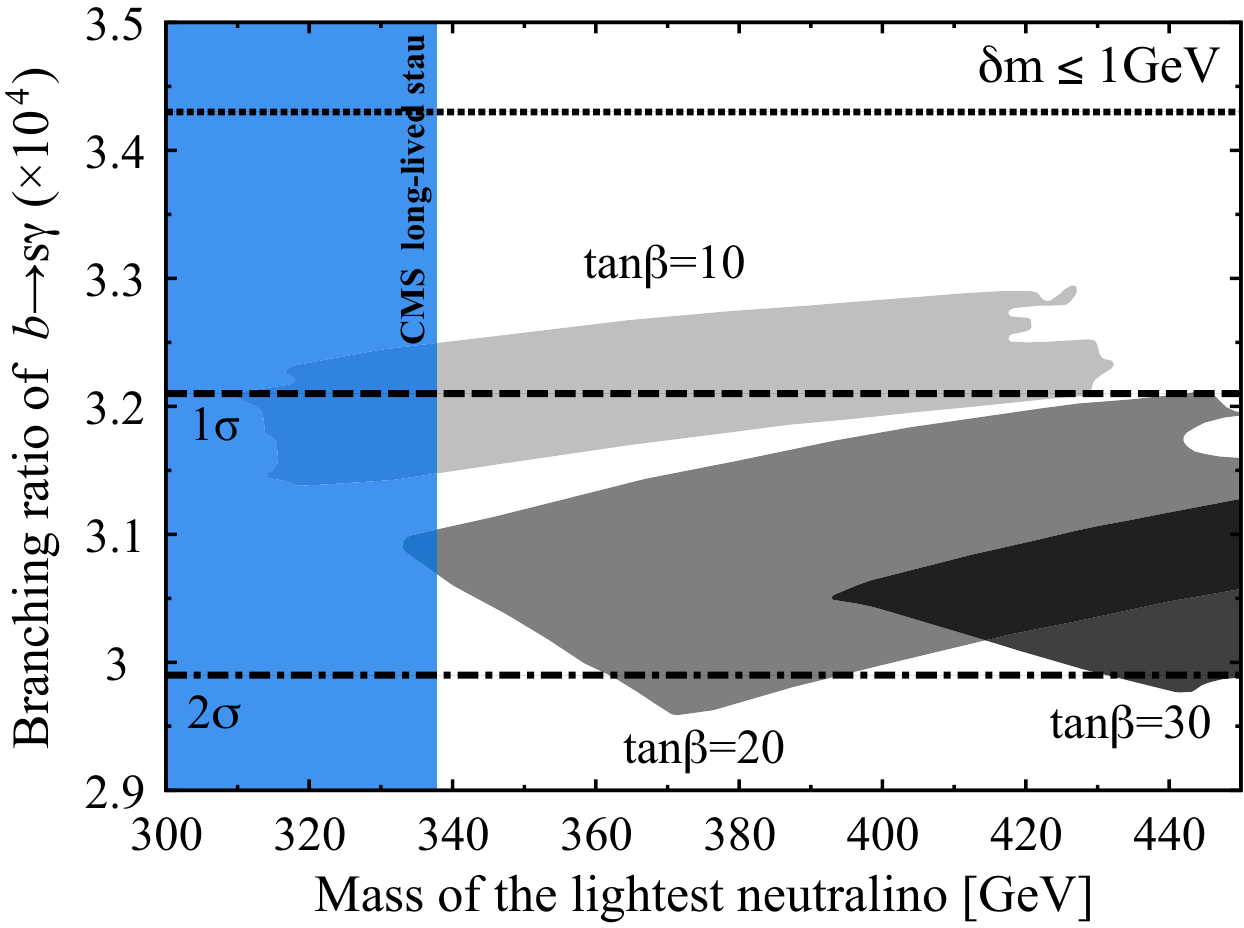} 
\end{center}
\caption{
Branching ratios of the decay processes of 
$B_s \rightarrow \mu^+ \mu^-$ and 
$B \rightarrow X_s \gamma$.  We fix 
$\tan \beta$ to $10, 20,$ and $30$, and 
$\delta m \leq 1$GeV.  The experimental values 
are indicated with the horizontal dotted lines.
}
\label{fig:b_physics}
\end{figure}

\subsection{Rare decays of B mesons}

We show that the branching ratios of $B_s \rightarrow 
\mu^+ \mu^-$ and $B \rightarrow X_s \gamma$ are 
consistent with the experimental results.

The first evidence for the decay $B_s \rightarrow \mu^+ \mu^-$ 
has been discovered by the LHCb collaboration \cite{Aaij:2012nna},
and the branching ratio has been measured as 
\begin{equation}
 \mathrm{BR}(B_s \rightarrow \mu^+ \mu^-) =
(3.2^{+1.5}_{-1.2}) \times 10^{-9}.
\end{equation}
It is also reported by \cite{Amhis:2012bh} that the other important rare 
decay $B \rightarrow X_s \gamma$ is strongly suppressed as 
\begin{equation}
 \mathrm{BR}(B \rightarrow X_s \gamma) =
(3.43 \pm {0.21} \pm {0.07}) \times 10^{-4}.
\end{equation}
Figure~\ref{fig:b_physics} shows the branching ratios of the two rare 
decays for $\delta m \leq 1$GeV.  The top panel is for $B_s \rightarrow 
\mu^+ \mu^-$ and the bottom is for $B_s \rightarrow X_s \gamma$. 
One can see that the branchig ratios are within the $3 \sigma$ in our 
allowed region, and hence are consistent with the experiments.

\subsection{Direct detection of neutralino dark matter} \label{sec:DD}  

\begin{table}[tp]
\begin{center}
\begin{tabular}{l|lll} \hline
$N$ ~ 
& ~ $f_{T_u}^{(N)}$ ~~~~~~~ 
& $f_{T_d}^{(N)}$ ~~~~~~~ 
& $f_{T_s}^{(N)}$ ~~~~~~~
\\ \hline \hline
$p$ 
& ~0.0153 
& 0.0191
& 0.0447
\\
$n$ 
& ~0.011
& 0.0273
& 0.0447
\\ \hline
\end{tabular}
\label{table:DD}
\caption{The mass fraction of light quarks in a proton $p$ and a neutron 
$n$~\cite{Belanger:2013oya, Beringer:2012}.}
\end{center}
\end{table}

\begin{figure}[tbp]
\begin{center}
\includegraphics[width=80mm]{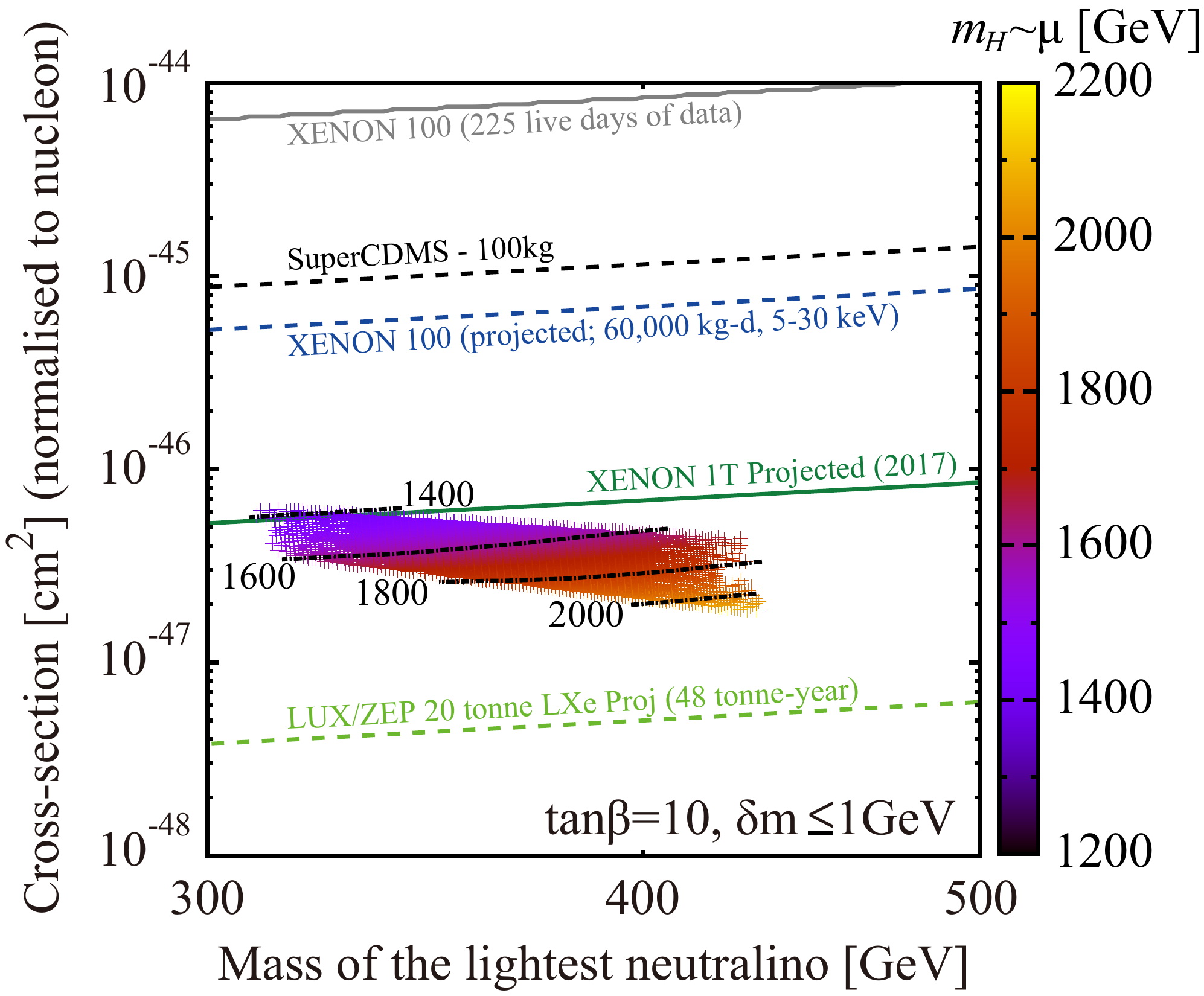}
\end{center}
\begin{center}
\includegraphics[width=80mm]{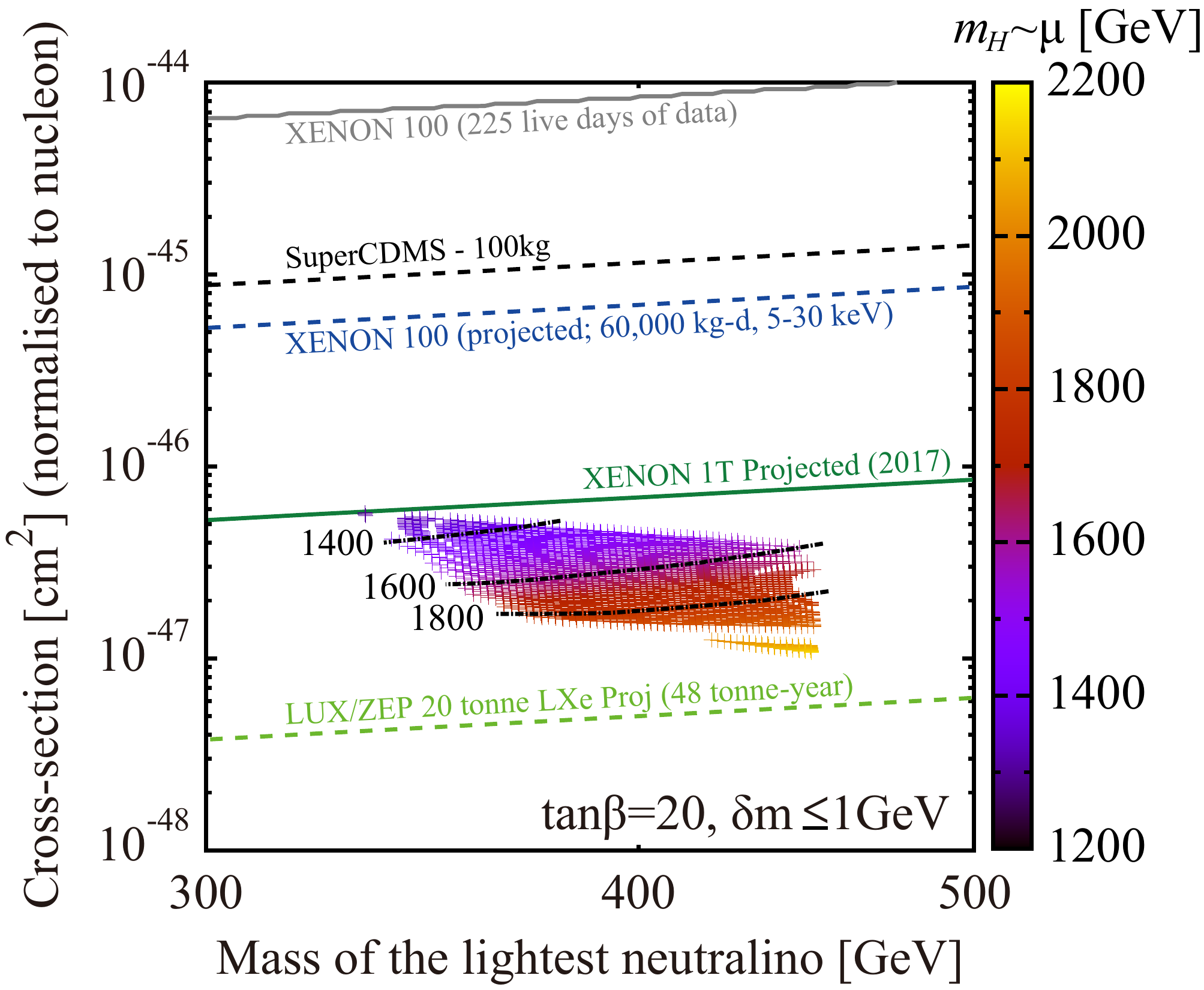}
\end{center}
\begin{center}
\includegraphics[width=80mm]{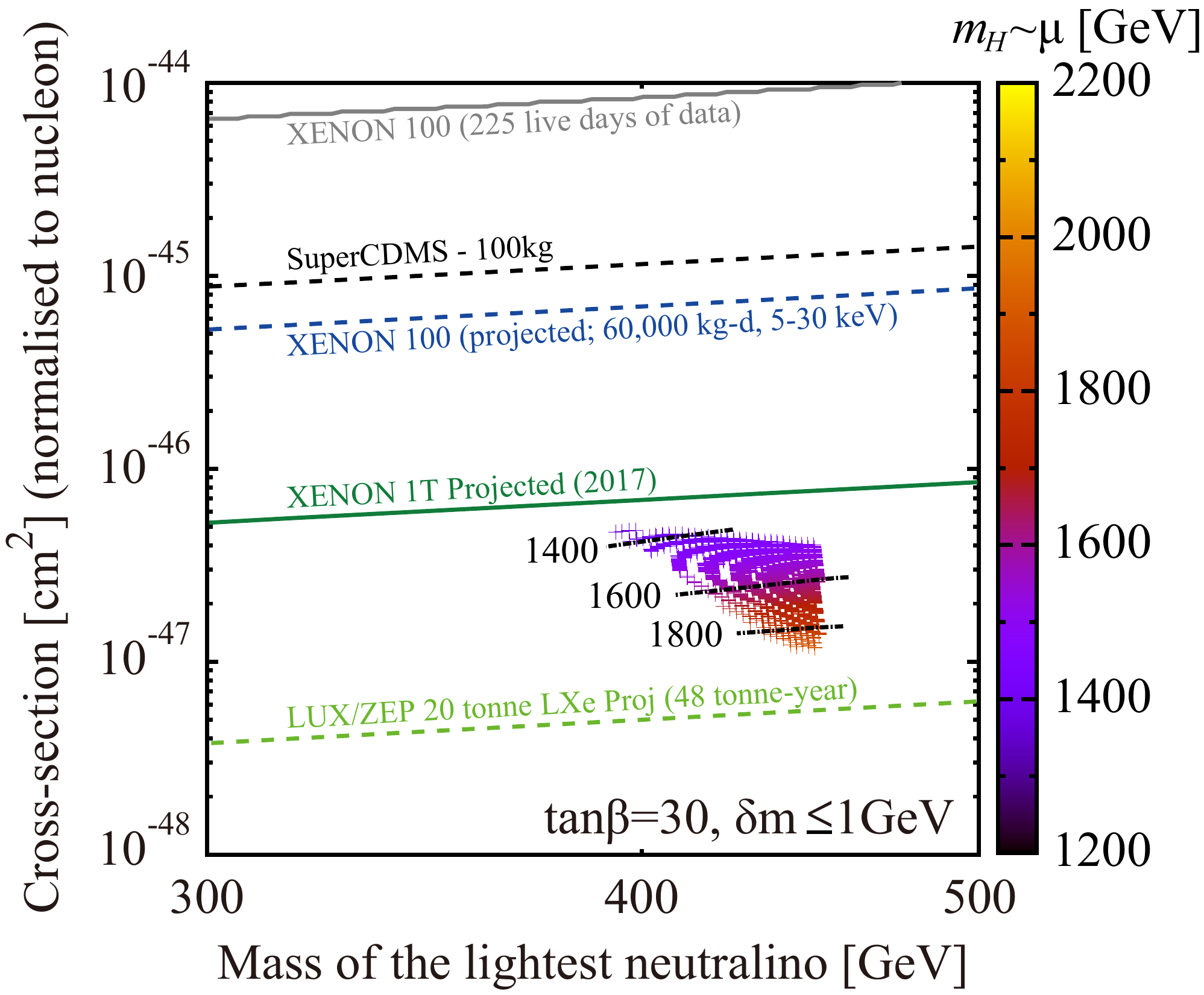}
\end{center}
\caption{Scatter plots of SI neutralino-nucleon cross section as 
a function of the mass of the neutralino dark matter. Each curves 
are shown for the current and future limit. A gradation of colors 
corresponds to the mass of the heavy Higgs boson. }
\label{fig:DD}
\end{figure}

One of the promising approaches to the neutralino dark matter is its 
direct detection. The scenario we are discussing can be examined by 
the direct detection measurements combining other measurements 
for the Higgs boson and the neutralino LSP.

The spin-independent (SI) scatterings are given by the Higgs and squarks 
exchanges. The squark exchange contributions are suppressed by heavy 
masses of squarks $m_{\tilde q}^{-4}$, where $m_{\tilde q} \simeq
2\text{TeV}$ in the scenario (see Fig.~\ref{fig:msquarksmgluino}), and 
hence the SI scattering is dominated by the Higgs exchange contributions. 
In order to examine the scenario, we have to analyze the dependence of 
the SI scatterings on the measurements for the Higgs boson and neutralinos. 
The SI scattering cross section of the neutralino and target nucleon ($T$) is 
given by~\cite{Jungman:1995df}, 
\begin{equation}
\begin{split}
   \sigma_\text{SI} = \frac{4}{\pi} 
   \Bigl( \frac{m_{\tilde \chi _{1} ^0} m_T}
   {m_{\tilde \chi _{1} ^0} + m_T} \Bigr)^2 
   (n_p f_p + n_n f_n)^2, 
\end{split} 
\end{equation}
where $m_T$ is the mass of the target nucleus. The Symbol $n_p (n_n)$ is 
the number of proton (neutron) in the target nucleus, and the effective coupling 
of the neutralino to a proton $f_p$ is given as 
\begin{equation}
\begin{split}
   f_p &= \sum_q f_q \langle p|\bar qq |p \rangle 
   \\
   &= \sum_{q=u,d,s} \frac{f_q}{m_q} m_p f_{T_q}^{(p)} 
   + \frac{2}{27} f_{T_G} \sum_{q=c,b,t} \frac{f_q}{m_q} m_p, 
\end{split} 
\end{equation}
where $f_{Tq}^{(p)} \equiv \langle p|m_q \bar q q | p \rangle/m_p$ is 
the mass fraction of light quarks in a proton (listed in Table~I), and 
$f_{T_G} = 1 - \sum_{u,d,s} f_{T_q}^{(p)}$. For the neutron, $f_n$ is 
derived with the same manner. The effective coupling of the neutralino to 
quarks $f_q$ is calculated as follows, 
\begin{equation}
\begin{split}
   f_q = m_q \frac{g_2^2}{4m_W} \Bigl( 
   \frac{C_{h \tilde \chi _{1} ^{0} \tilde \chi _{1} ^{0}} C_{hqq}}{m_h^2} + 
   \frac{C_{H \tilde \chi _{1} ^{0} \tilde \chi _{1} ^{0}} C_{Hqq}}{m_H^2} \Bigr). 
\label{f_q}   
\end{split} 
\end{equation}
Here $C_{hqq}$ and $C_{Hqq}$ are the Yukawa couplings of the lighter 
and the heavier Higgs bosons and quarks (details formulae are given in 
Ref.~\cite{Jungman:1995df}). For the bino-like neutralino LSP ($M_1 \ll 
M_2, \mu$), the coupling of the neutralino and the Higgs bosons, 
$C_{h \tilde \chi_1^0 \tilde \chi_1^0}$ and $C_{H \tilde \chi_1^0 
\tilde \chi_1^0}$, is perturbatively calculated as follows, 
\begin{equation}
\begin{split}
   &C_{h \tilde \chi_1^0 \tilde \chi_1^0} \simeq 
   \frac{m_Z \sin \theta_W \tan \theta_W}{M_1^2-\mu^2} 
   [M_1 \sin\beta + \mu \cos\beta], 
   \\&
   C_{H \tilde \chi_1^0 \tilde \chi_1^0} \simeq 
   \frac{m_Z \sin \theta_W \tan \theta_W}{M_1^2-\mu^2} 
   \mu \sin\beta.
\label{h_chi_chi}   
\end{split} 
\end{equation}
In the scenario, the mass of the heavier CP-even Higgs boson is much larger 
than the mass of the lighter Higgs boson (see Fig.~\ref{fig:mheavyhiggs}), 
and therefore the contribution from $H$ exchange is negligible.

Figure~\ref{fig:DD} shows the Spin-Independent (SI) scattering cross section 
in the allowed region as a function of the mass of the neutralino dark matter 
for $\delta m \leq 1$~GeV, and $\tan\beta$ is taken to be $10$, $20$, and 
$30$ from an upper panel to a lower one. Curves represent the current 
limit~\cite{Aprile:2012nq} and future limit~\cite{Aprile:2012zx, Akerib:2012ak}. 
Colors correspond to the mass of the heavier CP-even Higgs boson.

In Fig.~\ref{fig:DD}, we can see that the SI scattering cross section is 
smaller as $\tan\beta$ and/or the neutralino mass is larger. Such behavior 
can be easily understood by Eq.\eqref{h_chi_chi}. We can also see from 
the contour lines that the lighter mass of the heavy Higgs boson leads larger 
SI cross sections for all $\tan\beta$.  
This is understood as follows. 
The mass of the heavy Higgs boson is determined by the condition of the 
radiative electroweak symmetry breaking in the scenario, and is almost equal 
to $m_H \simeq \mu$ (see the discussion in Subsec.~\ref{sec:spectrum}). 
As is given in Eq.~(\ref{h_chi_chi}), smaller $\mu$ makes the neutralino mixing to 
be larger value, and hence large couplings between the neutralino dark matter 
and the Higgs bosons are obtained. Thus a clear correlation of the mass of the 
heavy Higgs boson and the cross section is appeared.

In the end, one can see that future direct detection experiments will reach to 
the sensitivity to detect the neutralino dark matter in the scenario. The mass 
of the heavy Higgs boson can be estimated from the measurements of the 
cross section and the mass of the neutralino at the direct detection and the 
LHC experiments. Thus the interplay of both experiments will play important 
roles to examine our scenario.

\section{Direct search at the LHC} \label{sec:LHCexp} 

We here estimate the number of events of the long-lived stau and 
the neutralino at the LHC with the center of mass energy $14$TeV, 
and consider the feasibility of the verification of the scenario.

We choose five sets of the CMSSM parameters in the allowed region for 
$\tan \beta = 20$.  The parameters of the sample points and the mass 
spectrum of the SUSY particles are shown in Table~\ref{table:IP_MS}.  
These points correspond to the mass of the neutralino, $350,~375,~400,
~425$, and $450$GeV, respectively.  
We calculate the branching ratios and the pair production cross sections 
of SUSY particles with CalcHEP 3.4~\cite{Belyaev:2012qa}. 
The branching ratios are shown in Tables~\ref{table:BR1} and 
\ref{table:BR2}. The first column is the parent particles and the 
second one is the final states. The production cross sections of SUSY 
particles are shown in Tables~\ref{table:CS1} and \ref{table:CS2}. 
The first column $\sigma(a,b)$ is the production cross sections of particles 
$a$ and $b$ from $p$-$p$ collision with the center of mass energy $14$TeV. 
The last raw of Table~\ref{table:CS2}, $\sigma$(all SUSY), is the sum of the 
production cross sections of SUSY particles.

Using the numbers given in the Tables, we calculate the number of the 
stau from SUSY cascade decays and direct productions. The branching 
fraction of the cascade decays of a SUSY particle $\tilde \psi$ to the stau 
is denoted as $\mathrm{BR}(\tilde \psi \to \tilde \tau_1)$. 
The symbol $\sigma (\tilde \psi)$ represents the sum of the cross sections 
of the SUSY particles. Then the effective total production cross section 
of the stau is 
\begin{equation}
\begin{split}
   \sigma(\tilde \tau _{1}^{(*)}) = \sum_{\tilde \psi}^\text{all SUSY} 
   \sigma(\tilde \psi) \times \mathrm{BR} (\tilde \psi \to \tilde \tau_1^{(*)}). 
\end{split} 
\end{equation}
Since the long-lived staus penetrate the LHC detectors, the number of 
missing energy is calculated as follows, 
\begin{equation}
\begin{split}
   N(\tilde \chi_1^0) = \bigl\{ \sigma(\text{all SUSY}) - \sigma(\tilde \tau_1) 
   - \sigma(\tilde \tau_1^*) \bigr\} \times \mathcal{L}_\text{int}, 
\end{split} 
\end{equation}
where $\mathcal{L}_\text{int}$ is the integrated luminosity. Assuming an 
integrated luminosity is $\mathcal{L}_\text{int} = 100\text{fb}^{-1}$, 
the expected number of the stau and the neutralino is 3900 and 3700 for 
$m_{\tilde \tau_1} = 350$GeV, and 710 and 980 for $m_{\tilde \tau_1} 
= 450$GeV~(see Table~\ref{table:SCSBR}).  
These numbers will be enough to examine the scenario\footnote{The 
efficiency to identify the stau and the neutralino with $350$ and $450$GeV 
is about $\mathcal{O}(10)\%$ in~\cite{Chatrchyan:2013oca}. 
Thus it is possible to examine our scenario with $\mathcal{L}_\mathrm{int} 
=100$fb$^{-1}$. Detailed simulation is beyond the scope of this paper and 
is left for our future work.}.

In Table~\ref{table:CS2}, one can see that the production cross sections 
of stops are comparable to those of gluinos and light squarks. Those are 
obtained as 4.40 fb~for $m_{\tilde \tau_1} = 350$GeV and 2.66fb for 
$m_{\tilde \tau_1} = 450$GeV. 
It is important to emphasize here that such production cross sections of 
stops are the results of our constraint that leads to relatively heavy SUSY 
spectrum. Therefore early discovery of the stop at the 14 TeV run is one 
of the predictions in the scenario.
\clearpage
\begin{widetext}

\begin{table}
\begin{center}
\tabcolsep=4mm
\begin{tabular}{lccccc} \hline \hline
Input Parameters
&
Point 1 [GeV] & Point 2 [GeV] & Point 3 [GeV] & Point 4 [GeV] & Point 5 [GeV]
\\[0.7mm] \hline
$M _{1/2}$
&
818.6 & 878.0 & 932.8 & 986.0 & 1038.0
\\[0.7mm] 
$m _{0}$
&
452.0 & 517.7 & 557.7 & 601.7 & 639.7
\\[0.7mm] 
$A _{0}$
&
-2264.7 & -2683.6 & -2918.4 & -3177.9 & -3397.0
\\[0.7mm] \hline
Particle
&
&
&
&
&
\\[0.7mm] \hline
$h$
&
123.8 & 124.4 & 124.6 & 124.8 & 124.9
\\[0.7mm] 
$H$
&
1494.6 & 1659.5 & 1775.7 & 1894.0 & 2002.0
\\[0.7mm] 
$A$
& 
1495.1 & 1660.3 & 1776.5 & 1895.1 & 2003.3
\\[0.7mm]
$H^{\pm}$
&
1497.0 & 1662.0 & 1778.1 & 1896.5 & 2004.6
\\[0.7mm] 
$\tilde g$
&
1822.4 & 1945.4 & 2057.8 & 2166.7 & 2272.6
\\[0.7mm] 
$\tilde \chi ^{\pm} _{1}$
&
665.6 & 716.3 & 762.4 & 807.4 & 851.2
\\[0.7mm] 
$\tilde \chi ^{\pm} _{2}$
&
1470.6 & 1638.9 & 1753.5 & 1873.7 & 1981.4
\\[0.7mm] 
$\tilde \chi ^{0} _{1}$
&
349.3 & 376.3 & 400.9 & 425.0 & 448.5
\\[0.7mm] 
$\tilde \chi ^{0} _{2}$
&
665.3 & 716.1 & 762.4 & 807.4 & 851.2
\\[0.7mm] 
$\tilde \chi ^{0} _{3}$
&
1466.8 & 1635.3 & 1750.1 & 1870.5 & 1978.4
\\[0.7mm] 
$\tilde \chi ^{0} _{4}$
&
1469.7 & 1638.4 & 1753.1 & 1873.3 & 1981.0
\\[0.7mm]
$\tilde e _{L}$
&
709.6 & 781.6 & 834.9 & 890.0 & 940.6
\\[0.7mm] 
$\tilde e _{R}$
&
547.5 & 614.2 & 658.7& 706.2 & 748.6
\\[0.7mm] 
$\tilde \nu _{e}$
&
749.2 & 777.4 & 830.9 & 886.2 & 937.0
\\[0.7mm] 
$\tilde \tau _{1}$
&
350.3 & 377.0 & 401.0 & 425.6 & 449.1
\\[0.7mm] 
$\tilde \tau _{2}$
&
656.0 & 713.7 & 759.3 & 805.9 & 849.4
\\[0.7mm] 
$\tilde \nu _{\tau}$
&
642.7 & 701.2 & 747.4 & 794.6 & 838.7
\\[0.7mm] 
$\tilde u _{L}$
&
1710.9 & 1834.4 & 1942.2 & 2048.0 & 2149.7
\\[0.7mm] 
$\tilde u _{R}$
&
1646.6 & 1765.6 & 1869.0 & 1970.6 & 2068.0
\\[0.7mm]
$\tilde d _{L}$
&
1712.6 & 1835.9 & 1943.7 & 2049.4 & 2151.0
\\[0.7mm] 
$\tilde d _{R}$
&
1639.9 & 1758.3 & 1861.1 & 1962.1 & 2059.0
\\[0.7mm] 
$\tilde t _{1}$
&
945.8 & 936.5 & 968.6 & 987.7 & 1016.3
\\[0.7mm]
$\tilde t _{2}$
&
1432.8 & 1497.5 & 1573.1 & 1641.3 & 1710.8
\\[0.7mm] 
$\tilde b _{1}$
&
1384.6 & 1452.8 & 1528.4 & 1598.2 & 1669.3
\\[0.7mm] 
$\tilde b _{2}$
&
1561.6 & 1665.8 & 1760.5 & 1852.6 & 1942.0
\\[0.7mm] \hline \hline
\end{tabular}
\caption{Input parameters and the mass spectrum of sample points. 
The value of $\tan \beta$ is fixed to 20 and sign$(\mu) > 0$.  
We choose these parameters as the $m _{\tilde \chi _{1} ^{0}}$ is fixed to 
$350,~375,~400,~425$, and $450$~GeV, respectively.}
\label{table:IP_MS}
\end{center}
\end{table}

\begin{table}
\begin{center}
\tabcolsep=2mm
\begin{tabular}{lcccccc} \hline \hline
Particle
&
Final States
&
Point1 [\%] & Point2 [\%] & Point3 [\%] & Point4 [\%] & Point5 [\%]
\\[0.7mm] \hline
$\tilde g$
&
$\bar{t} , \tilde t _{1}$ 
&
21.4 & 22.6 & 22.9 & 23.3 & 23.6
\\[0.7mm]
&
$t , \tilde t _{1} ^{\ast}$
&
21.4 & 22.6 & 22.9 & 23.3 & 23.6 
\\[0.7mm]
&
$\bar{b} , \tilde b _{1}$
&
8.3 & 8.4 & 8.4 & 8.5 & 8.5
\\[0.7mm]
&
$b , \tilde b _{1} ^{\ast}$
&
8.3 & 8.4 & 8.4 & 8.5 & 8.5 
\\[0.7mm]
&
$\bar{t} , \tilde t _{2}$
&
7.9 & 8.2 & 8.3 & 8.3 & 8.4 
\\[0.7mm] 
&
$t , \tilde t _{2} ^{\ast}$
&
7.9 & 8.2 & 8.3 & 8.3 & 8.4 
\\[0.7mm] 
&
$\bar{b} , \tilde b _{2}$
&
3.3 & 3.1 & 3.0 & 3.0 & 2.9
\\[0.7mm] 
&
$b , \tilde b _{2} ^{\ast}$
&
3.3 & 3.1 & 3.0 & 3.0 & 2.9 
\\[0.7mm] \hline
$\tilde u _{L}$
&
$d , \tilde \chi ^{+} _{1}$
&
65.8 & 65.8 & 65.8 & 65.8 & 65.7
\\[0.7mm] 
&
$u , \tilde \chi ^{0} _{2}$
&
32.8 & 32.9 & 32.9 & 32.9 & 32.9
\\[0.7mm] 
&
$u , \tilde \chi ^{0} _{1}$
&
1.4 & 1.4 & 1.4 & 1.4 & 1.4
\\[0.7mm] \hline
$\tilde d _{L}$
&
$u , \tilde \chi ^{-} _{1}$
&
65.6 & 65.6 & 65.7 & 65.7 & 65.7
\\[0.7mm] 
&
$d , \tilde \chi ^{0} _{2}$
&
32.9 & 32.9 & 32.9 & 32.9 & 32.9
\\[0.7mm] 
&
$d , \tilde \chi ^{0} _{1}$
&
1.5 & 1.5 & 1.5 & 1.5 & 1.5
\\[0.7mm] \hline
$\tilde t _{1}$
&
$t , \tilde \chi ^{0} _{1}$
&
78.4 & 86.3 & 89.4 & 92.7 & 94.8
\\[0.7mm] 
&
$b , \tilde \chi ^{+}  _{1}$
&
15.8 & 10.5 & 8.3 & 6.0 & 4.6
\\[0.7mm] 
&
$t , \tilde \chi ^{0}  _{2}$
&
5.9 & 3.2 & 2.4 & 1.3 & 0.6
\\[0.7mm] \hline
$\tilde t _{2}$
&
$Z , \tilde t  _{1}$
&
41.7 & 43.5 & 43.7 & 44.1 & 44.3
\\[0.7mm] 
&
$h , \tilde t  _{1}$
&
29.1 & 32.0 & 33.2 & 34.4 & 35.3
\\[0.7mm] 
&
$b , \tilde \chi ^{+} _{1}$
&
19.1 & 16.0 & 15.1 & 14.0 & 13.4
\\[0.7mm]
&
$t , \tilde \chi ^{0}  _{2}$
&
9.1 & 7.7 & 7.2 & 6.7 & 6.4
\\[0.7mm] \hline
$\tilde b _{1}$
&
$W ^{-} , \tilde t _{1}$
&
69.9 & 83.7 & 76.9 & 78.7 & 79.8
\\[0.7mm]
&
$t , \tilde \chi ^{-} _{1}$
&
19.3 & 15.9 & 14.9 & 13.7 & 13.0
\\[0.7mm] 
&
$b , \tilde \chi ^{0} _{2}$
&
10.2 & 8.4 & 7.8 & 7.9 & 6.8
\\[0.7mm] \hline
$\tilde b _{2}$
&
$W ^{-} , \tilde t  _{1}$
&
44.0 & 39.7 & 36.5 & 33.4 & 30.9
\\[0.7mm] 
&
$h , \tilde \chi ^{0} _{1}$
&
31.6 & 30.2 & 30.2 & 29.8 & 29.7
\\[0.7mm] 
&
$h , \tilde b _{1}$
&
9.4 & 12.6 & 13.8 & 15.0 & 15.8
\\[0.7mm] 
&
$Z , \tilde b _{1}$
& 
5.7 & 7.7 & 8.5 & 9.4 & 9.9
\\[0.7mm] 
&
$W ^{-} , \tilde t _{2}$
&
3.3 & 6.6 & 8.3 & 10.2 & 11.7
\\[0.7mm] 
&
$t , \tilde \chi ^{-} _{1}$
&
2.9 & 2.1 & 1.8 & 1.5 & 1.3
\\[0.7mm] 
&
$b , \tilde \chi ^{0} _{2}$
&
1.5 & 1.1 & 0.9 & 0.8 & 0.7
\\[0.7mm] \hline
$\tilde \nu _{\tau}$
&
$W ^{+} , \tilde \tau _{1}$
&
81.8 & 83.0 & 83.4 & 83.8 & 84.0
\\[0.7mm] 
&
$\nu _{\tau}, \tilde \chi ^{0} _{1}$
&
18.2 & 17.0 & 16.6 & 16.2 & 16.0
\\[0.7mm] \hline
$\tilde \tau _{2}$
&
$Z, \tilde \tau _{1}$
&
41.3 & 41.8 & 41.9 & 42.0 & 42.0
\\[0.7mm]
&
$h, \tilde \tau _{1}$
&
39.0 & 39.9 & 40.5 & 40.9 & 41.2
\\[0.7mm]
&
$\tau, \tilde \chi _{1} ^{0}$
&
19.7 & 18.3 & 17.7 & 14.2 & 16.8
\\[0.7mm] \hline
$\tilde \chi ^{0} _{2}$
&
$\bar {\tau}, \tilde \tau _{1}$
&
40.1 & 45.0 & 45.0 & 46.1 & 46.2
\\[0.7mm] 
&
$\tau, \tilde \tau _{1} ^{\ast}  $
&
40.1 & 45.0 & 45.0 & 46.1 & 46.2
\\[0.7mm] 
&
$\bar{\nu} _{\tau}, \tilde{\nu} _{\tau}$
&
7.3 & 3.6 & 3.6 & 2.6 & 2.5 
\\[0.7mm] 
&
$\nu _{\tau}, \tilde \nu ^{\ast}$
&
7.3 & 3.6 & 3.6 & 2.6 & 2.5 
\\[0.7mm] 
&
$h, \tilde \chi ^{0} _{1}$
&
2.5 & 2.5 & 2.5 & 2.5 & 2.4
\\[0.7mm] \hline \hline
\end{tabular}
\caption{Branching ratios of the SUSY particles on the sample points.}
\label{table:BR1}
\end{center}
\end{table}

\begin{table}
\begin{center}
\tabcolsep=2mm
\begin{tabular}{lcccccc} \hline \hline
Particle
&
Final States
&
Point1 [\%] & Point2 [\%] & Point3 [\%] & Point4 [\%] & Point5 [\%]
\\[0.7mm] \hline 
$\tilde \chi ^{0} _{3}$   
&
$\bar t, \tilde t _{1}$
& 
22.6 & 26.7 & 26.7 & 26.9 & 27.0
\\[0.7mm] 
&
$t, \tilde t _{1} ^{\ast}$
& 
22.6 & 26.7 & 26.7 & 26.9 & 27.0
\\[0.7mm] 
&
$W ^{-}, \tilde \chi ^{+} _{1}$
& 
15.0 & 12.6 & 11.6 & 10.7 & 10.3
\\[0.7mm] 
&
$W ^{+}, \tilde \chi ^{-} _{1}$
& 
15.0 & 12.6 & 11.6 & 10.7 & 10.3
\\[0.7mm] 
&
$Z, \tilde \chi ^{0} _{2}$
& 
13.2 & 11.0 & 10.1 & 9.3 & 8.9
\\[0.7mm] 
&
$Z, \tilde \chi ^{0} _{1}$
& 
4.0 & 3.3 & 3.0 & 2.8 & 2.7
\\[0.7mm] 
&
$h, \tilde \chi _{2} ^{0}$
& 
1.6 & 1.4 & 1.3 & 1.3 & 1.2
\\[0.7mm] 
&
$\bar t, \tilde t _{2}$
& 
| & | & 1.8 & 3.1 & 3.7
\\[0.7mm] 
&
$t, \tilde t _{2} ^{\ast}$
& 
| & | & 1.8 & 3.1 & 3.7
\\[0.7mm] \hline
$\tilde \chi ^{0} _{4}$   
&
$\bar t, \tilde t _{1}$
& 
27.0 & 29.8 & 29.9 & 30.0 & 29.9
\\[0.7mm] 
&
$t, \tilde t _{1} ^{\ast}$
& 
27.0 & 29.8 & 29.9 & 30.0 & 29.9
\\[0.7mm] 
&
$W ^{-}, \tilde \chi ^{+} _{1}$
& 
12.3 & 10.7 & 10.3 & 9.8 & 9.5
\\[0.7mm] 
&
$W ^{+}, \tilde \chi ^{-} _{1}$
& 
12.3 & 10.7 & 10.3 & 9.8 & 9.5
\\[0.7mm] 
&
$h, \tilde \chi ^{0} _{2}$
& 
11.4 & 9.9 & 9.6 & 9.1 & 8.9
\\[0.7mm] 
&
$h, \tilde \chi ^{0} _{1}$
& 
3.1 & 2.6 & 2.5 & 2.4 & 2.3
\\[0.7mm] 
&
$Z, \tilde \chi ^{0} _{2}$
& 
1.3 & 1.2 & 1.2 & 1.1 & 1.1
\\[0.7mm] \hline
$\tilde \chi ^{+} _{1}$   
&
$\nu _{\tau}, \tilde \tau _{1} ^{\ast}$
& 
79.3 & 89.1 & 89.2 & 91.3 & 91.5
\\[0.7mm]
&
$\bar \tau , \tilde \nu _{\tau}$
& 
15.0 & 7.5 & 7.3 & 5.4 & 5.2
\\[0.7mm]
&
$W ^{+}, \tilde \chi _{1} ^{0}$
& 
3.1 & 3.2 & 3.1 & 3.2 & 3.2
\\[0.7mm] \hline
$\tilde \chi ^{+} _{2}$
&
$\bar{b}, \tilde t _{1}$
& 
48.4 & 53.0 & 53.3 & 53.9 & 54.1
\\[0.7mm]
&
$h, \tilde \chi _{1} ^{+}$
& 
14.2 & 11.9 & 11.3 & 10.6 & 10.3
\\[0.7mm]
&
$Z, \tilde \chi _{1} ^{+}$
& 
13.8 & 11.5 & 10.9 & 10.2 & 9.9
\\[0.7mm] 
&
$W ^{+}, \tilde \chi _{2} ^{0}$
& 
14.0 & 11.7 & 11.1 & 10.4 & 10.0
\\[0.7mm]
&
$W ^{+}, \tilde \chi _{1} ^{0}$
& 
4.7 & 3.9 & 3.7 & 3.5 & 3.4
\\[0.7mm] 
&
$t, \tilde b _{1} ^{\ast}$
& 
4.8 & 3.4 & 5.1 & 6.8 & 7.8
\\[0.7mm] 
&
$\nu _{\tau}, \tilde \tau _{1} ^{\ast}$
& 
2.1 & 1.8 & 1.8 & 1.7 & 1.6
\\[0.7mm] 
&
$\bar \tau, \tilde \nu _{\tau}$
& 
1.9 & 1.6 & 1.5 & 1.4 & 1.4
\\[0.7mm] \hline
$H$   
&
$b, \bar b$
&
63.3 & 61.1 & 60.4 & 59.6 & 59.2 
\\[0.7mm]
&
$\tau, \bar \tau$
&
11.6 & 11.4 & 11.4 & 11.3 & 11.3
\\[0.7mm]
&
$\tilde \tau _{1}, \tilde \tau _{2} ^{\ast}$
&
10.0 & 11.3 & 11.7 & 12.2 & 12.6
\\[0.7mm]
&
$\tilde \tau _{1} ^{\ast}, \tilde \tau _{2}$
&
10.0 & 11.3 & 11.7 & 12.2 & 12.6
\\[0.7mm]
&
$\tilde \tau _{1} ^{\ast}, \tilde \tau _{1}$
&
1.9 & 1.8 & 1.7 & 1.6 & 1.4
\\[0.7mm] 
&
$t, \bar t$
&
1.6 & 1.6 & 1.6 & 1.6 & 1.6
\\[0.7mm] \hline
$A$    
&
$b, \bar b$
&
62.4 & 60.8 & 60.2 & 59.5 & 59.1 
\\[0.7mm]
&
$\tau, \bar \tau$
&
11.6 & 11.4 & 11.3 & 11.3 & 11.3
\\[0.7mm]
&
$\tilde \tau _{1}, \tilde \tau _{2} ^{\ast}$
&
11.5 & 12.7 & 13.1 & 13.5 & 13.7 
\\[0.7mm]
&
$\tilde \tau _{1} ^{\ast}, \tilde \tau _{2}$
&
11.5 & 12.7 & 13.1 & 13.5 & 13.7 
\\[0.7mm]
&
$t, \bar t$
&
1.6 & 1.5 & 1.5 & 1.5 & 1.5 
\\[0.7mm] \hline
$H ^{+}$   
&
$\bar{b}, t$
& 
62.9 & 63.9 & 64.8 & 65.2 & 66.2
\\[0.7mm] 
&
$\tilde{\tau} _{1} , \tilde \nu _{\tau}$
& 
25.9 & 24.9 & 23.9 & 23.4 & 22.2
\\[0.7mm]
&
$\bar{\tau}, \nu _{\tau}$
& 
10.3 & 10.5 & 10.6 & 10.8 & 11.0
\\[0.7mm] \hline \hline
\end{tabular}
\caption{Branching ratios of the SUSY particles on the sample points.  }
\label{table:BR2}
\end{center}
\end{table}

\begin{table}
\begin{center}
\tabcolsep=4mm
\begin{tabular}{lccccc} \hline \hline
Cross Section
&
Point1 [fb] & Point2 [fb] & Point3 [fb] & Point4 [fb] & Point5 [fb]
\\[0.7mm] \hline
$\sigma(\tilde u _{L}, \tilde u _{L})$
&
2.915 & 1.873 & 1.277& 0.879& 0.614
\\[0.7mm] 
$\sigma(\tilde u _{L}, \tilde u _{R})$
&
1.672 & 1.024 & 0.668 & 0.441 & 0.296
\\[0.7mm] 
$\sigma(\tilde u _{R}, \tilde u _{R})$
&
2.970 & 1.926 & 1.327 & 0.923 & 0.652
\\[0.7mm] 
$\sigma(\tilde d _{L}, \tilde d _{L})$
&
0.377 & 0.225 & 0.144 & 0.095 & 0.061
\\[0.7mm] 
$\sigma(\tilde d _{L}, \tilde d _{R})$
&
0.194 & 0.110 & 0.068 & 0.042 & 0.026
\\[0.7mm] 
$\sigma(\tilde d _{R}, \tilde d _{R})$
&
0.381 & 0.230 & 0.149 & 0.098 & 0.065
\\[0.7mm] 
$\sigma(\tilde u _{L} , \tilde d _{L})$
&
3.243 & 2.016 & 1.335 & 0.894 &0.608
\\[0.7mm] 
$\sigma(\tilde u _{L} , \tilde d _{R})$
&
0.557 & 0.329 & 0.208 & 0.133 & 0.087
\\[0.7mm] 
$\sigma(\tilde u _{R} , \tilde d _{L})$
&
0.551 & 0.325 & 0.205 & 0.131 & 0.086
\\[0.7mm] 
$\sigma(\tilde u _{R} , \tilde d _{R})$
&
2.680 & 1.680 & 1.124 & 0.759 & 0.522
\\[0.7mm]
$\sigma(\tilde g , \tilde u _{L})$
&
2.735 & 1.506 & 0.899 & 0.537 & 0.330
\\[0.7mm] 
$\sigma(\tilde g , \tilde u _{R})$
&
3.156 & 1.750 & 1.041 & 0.633 & 0.391
\\[0.7mm] 
$\sigma(\tilde g , \tilde d _{L})$
&
0.826 & 0.440 & 0.252 & 0.148 & 0.088
\\[0.7mm] 
$\sigma(\tilde g , \tilde d _{R})$
&
0.981 & 0.527 & 0.305 & 0.180 & 0.109
\\[0.7mm] 
$\sigma(\tilde g , \tilde g)$
&
0.440 & 0.219 & 0.118 & 0.065 & 0.037
\\[0.7mm] 
$\sigma(\tilde u _{L} , \tilde u _{L} ^{\ast})$
&
0.059 & 0.032 & 0.019 & 0.012 & 0.006 
\\[0.7mm]  
$\sigma(\tilde u _{L} , \tilde u _{R} ^{\ast})$
&
0.220 & 0.126 & 0.078 & 0.049 & 0.031 
\\[0.7mm] 
$\sigma(\tilde u _{R}, \tilde u _{L} ^{\ast})$
&
0.220 & 0.126 & 0.078 & 0.049 & 0.031 
\\[0.7mm] 
$\sigma(\tilde u _{R} , \tilde u _{R} ^{\ast})$
&
0.084 & 0.047 & 0.027 & 0.016 & 0.011 
\\[0.7mm] 
$\sigma(\tilde d _{L} , \tilde d _{L} ^{\ast})$
&
0.037 & 0.019 & 0.011 & 0.006 & 0.003
\\[0.7mm]  
$\sigma(\tilde d _{L} , \tilde d _{R} ^{\ast})$
&
0.080 & 0.042 & 0.024 & 0.014 & 0.008
\\[0.7mm]  
$\sigma(\tilde d _{R} , \tilde d _{L} ^{\ast})$
&
0.080 & 0.042 & 0.024 & 0.014 & 0.008
\\[0.7mm]  
$\sigma(\tilde d _{R} , \tilde d _{R} ^{\ast})$
&
0.052 & 0.027 & 0.016 & 0.008 & 0.003
\\[0.7mm]  
$\sigma(\tilde u _{L} , \tilde d _{R} ^{\ast})$
&
0.254 & 0.139 & 0.083 & 0.050 & 0.030
\\[0.7mm] 
$\sigma(\tilde u _{R} , \tilde d _{L} ^{\ast})$
&
0.249 & 0.136 & 0.081 & 0.048 & 0.029
\\[0.7mm] 
$\sigma(\tilde u _{L} , \tilde d _{L} ^{\ast})$
&
0.035 & 0.018 & 0.010 & 0.005 & 0.003
\\[0.7mm]
$\sigma(\tilde u _{R} , \tilde d _{R} ^{\ast})$
&
0.056 & 0.029 & 0.017 & 0.009 & 0.006
\\[0.7mm] 
$\sigma(\tilde d _{L} , \tilde u _{L} ^{\ast})$
&
0.010 & 0.005 & 0.003 & 0.002 & 0.001 
\\[0.7mm]
$\sigma(\tilde d _{R} , \tilde u _{L} ^{\ast})$
&
0.069 & 0.038 & 0.023 & 0.014 & 0.009
\\[0.7mm]
$\sigma(\tilde d _{L} , \tilde u _{R} ^{\ast})$
&
0.069 & 0.037 & 0.022 & 0.014 & 0.008
\\[0.7mm] 
$\sigma(\tilde d _{R} , \tilde u _{R} ^{\ast})$
&
0.015 & 0.008 & 0.005 & 0.003 & 0.002
\\[0.7mm] 
$\sigma(\tilde g, \tilde u _{L} ^{\ast})$
&
0.051 & 0.025 & 0.014 & 0.008 & 0.004
\\[0.7mm] 
$\sigma(\tilde g, \tilde u _{R} ^{\ast})$
&
0.061 & 0.030 & 0.017 & 0.009 & 0.005
\\[0.7mm] 
$\sigma(\tilde g, \tilde d _{L} ^{\ast})$
&
0.045 & 0.021 & 0.011 & 0.006 & 0.003
\\[0.7mm] 
$\sigma(\tilde g, \tilde d _{R} ^{\ast})$
&
0.056 & 0.027 & 0.014 & 0.007 & 0.004
\\[0.7mm] \hline \hline
\end{tabular}
\caption{Cross sections of SUSY particles on the sample points.
We assume the energy in the center of mass system as 14 TeV at LHC experiment. 
}
\label{table:CS1}
\end{center}
 \end{table}

\begin{table}
\begin{center}
\tabcolsep=4mm
\begin{tabular}{lccccc} \hline \hline
Cross Section
&
Point1 [fb] & Point2 [fb] & Point3 [fb] & Point4 [fb] & Point5 [fb]
\\[0.7mm] \hline
$\sigma(\tilde t _{1}, \tilde t _{1} ^{\ast})$
& 4.399 & 4.704 & 3.662 & 3.245 & 2.655 
\\[0.7mm] 
$\sigma(\tilde t _{2}, \tilde t _{2} ^{\ast})$
& 0.180 & 0.129 & 0.085 & 0.058 & 0.039
\\[0.7mm] 
$\sigma(\tilde b _{1}, \tilde b _{1} ^{\ast})$
& 0.252 & 0.169 & 0.108 & 0.074 & 0.050
\\[0.7mm] 
$\sigma(\tilde b _{2}, \tilde b _{2} ^{\ast})$
& 0.089 & 0.050 & 0.030 & 0.018 & 0.011
\\[0.7mm] 
$\sigma(\tilde \chi ^{0} _{2}, \tilde \chi ^{0} _{2})$
& 0.051 & 0.035 & 0.024 & 0.018 & 0.013
\\[0.7mm] 
$\sigma(\tilde \chi ^{0} _{1}, \tilde \chi ^{0} _{1})$
& 0.103 & 0.037 & 0.028 & 0.022 & 0.018
\\[0.7mm] 
$\sigma(\tilde \chi ^{0} _{2}, \tilde g)$
& 0.101 & 0.062 & 0.040 & 0.026 & 0.017
\\[0.7mm] 
$\sigma(\tilde \chi ^{0} _{1}, \tilde g)$
& 0.114 & 0.073 & 0.049 & 0.034 & 0.023
\\[0.7mm] 
$\sigma(\tilde \chi ^{+} _{1}, \tilde \chi ^{-} _{1})$
& 1.229 & 0.861 & 0.629 & 0.470 & 0.355
\\[0.7mm] 
$\sigma(\tilde \chi ^{+} _{1}, \tilde \chi ^{0} _{2})$
& 3.514 & 2.499 & 1.858 & 1.404 & 1.075
\\[0.7mm] 
$\sigma(\tilde \chi ^{-} _{1}, \tilde \chi ^{0} _{2})$
& 1.232 & 0.852 & 0.616 & 0.455 & 0.341
\\[0.7mm] 
$\sigma(\tilde \chi ^{+} _{1}, \tilde \chi ^{0} _{1})$
& 0.023 & 0.017 & 0.012 & 0.005 & 0.003
\\[0.7mm] 
$\sigma(\tilde \chi ^{-} _{1}, \tilde \chi ^{0} _{1})$
& 0.008 & 0.005 & 0.004 & 0.003 & 0.002
\\[0.7mm] 
$\sigma(\tilde \chi ^{+} _{1}, \tilde g)$
& 0.344 & 0.209 & 0.134 & 0.087 & 0.058
\\[0.7mm] 
$\sigma(\tilde \chi ^{-} _{1}, \tilde g)$
& 0.094 & 0.057 & 0.037 & 0.024 & 0.016
\\[0.7mm]
$\sigma(\tilde \tau _{1}, \tilde \tau ^{\ast} _{1})$
& 0.422 & 0.313 & 0.241 & 0.188 & 0.149
\\[0.7mm] \hline
$\sigma(\text{all SUSY})$
& 37.730 & 25.268 & 17.277 & 12.445 & 8.456
\\[0.7mm] \hline \hline
\end{tabular}
\caption{Cross sections of SUSY particles on the sample points.
We assume the energy in the center of mass system as 14 TeV at LHC experiment. 
}
\label{table:CS2}
\end{center}
 \end{table}

\begin{table}
\begin{center}
\tabcolsep=2mm
\begin{tabular}{lccccc} \hline \hline
Cross Section $\times$Branching Ratio
& Point1 [fb] & Point2 [fb] & Point3 [fb] & Point4 [fb] & Point5 [fb]
\\[0.7mm] \hline
$\sigma (\tilde g ) \times \text{BR}(\tilde g \to \tilde \tau _{1})$
&
1.105 & 0.452 & 0.224 & 0.108 & 0.060
\\[0.7mm] 
$\sigma (\tilde g ) \times \text{BR}(\tilde g \to \tilde \tau _{1} ^{\ast})$
&
1.105 & 0.452 & 0.224 & 0.108 & 0.060
\\[0.7mm] 
$\sigma (\tilde u _{L}) \times \text{BR} (\tilde u _{L} \to \tilde \tau _{1})$
&
3.391 & 1.779 & 1.161 & 0.733 & 0.284
\\[0.7mm] 
$\sigma (\tilde u _{L}) \times \text{BR}(\tilde u _{L} \to \tilde \tau _{1} ^{\ast})$
&
11.021 & 7.023 & 4.600 & 3.066 & 1.195
\\[0.7mm] 
$\sigma (\tilde d _{L}) \times \text{BR}(\tilde d _{L} \to \tilde \tau _{1})$
&
4.203 & 2.618 & 1.682 & 1.103 & 0.730
\\[0.7mm] 
$\sigma (\tilde d _{L}) \times \text{BR}(\tilde d _{L} \to \tilde \tau _{1} ^{\ast})$
&
1.297 & 0.665 & 0.425 & 0.264 & 0.174
\\[0.7mm]
$\sigma (\tilde t _{1} ) \times \text{BR}(\tilde t _{1} \to \tilde \tau _{1})$
&
0.204& 0.103 & 0.060 & 0.029 & 0.013
\\[0.7mm] 
$\sigma (\tilde t _{1}) \times \text{BR}(\tilde t _{1} \to \tilde \tau _{1} ^{\ast})$
&
0.755 & 0.544 & 0.330 & 0.207 & 0.125
\\[0.7mm] 
$\sigma (\tilde t _{2}) \times \text{BR}(\tilde t _{2} \to \tilde \tau _{1})$
& 
0.019 & 0.008 & 0.005 & 0.003 & 0.002
\\[0.7mm] 
$\sigma (\tilde t _{2}) \times \text{BR}(\tilde t _{2} \to \tilde \tau _{1} ^{\ast})$
&
0.064 & 0.036 & 0.021 & 0.013 & 0.008
\\[0.7mm] 
$\sigma (\tilde b _{1}) \times \text{BR}(\tilde b _{1} \to \tilde \tau _{1})$
&
0.053 & 0.029 & 0.017 & 0.010 & 0.006
\\[0.7mm] 
$\sigma (\tilde b _{1}) \times \text{BR}(\tilde b _{1} \to \tilde \tau _{1} ^{\ast})$
&
0.036 & 0.018 & 0.008 & 0.004 & 0.002
\\[0.7mm]
$\sigma (\tilde b _{2}) \times \text{BR}(\tilde b _{2} \to \tilde \tau _{1})$
&
0.008 & 0.004 & 0.002 & 0.001 & 0.001
\\[0.7mm] 
$\sigma (\tilde b _{2}) \times \text{BR}(\tilde b _{2} \to \tilde \tau _{1} ^{\ast})$
&
0.010 & 0.005 & 0.002 & 0.001 & 0.001
\\[0.7mm]
$\sigma (\tilde u _{L} ^{\ast}) \times \text{BR}(\tilde u _{L} ^{\ast} \to \tilde \tau _{1})$
&
0.161 & 0.092 & 0.056 & 0.035 & 0.021
\\[0.7mm]
$\sigma (\tilde u _{L} ^{\ast}) \times \text{BR}(\tilde u _{L} ^{\ast} \to \tilde \tau _{1} ^{\ast})$
&
0.099 & 0.047 & 0.028 & 0.017 & 0.010
\\[0.7mm]
$\sigma (\tilde d _{L} ^{\ast}) \times \text{BR}(\tilde d _{L} ^{\ast} \to \tilde \tau _{1})$
&
0.107 & 0.049 & 0.028 & 0.015 & 0.009
\\[0.7mm]
$\sigma (\tilde d _{L} ^{\ast}) \times \text{BR}(\tilde d _{L} ^{\ast} \to \tilde \tau _{1} ^{\ast})$
&
0.348 & 0.191 & 0.111 & 0.065 & 0.037
\\[0.7mm] 
$\sigma (\tilde t _{1} ^{\ast} ) \times \text{BR}(\tilde t _{1} ^{\ast} \to \tilde \tau _{1})$
&
0.755 & 0.544 & 0.330 & 0.207 & 0.125
\\[0.7mm]
$\sigma (\tilde t _{1} ^{\ast} ) \times \text{BR}(\tilde t _{1} ^{\ast} \to \tilde \tau _{1} ^{\ast})$
&
0.204 & 0.103 & 0.060 & 0.029 & 0.013
\\[0.7mm]
$\sigma (\tilde t _{2} ^{\ast} ) \times \text{BR}(\tilde t _{2} ^{\ast} \to \tilde \tau _{1})$
&
0.064 & 0.036 & 0.021 & 0.013 & 0.008
\\[0.7mm]
$\sigma (\tilde t _{2} ^{\ast}) \times \text{BR}(\tilde t _{2} ^{\ast} \to \tilde \tau _{1} ^{\ast})$
&
0.019 & 0.008 & 0.005 & 0.003 & 0.002
\\[0.7mm]
$\sigma (\tilde b _{1} ^{\ast}) \times \text{BR}(\tilde b _{1} ^{\ast} \to \tilde \tau _{1})$
&
0.036 & 0.018 & 0.008 & 0.004 & 0.002
\\[0.7mm]
$\sigma (\tilde b _{1} ^{\ast}) \times \text{BR}(\tilde b _{1} ^{\ast} \to \tilde \tau _{1} ^{\ast})$
&
0.053 & 0.029 & 0.017 & 0.010 & 0.006
\\[0.7mm]
$\sigma (\tilde b _{2} ^{\ast}) \times \text{BR}(\tilde b _{2} ^{\ast} \to \tilde \tau _{1})$
&
0.010 & 0.005 & 0.002 & 0.001 & 0.001
\\[0.7mm]
$\sigma (\tilde b _{2} ^{\ast}) \times \text{BR}(\tilde b _{2} ^{\ast} \to \tilde \tau _{1} ^{\ast})$
&
0.008 & 0.004 & 0.002 & 0.001 & 0.001
\\[0.7mm]
$\sigma (\tilde \chi ^{0} _{2} ) \times \text{BR}(\tilde \chi ^{0} _{2} \to \tilde \tau _{1})$
&
2.281 & 1.672 & 1.229 & 0.928 & 0.705
\\[0.7mm] 
$\sigma (\tilde \chi ^{0} _{2}) \times \text{BR}(\tilde \chi ^{0} _{2} \to \tilde \tau _{1} ^{\ast})$
&
2.281 & 1.672 & 1.229 & 0.928 & 0.705
\\[0.7mm] 
$\sigma (\tilde \chi ^{+} _{1} ) \times \text{BR}(\tilde \chi ^{+} _{1} \to \tilde \tau _{1})$
&
0.627 & 0.223 & 0.162 & 0.090 & 0.065
\\[0.7mm] 
$\sigma (\tilde \chi ^{+} _{1} ) \times \text{BR}(\tilde \chi ^{+} _{1} \to \tilde \tau _{1} ^{\ast})$
&
4.679 & 3.418 & 2.510 & 1.885 & 1.430
\\[0.7mm] 
$\sigma (\tilde \chi ^{-} _{1}) \times \text{BR}(\tilde \chi ^{-} _{1} \to \tilde \tau _{1})$
&
2.347 & 1.692 & 1.226 & 0.913 & 0.685
\\[0.7mm] 
$\sigma (\tilde \chi ^{-} _{1}) \times \text{BR}(\tilde \chi ^{-} _{1} \to \tilde \tau _{1} ^{\ast})$
&
0.314 & 0.110 & 0.079 & 0.043 & 0.031
\\[0.7mm] 
$\sigma (p \to \tilde \tau _{1})$
&
0.422 & 0.313 & 0.241 & 0.188 & 0.149
\\[0.7mm] 
$\sigma (p \to \tilde \tau _{1} ^{\ast})$
&
0.422 & 0.313 & 0.241 & 0.188 & 0.149
\\[0.7mm]  \hline
Total : $\sigma (\tilde \tau _{1})$
&
15.954 & 9.728 & 7.739 & 4.415 & 2.885
\\[0.7mm] 
Total : $\sigma (\tilde \tau _{1} ^{\ast})$
&
22.715 & 16.310 & 9.893 & 6.831 & 3.948
\\[0.7mm] \hline
Number of produced $\tilde \tau ^{(\ast)} _{1}$
&
&
&
&
&
\\[0.7mm] \hline
$N(\tilde \tau _{1})$
&
1595 & 972 & 773 & 441 & 288
\\[0.7mm] 
$N(\tilde \tau ^{\ast} _{1})$
&
2271 & 1631 & 989 & 683 & 394
\\[0.7mm] \hline 
Number of produced $\tilde \chi _{1} ^{0}$
&
&
&
&
&
\\[0.7mm] \hline
$N(\tilde \chi _{1} ^{0} )$
&
3679 & 2449 & 1692 & 1364 & 1007
\\[0.7mm] \hline \hline
\end{tabular}
\caption{Summary of the cross sections and 
the branching ratios, and number of the produced stau and neutralino.  
We assume the energy in the center of mass system as 14 TeV at LHC experiment. 
In the estimation of number of produced stau and neutralino, 
we assume the luminosity as $100~\text{fb} ^{-1}$.  
}
\label{table:SCSBR}
\end{center}
\end{table}

\end{widetext}
\clearpage

\section{Summary}  \label{sec:sum} 

We have studied the scenario of the CMSSM in which the so-called 
lithium-7 problem can be solved via internal conversion processes with 
long-lived staus. 
For the abundance of lithium-7 to be reduced to the observed one, we 
imposed the following conditions, that (i) the mass difference of the stau 
NLSP and the neutralino LSP is smaller than $0.1~(1)$GeV, and (ii) the 
yield value of the stau NLSP is larger than $10^{-13}$. 
The first condition that guarantees enough long lifetime of the stau 
constrains the stau mass for a fixed neutralino mass, hence the scalar 
soft mass $m_0$ and the trilinear coupling $A_0$, while the second one 
that guarantees the sufficient reduction of the lithium-7 constrains the 
upper bound on the neutralino mass or the gaugino soft mass $M_{1/2}$. 
We analyzed the parameter space as well as the SUSY spectrum of the 
CMSSM by taking the recent results on the Higgs mass, SUSY searches and 
the dark matter abundance into account.

In Sec.~III, we have shown the allowed region on the $A_0$-$m_0$ and 
$m_0$-$M_{1/2}$ plane for $\delta m \leq 0.1$ and $1$GeV varying 
$\tan \beta=10,~20$ and $30$, respectively. We found a linear relation 
between $A_0$ and $m_0$ in all cases as given in Eq.\eqref{eq:linearrelation}. 
The relation originates from the tightly degenerate mass of the stau and the 
relic abundance of the neutralino. It was also found that the allowed region 
on $A_0$-$m_0$ plane shifts to higher scale as $\tan \beta$ increases. 
Thus the SUSY spectrum is heavier for large $\tan \beta$. 
On the other hand, as seen in Fig.\ref{fig:tb20_cut_m0m12mh_dm1},  
$M_{1/2}$ is constrained between $750 (950)$ and $1050$GeV for 
$\tan \beta=10 (30)$. The upper bound comes from the yield value of the 
stau while the lower one from the small mass difference and the dark matter 
abundance.  
Thus, the conditions required to solve the lithium-7 problem play an important 
role to determine the allowed region of the CMSSM parameters. Notably, these 
results naturally lead to heavy SUSY spectrum. The bounds on $M_{1/2}$ lead 
to the stau mass between $310 (400)$ and $450$GeV. Such heavy staus evade 
the present bound from direct searches at LHC.

In Figs.\ref{fig:msquarksmgluino}-\ref{fig:mheavyhiggs}, we have shown the 
SUSY spectrum and the heavier Higgs mass in the allowed region. One can see 
that the whole spectrum is relatively heavy so that it is consistent with the null 
results of the SUSY searches at the LHC experiment. 
Among the spectrum, one of the important predictions is the masses of the 
gluino and the $1$st/$2$nd generation squarks. In Fig.\ref{fig:msquarksmgluino}, 
it was shown that the the masses are clearly correlated with the neutralino mass. 
This is the direct consequences from the linear relation of $A_0$ and $m_0$. 
Thus in our scenario, once one of them is determined, the others can be predicted. 
It is important to emphasize here that the masses of gluino and squarks are 
constrained between $1.6 (1.8)$ and $2.3$TeV for $\tan\beta=10 (30)$. 
These gluino and squark masses are out of reach at $8$ TeV LHC run but really in 
reach of $14$TeV LHC run. The expected numbers of the long-lived staus (nearly 
degenerate neutralino) to $100$fb$^{-1}$ are about $3900~(3700)$ for the 
light neutralino mass and $710~(980)$ for the heavy one, respectively. 
With these numbers, the long-lived stau and the nearly degenerate neutralino 
will be identified easily. Thus our scenario explains why SUSY has not been found 
yet, and at the same time predicts the early discovery of SUSY in the coming few 
years.
The other important prediction is the stop mass. The stops are also relatively 
heavy in our scenario. Such stop masses gives large radiative corrections to the 
Higgs mass, and in fact the Higgs mass is pushed up to $125$GeV in sizable 
regions of the parameter space. 
The production cross section of such a lighter stop pair is comparable to those 
of the gluino and the squarks, and the expected number of the stops to 
$100$fb$^{-1}$ are between $880$ and $540$. Thus we can expect that the 
lighter stops also will be found at the $14$TeV run of LHC.

The SUSY spectrum is consistent with the present results on muon anomalous 
magnetic moment within $3~\sigma$ and $b \rightarrow s + \gamma$ and 
$B_s \rightarrow \mu \mu$ within $1~\sigma$. The direct detection cross 
section of the neutralino is from $5 \times 10^{-47}$ to $10^{-47}$cm$^2$ 
and is in reach of XENON $1$ton and LUX/ZEP $20$ton.

In conclusion, our scenario is predictive and indeed testable in the coming 
$14$TeV LHC run. When SUSY is the solution of the lithium-7 problem and its 
breaking is controlled in the CMSSM framework, the time to discover SUSY is 
coming soon.

\section*{Acknowledgments}   

The work of Y.K. was financially supported by the Sasakawa Scientific
Research  Grant from The Japan Science Society.
This  work was supported in part by the Grant-in-Aid for the Ministry of Education,
Culture, Sports, Science, and Technology, Government of Japan,
No. 24340044 (J.S.). and No. 25003345 (M.Y.) and No. 23740190 (T.S.).



\begin{thebibliography}{99}  




\bibitem{2012gk}
  G.~Aad {\it et al.}  [ATLAS Collaboration],
  Phys.\ Lett.\ B {\bf 716} (2012) 1  [arXiv:1207.7214  [hep-ex]].  


\bibitem{2012gu}
  S.~Chatrchyan {\it et al.}  [CMS Collaboration],
  Phys.\ Lett.\ B {\bf 716} (2012) 30  [arXiv:1207.7235 [hep-ex]].  



\bibitem{King:1995vk} 
  S.~F.~King and P.~L.~White,
  Phys.\ Rev.\ D {\bf 52}, 4183 (1995)
  [hep-ph/9505326].


\bibitem{Buchmueller:2011ab} 
  O.~Buchmueller, R.~Cavanaugh, A.~De Roeck, M.~J.~Dolan, J.~R.~Ellis, H.~Flacher, S.~Heinemeyer and G.~Isidori {\it et al.},
  Eur.\ Phys.\ J.\ C {\bf 72}, 2020 (2012)
  [arXiv:1112.3564 [hep-ph]].

\bibitem{Kadastik:2011aa} 
  M.~Kadastik, K.~Kannike, A.~Racioppi and M.~Raidal,
  JHEP {\bf 1205}, 061 (2012)
  [arXiv:1112.3647 [hep-ph]].



\bibitem{Okada:1990vk}
  Y.~Okada, M.~Yamaguchi and T.~Yanagida,
  Prog.\ Theor.\ Phys.\  {\bf 85} (1991) 1.  

\bibitem{Ellis:1990nz}
  J.~R.~Ellis, G.~Ridolfi and F.~Zwirner,
  Phys.\ Lett.\ B {\bf 257} (1991) 83.  
  
\bibitem{Haber:1990aw}
  H.~E.~Haber and R.~Hempfling,
  Phys.\ Rev.\ Lett.\  {\bf 66} (1991) 1815.  
  
\bibitem{Casas:1994us}
  J.~A.~Casas, J.~R.~Espinosa, M.~Quiros and A.~Riotto,
  Nucl.\ Phys.\ B {\bf 436} (1995) 3   [Erratum-ibid.\ B {\bf 439} (1995) 466]  
  [hep-ph/9407389].  




\bibitem{Navarro:1995iw}
  J.~F.~Navarro, C.~S.~Frenk and S.~D.~M.~White,
  Astrophys.\ J.\  {\bf 462} (1996) 563  
  [astro-ph/9508025].  

\bibitem{Clowe:2006eq}
  D.~Clowe, M.~Bradac, A.~H.~Gonzalez, M.~Markevitch, S.~W.~Randall, 
  C.~Jones and D.~Zaritsky,
  Astrophys.\ J.\  {\bf 648} (2006) L109  [astro-ph/0608407].  

\bibitem{Springel:2008cc}
  V.~Springel, J.~Wang, M.~Vogelsberger, A.~Ludlow, A.~Jenkins, 
  A.~Helmi, J.~F.~Navarro and C.~S.~Frenk {\it et al.},
  Mon.\ Not.\ Roy.\ Astron.\ Soc.\  {\bf 391} (2008) 1685  
  [arXiv:0809.0898 [astro-ph]].  


\bibitem{BoylanKolchin:2009nc}
  M.~Boylan-Kolchin, V.~Springel, S.~D.~M.~White, A.~Jenkins and G.~Lemson,
  Mon.\ Not.\ Roy.\ Astron.\ Soc.\  {\bf 398} (2009) 1150  
  [arXiv:0903.3041 [astro-ph.CO]].  


\bibitem{Guo:2010ap}
  Q.~Guo, S.~White, M.~Boylan-Kolchin, G.~De Lucia, G.~Kauffmann, 
  G.~Lemson, C.~Li and V.~Springel {\it et al.},
  Mon.\ Not.\ Roy.\ Astron.\ Soc.\  {\bf 413} (2011) 101  
  [arXiv:1006.0106 [astro-ph.CO]].  


\bibitem{Hinshaw:2012fq}
  G.~Hinshaw, D.~Larson, E.~Komatsu, D.~N.~Spergel, C.~L.~Bennett, 
  J.~Dunkley, M.~R.~Nolta and M.~Halpern {\it et al.},
  arXiv:1212.5226 [astro-ph.CO].  



\bibitem{Feng:1999mn}
  J.~L.~Feng, K.~T.~Matchev and T.~Moroi,
  Phys.\ Rev.\ Lett.\  {\bf 84} (2000) 2322  [hep-ph/9908309].  

\bibitem{Feng:1999zg}
  J.~L.~Feng, K.~T.~Matchev and T.~Moroi,
  Phys.\ Rev.\ D {\bf 61} (2000) 075005  [hep-ph/9909334].  



\bibitem{Aprile:2011hi}
  E.~Aprile {\it et al.}  [XENON100 Collaboration],
  Phys.\ Rev.\ Lett.\  {\bf 107} (2011) 131302  
  [arXiv:1104.2549 [astro-ph.CO]].  

\bibitem{Buchmueller:2012hv}
  O.~Buchmueller, R.~Cavanaugh, M.~Citron, A.~De Roeck, M.~J.~Dolan, 
  J.~R.~Ellis, H.~Flacher and S.~Heinemeyer {\it et al.},
  Eur.\ Phys.\ J.\ C {\bf 72} (2012) 2243  
  [arXiv:1207.7315 [hep-ph]].  



\bibitem{Griest:1990kh}
  K.~Griest and D.~Seckel,
  Phys.\ Rev.\  D {\bf 43} (1991) 3191.

\bibitem{Edsjo:1997bg}
  J.~Edsjo and P.~Gondolo,
  Phys.\ Rev.\  D {\bf 56} (1997) 1879



\bibitem{Aparicio:2012iw}
  L.~Aparicio, D.~G.~Cerdeno and L.~E.~Ibanez,
  JHEP {\bf 1204} (2012) 126  [arXiv:1202.0822 [hep-ph]].  


\bibitem{Citron:2012fg} 
  M.~Citron, J.~Ellis, F.~Luo, J.~Marrouche, K.~A.~Olive and K.~J.~de Vries,
  Phys.\ Rev.\ D {\bf 87}, 036012 (2013)
  [arXiv:1212.2886 [hep-ph]].





\bibitem{Profumo:2004qt}
  S.~Profumo, K.~Sigurdson, P.~Ullio and M.~Kamionkowski,
  Phys.\ Rev.\  D {\bf 71} (2005) 023518
  [arXiv:astro-ph/0410714].

\bibitem{Jittoh:2005pq}
  T.~Jittoh, J.~Sato, T.~Shimomura and M.~Yamanaka,
  Phys.\ Rev.\  D {\bf 73} (2006) 055009
  [arXiv:hep-ph/0512197].




\bibitem{Jedamzik:2004er}
  K.~Jedamzik,
  Phys.\ Rev.\  D {\bf 70}, 063524 (2004)
  [arXiv:astro-ph/0402344].

\bibitem{Kawasaki:2004yh}
  M.~Kawasaki, K.~Kohri and T.~Moroi,
  Phys.\ Lett.\  B {\bf 625}, 7 (2005)
  [arXiv:astro-ph/0402490].

\bibitem{Kawasaki:2004qu}
  M.~Kawasaki, K.~Kohri and T.~Moroi,
  Phys.\ Rev.\  D {\bf 71}, 083502 (2005)
  [arXiv:astro-ph/0408426].

\bibitem{Pospelov:2006sc}
  M.~Pospelov,
  Phys.\ Rev.\ Lett.\  {\bf 98}, 231301 (2007)
  [arXiv:hep-ph/0605215].

\bibitem{Kohri:2006cn}
  K.~Kohri and F.~Takayama,
  Phys.\ Rev.\  D {\bf 76}, 063507 (2007)
  [arXiv:hep-ph/0605243].
  
\bibitem{Kaplinghat:2006qr}
  M.~Kaplinghat and A.~Rajaraman,
  Phys.\ Rev.\  D {\bf 74}, 103004 (2006)
  [arXiv:astro-ph/0606209].

\bibitem{Cyburt:2006uv}
  R.~H.~Cyburt, J.~R.~Ellis, B.~D.~Fields, K.~A.~Olive and V.~C.~Spanos,
  JCAP {\bf 0611}, 014 (2006)
  [arXiv:astro-ph/0608562].

\bibitem{Hamaguchi:2007mp}
  K.~Hamaguchi, T.~Hatsuda, M.~Kamimura, Y.~Kino and T.~T.~Yanagida,
  Phys.\ Lett.\  B {\bf 650} (2007) 268
  [arXiv:hep-ph/0702274].    

\bibitem{Bird:2007ge}
  C.~Bird, K.~Koopmans and M.~Pospelov,
  Phys.\ Rev.\  D {\bf 78}, 083010 (2008)
  [arXiv:hep-ph/0703096].

\bibitem{Kawasaki:2007xb}
  M.~Kawasaki, K.~Kohri and T.~Moroi,
  Phys.\ Lett.\  B {\bf 649}, 436 (2007)
  [arXiv:hep-ph/0703122].

\bibitem{Jittoh:2007fr}
  T.~Jittoh, K.~Kohri, M.~Koike, J.~Sato, T.~Shimomura and M.~Yamanaka,
  Phys.\ Rev.\  D {\bf 76} (2007) 125023
  [arXiv:0704.2914 [hep-ph]].

\bibitem{Jedamzik:2007cp}
  K.~Jedamzik,
  Phys.\ Rev.\  D {\bf 77}, 063524 (2008)
  [arXiv:0707.2070 [astro-ph]].

\bibitem{Cumberbatch:2007me} 
  D.~Cumberbatch, K.~Ichikawa, M.~Kawasaki, K.~Kohri, J.~Silk and G.~D.~Starkman,
  Phys.\ Rev.\ D {\bf 76}, 123005 (2007)
  [arXiv:0708.0095 [astro-ph]].

\bibitem{Pradler:2007is}
  J.~Pradler and F.~D.~Steffen,
  Phys.\ Lett.\  B {\bf 666}, 181 (2008)
  [arXiv:0710.2213 [hep-ph]].

\bibitem{Kawasaki:2008qe}
  M.~Kawasaki, K.~Kohri, T.~Moroi and A.~Yotsuyanagi,
  Phys.\ Rev.\  D {\bf 78}, 065011 (2008)
  [arXiv:0804.3745 [hep-ph]].

\bibitem{Jittoh:2008eq}
  T.~Jittoh, K.~Kohri, M.~Koike, J.~Sato, T.~Shimomura and M.~Yamanaka,
  Phys.\ Rev.\  D {\bf 78} (2008) 055007
  [arXiv:0805.3389 [hep-ph]]. 

\bibitem{Kusakabe:2008kf}
  M.~Kusakabe, T.~Kajino, T.~Yoshida, T.~Shima, Y.~Nagai and T.~Kii,
  Phys.\ Rev.\  D {\bf 79}, 123513 (2009)
  [arXiv:0806.4040 [astro-ph]].
  
\bibitem{Pospelov:2008ta}
  M.~Pospelov, J.~Pradler and F.~D.~Steffen,
  JCAP {\bf 0811}, 020 (2008)
  [arXiv:0807.4287 [hep-ph]].

\bibitem{Kamimura:2008fx}
  M.~Kamimura, Y.~Kino and E.~Hiyama,
  Prog.\ Theor.\ Phys.\  {\bf 121}, 1059 (2009)
  [arXiv:0809.4772 [nucl-th]].

\bibitem{Kohri:2008cf}
  K.~Kohri and Y.~Santoso,
  Phys.\ Rev.\  D {\bf 79}, 043514 (2009)
  [arXiv:0811.1119 [hep-ph]].


\bibitem{Bailly:2008yy}
  S.~Bailly, K.~Jedamzik and G.~Moultaka,
  Phys.\ Rev.\  D {\bf 80}, 063509 (2009)
  [arXiv:0812.0788 [hep-ph]].

\bibitem{Bailly:2009pe}
  S.~Bailly, K.~Y.~Choi, K.~Jedamzik and L.~Roszkowski,
  JHEP {\bf 0905}, 103 (2009)
  [arXiv:0903.3974 [hep-ph]].

\bibitem{Cyburt:2009pg}
  R.~H.~Cyburt, J.~Ellis, B.~D.~Fields, F.~Luo, K.~A.~Olive and V.~C.~Spanos,
  JCAP {\bf 0910}, 021 (2009)
  [arXiv:0907.5003 [astro-ph.CO]].

\bibitem{Jittoh:2010wh}
  T.~Jittoh, K.~Kohri, M.~Koike, J.~Sato, T.~Shimomura and M.~Yamanaka,
  Phys.\ Rev.\  D {\bf 82} (2010) 115030
  [arXiv:1001.1217 [hep-ph]].  

\bibitem{Kusakabe:2010cb}
  M.~Kusakabe, T.~Kajino, T.~Yoshida and G.~J.~Mathews,
  Phys.\ Rev.\  D {\bf 81}, 083521 (2010)
  [arXiv:1001.1410 [astro-ph.CO]].

\bibitem{Pospelov:2010cw}
  M.~Pospelov and J.~Pradler,
  Phys.\ Rev.\  D {\bf 82}, 103514 (2010)
  [arXiv:1006.4172 [hep-ph]].   

\bibitem{Pospelov:2010hj}
  M.~Pospelov and J.~Pradler,
  Ann.\ Rev.\ Nucl.\ Part.\ Sci.\  {\bf 60} (2010) 539  
  [arXiv:1011.1054 [hep-ph]].  

\bibitem{Kawasaki:2010yh}
  M.~Kawasaki and M.~Kusakabe,
  Phys.\ Rev.\  D {\bf 83}, 055011 (2011)
  [arXiv:1012.0435 [hep-ph]].

\bibitem{Jittoh:2011ni}
  T.~Jittoh, K.~Kohri, M.~Koike, J.~Sato, K.~Sugai, M.~Yamanaka and K.~Yazaki,
  Phys.\ Rev.\  D {\bf 84} (2011) 035008
  [arXiv:1105.1431 [hep-ph]].

\bibitem{Kusakabe:2012ds}
  M.~Kusakabe, A.~B.~Balantekin, T.~Kajino and Y.~Pehlivan,
  Phys.\ Lett.\ B {\bf 718} (2013) 704  [arXiv:1202.5603 [astro-ph.CO]].  

\bibitem{Kohri:2012gc}
  K.~Kohri, S.~Ohta, J.~Sato, T.~Shimomura and M.~Yamanaka,
  Phys.\ Rev.\ D {\bf 86} (2012) 095024  [arXiv:1208.5533 [hep-ph]].  

\bibitem{Kusakabe:2013tra}
  M.~Kusakabe, K.~S.~Kim, M.~-K.~Cheoun, T.~Kajino and Y.~Kino,
  arXiv:1305.6155 [astro-ph.CO].



\bibitem{Monaco:2010mm}
  L.~Monaco, P.~Bonifacio, L.~Sbordone, S.~Villanova and E.~Pancino,
  Astron.\ Astrophys.\  {\bf 519} (2010) L3.
  arXiv:1008.1817 [astro-ph.GA].  
  


\bibitem{Coc:2011az}
  A.~Coc, S.~Goriely, Y.~Xu, M.~Saimpert and E.~Vangioni,
  Astrophys.\ J.\  {\bf 744} (2012) 158  
  [arXiv:1107.1117 [astro-ph.CO]].  



\bibitem{Spite:1982dd}
  F.~Spite and M.~Spite,
  Astron.\ Astrophys.\  {\bf 115} (1982) 357.  




\bibitem{CMS:aya} 
  [CMS Collaboration],
  CMS-PAS-HIG-12-045.
{ATLAS:2012klq}{CMS:aya}
   
\bibitem{ATLAS:2012klq} 
  [ATLAS Collaboration],
  ATLAS-CONF-2012-170.


\bibitem{Allanach:2001hm} 
  B.~C.~Allanach,
  eConf C {\bf 010630}, P319 (2001)
  [hep-ph/0110227].

\bibitem{Djouadi:2002nh} 
  A.~Djouadi,
  hep-ph/0211357.

\bibitem{Allanach:2003jw} 
  B.~C.~Allanach, S.~Kraml and W.~Porod,
  JHEP {\bf 0303}, 016 (2003)
  [hep-ph/0302102].

\bibitem{Allanach:2004rh} 
  B.~C.~Allanach, A.~Djouadi, J.~L.~Kneur, W.~Porod and P.~Slavich,
  JHEP {\bf 0409}, 044 (2004)
  [hep-ph/0406166].



\bibitem{Belanger:2013oya}
  G.~Belanger, F.~Boudjema, A.~Pukhov and A.~Semenov,
  arXiv:1305.0237 [hep-ph].



\bibitem{Porod:2003um}
  W.~Porod,
  Comput.\ Phys.\ Commun.\  {\bf 153} (2003) 275
  [hep-ph/0301101].

\bibitem{Porod:2011nf}
  W.~Porod and F.~Staub,
  Comput.\ Phys.\ Commun.\  {\bf 183} (2012) 2458
  [arXiv:1104.1573 [hep-ph]].


\bibitem{Heinemeyer:1998yj}
  S.~Heinemeyer, W.~Hollik and G.~Weiglein,
  Comput.\ Phys.\ Commun.\  {\bf 124} (2000) 76
  [hep-ph/9812320].

\bibitem{Heinemeyer:1998np}
  S.~Heinemeyer, W.~Hollik and G.~Weiglein,
  Eur.\ Phys.\ J.\ C {\bf 9} (1999) 343
  [hep-ph/9812472].

\bibitem{Degrassi:2002fi}
  G.~Degrassi, S.~Heinemeyer, W.~Hollik, P.~Slavich and G.~Weiglein,
  Eur.\ Phys.\ J.\ C {\bf 28} (2003) 133
  [hep-ph/0212020].

\bibitem{Frank:2006yh}
  M.~Frank, T.~Hahn, S.~Heinemeyer, W.~Hollik, H.~Rzehak and G.~Weiglein,
  JHEP {\bf 0702} (2007) 047
  [hep-ph/0611326].






\bibitem{Martin:1997ns} 
  S.~P.~Martin,
  In *Kane, G.L. (ed.): Perspectives on supersymmetry II* 1-153
  [hep-ph/9709356].




\bibitem{Pradler:2008qc} 
  J.~Pradler and F.~D.~Steffen,
  Nucl.\ Phys.\ B {\bf 809}, 318 (2009)
  [arXiv:0808.2462 [hep-ph]].




\bibitem{Chatrchyan:2013oca} 
  S.~Chatrchyan {\it et al.}  [CMS Collaboration],
  arXiv:1305.0491 [hep-ex].




\bibitem{ATLAS:2013pla} 
  [ATLAS Collaboration],
 ATLAS-CONF-2013-037 .




\bibitem{Bennett:2006fi}
  G.~W.~Bennett {\it et al.}  [Muon G-2 Collaboration],
  Phys.\ Rev.\ D {\bf 73} (2006) 072003
  [hep-ex/0602035].

\bibitem{Hagiwara:2011af}
  K.~Hagiwara, R.~Liao, A.~D.~Martin, D.~Nomura and T.~Teubner,
  J.\ Phys.\ G {\bf 38} (2011) 085003
  [arXiv:1105.3149 [hep-ph]].

%

\bibitem{Aaij:2012nna}
  RAaij {\it et al.}  [LHCb Collaboration],
  Phys.\ Rev.\ Lett.\  {\bf 110} (2013) 021801
  [arXiv:1211.2674 [hep-ex]].

\bibitem{Amhis:2012bh}
  Y.~Amhis {\it et al.}  [Heavy Flavor Averaging Group Collaboration],
  arXiv:1207.1158 [hep-ex].



\bibitem{Beringer:2012}
  J.~beringer {\it et al.}  [Partticle Data Group Collaboration], 
   Phys.\ Rev.\ D {\bf 86} (2012) 010001
   
\bibitem{Jungman:1995df}
  G.~Jungman, M.~Kamionkowski and K.~Griest,
  Phys.\ Rept.\  {\bf 267} (1996) 195  [hep-ph/9506380].  

\bibitem{Aprile:2012nq}
  E.~Aprile {\it et al.}  [XENON100 Collaboration],
  Phys.\ Rev.\ Lett.\  {\bf 109} (2012) 181301  
  [arXiv:1207.5988 [astro-ph.CO]].  


\bibitem{Aprile:2012zx}
  E.~Aprile [XENON1T Collaboration],
  arXiv: 1206.6288 [astro-ph.IM].  


\bibitem{Akerib:2012ak}
  D.~S.~Akerib {\it et al.}  [LUX Collaboration],
  Astropart.\ Phys.\  {\bf 45} (2013) 34  [arXiv:1210.4569 [astro-ph.IM]].  






\bibitem{Belyaev:2012qa}
  A.~Belyaev, N.~D.~Christensen and A.~Pukhov,
  Comput.\ Phys.\ Commun.\  {\bf 184} (2013) 1729
  [arXiv:1207.6082 [hep-ph]].



 











  

\end{thebibliography}
\end{document}